\documentclass[a4paper,usedcolumn,usenatbib]{mnras}

\usepackage[utf8]{inputenc}
\usepackage{ae,aecompl}

\usepackage{graphicx}	% Including figure files
\usepackage{amsmath}	% Advanced maths commands
\usepackage{amssymb}	% Extra maths symbols
\usepackage{xcolor}

\usepackage{longtable}
\usepackage{ltxtable}
\usepackage{supertabular,booktabs}

\usepackage{booktabs}
\usepackage[detect-all]{siunitx}

\usepackage[labelfont=bf,small]{caption}
\usepackage{hyperref} 
\usepackage{graphicx}	% Including figure files
\usepackage{amsmath}	% Advanced maths commands
\usepackage{amssymb}	% Extra maths symbols
\usepackage{float}

\usepackage{mathtools}
\usepackage[export]{adjustbox}

\newcommand\eg{{\it e.g.}\ }
\newcommand\ie{{\it i.e.}\ }

\newcommand{\LCDM}{$\Lambda$CDM }

\title[EasyCritics II. Testing its efficiency and new SL candidates]
{EasyCritics II. Testing its efficiency: new gravitational lens candidates in CFHTLenS}

\author[Carrasco et al.]{Mauricio Carrasco,$^{1}$
Sebastian Stapelberg,$^{1}$
Matteo Maturi,$^{1}$
\newauthor
Matthias Bartelmann,$^{1}$
Gregor Seidel,$^{2}$
Thomas Erben$^{3}$
\\
$^{1}$Zentrum für Astronomie, Institut für Theoretische Astrophysik, Philosophenweg 12, 69120, Heidelberg, Germany.\\
$^{2}$Max-Planck-Institute for Astronomy, K\"onigstuhl 17, D-69117 Heidelberg, Germany.\\
$^{3}$Argelander Institute for Astronomy, University of Bonn, Auf dem H\"ugel 71, D-53121 Bonn, Germany.\\
}

\date{}%{Accepted XXX. Received YYY; in original form ZZZ}

\pubyear{2017}

\begin{document}
\label{firstpage}
\pagerange{\pageref{firstpage}--\pageref{lastpage}}
\maketitle

\begin{abstract}
We report the results of \textit{EasyCritics}, a fully automated algorithm for the efficient search of strong-lensing (SL) regions 
in wide-field surveys, applied to the Canada-France-Hawaii Telescope Lensing Survey (CFHTLenS).
By using only the photometric information of the brightest elliptical galaxies distributed over a wide redshift range ($\smash{0.2 \lesssim z \lesssim 0.9}$)
and without requiring the identification of arcs,
our algorithm produces lensing potential models 
and catalogs of critical curves of the entire survey area. 
We explore several parameter set configurations in order to 
test the efficiency of our approach. 
In a specific configuration, \textit{EasyCritics} generates only $\sim1200$ possibly super-critical regions in the CFHTLS area, 
drastically reducing the effective area for inspection from $154$ sq. deg to $\sim0.623$ sq. deg,
\ie by more than two orders of magnitude.
Among the pre-selected SL regions, we identify 32 of the 44 previously known lenses on the group and cluster scale, 
and discover 9 new promising lens candidates.
The detection rate can be easily improved to $\sim82\%$
 by a simple modification in the parameter set, but at the expense of increasing the total number of possible SL candidates.
Note that \textit{EasyCritics} is fully complementary to other arc-finders since 
we characterize lenses instead of directly identifying arcs. 
 Although future comparisons against numerical simulations are required for fully assessing
  the efficiency of \textit{EasyCritics}, the algorithm seems very promising for 
  upcoming surveys covering $\smash{10^{4}}$  sq. deg, such as the \textit{Euclid} mission and \textit{LSST}, 
  where the pre-selection of candidates for any kind of SL analysis will be indispensable due to the expected enormous data volume.
\end{abstract}


%%%%%%%%%%%%%%%%%%%%%%%%%%%%%%%%%%%%%%%%%%%%%%%%%%

%%%%%%%%%%%%%%%%% BODY OF PAPER %%%%%%%%%%%%%%%%%%

\section{Introduction} \label{sec:intro}

Gravitational lensing by clusters of galaxies is among the main 
cosmological tools to access the nature of dark matter (DM) and the far universe. 
Strong lensing (SL)
signatures probe the inner matter distribution of galaxy clusters,
 allowing robust mass reconstruction and magnification 
studies of the clusters core \citep[e.g.][]{Kneib2003, 
Broadhurst2005a,Broadhurst2005b, Zitrin2012CLASH0329, Coe12,Limousin2012_M0717, 
Jauzac2015}. 
The apparent flux magnification 
turns galaxy clusters into gravitational telescopes, which can be used to
study very high-redshift galaxies
which would otherwise be too faint to be observed 
\citep[e.g.][]{Stark07,Richard08,Bouwens09, Bradac09, Hall12, Zheng2012NaturZ,Coe13}. 
Moreover, the abundance of strongly lensed background galaxies,
 which appear as gravitational arcs and multiple images, can be compared against predictions of 
the lensing efficiency of cluster-scale halos in 
simulations to test the current cosmological framework 
\citep[][]{ Bartelmann98,Bartelmann2003_arc_st,Meneghetti2013_arc_st}.
Several theoretical and observational studies have found that the $\Lambda$CDM
cosmological models underestimate the number of giant arcs on the 
sky by perhaps as much as an order of magnitude 
\citep[known as the ``arc statistics problem'';][]{Bartelmann98, Gladders03, Li06}.
Besides, a long series of theoretical works have explored 
a variety of astrophysical effects of both the lenses and 
the sources, which can mitigate the tension of the arc statistics
problem \cite[e.g.][]{Dalal+2004arcs, Wambsganss2004_arc_st,
Meneghetti2003, Torri2004, Puchwein05,
meneghetti2007,Wambsganss2008_arc_st, Mead10}; 
however, the discrepancy is not yet solved.
It is important to notice that all studies conducted so far suffer from the lack
 of systematic arc surveys and from the limited abundance of giant arcs \citep{Bayliss12},
 making the available sample not uniform \citep{Meneghetti2011,Meneghetti2013_arc_st},
 as well as by the fact that different approaches have been used.
Therefore, enlarging and standardizing the sample of giant arcs
by using a well-characterized selection function and 
common comparison methods, is mandatory
for explaining or alleviating this controversy.

In recent years, automated algorithms have been 
developed for the search of gravitational arcs and multiple images \citep{Lenzen2004_arcfinder,
Horesh2005_arcfinder, Alard2006_arcfinder,Seidel2007Arcfinder,
Joseph2014_arcfinder, More_2012_arcfinder, Gavazzi_2014_ringfinder};
however these approaches suffer from strong contamination and require
a large amount of human intervention 
\citep{More_2012_arcfinder, Limousin09_SL_Ggroup, Cabanac07, Maturi14}. 
Moreover, most of the detections are just candidates and only few hundred cases have been spectroscopically confirmed 
\citep[e.g.][]{Bayliss11b, Oguri12, Carrasco17_spec}.
As a matter of fact, the search for SL systems
has been conducted almost exclusively by visual inspection
of several hundred thousand images.
An attempt to deal with such a large amount of eye-balling has been based on citizen science, 
the SPACE WARPS project \citep[SW;][]{Marshall_2016_SWII, More_2016_SWI}.
Through crowd-sourced visual inspection, this program 
yields high purity and completeness samples.
SW has recovered about $65\%$ of known lenses previously discovered on
the Canada--France--Hawaii Telescope
Legacy Survey (CFHTLS\footnote{http://www.cfht.hawaii.edu/Science/CFHTLS/}), 
by dividing its $\sim160$ sq. deg into 
$\sim430 000$ overlapping 82 by 82 arcseconds tiles and 
performing more than 11 million image classifications over the course
of 8 months, with the help of $\sim37 000$ volunteers.
Despite these results, SW depends on the number of citizen 
scientists and their performance, making it less competitive for the 
upcoming wide-field surveys ($10^4$ sq. deg area) where there will be of the order of 
$10^7$ images to visually inspect and classify (assuming that the same tiling strategy is used).
Performing those tasks in reasonable time thus requires to either increase
the number of volunteers by orders of magnitude
or 
change completely the approach and rely on robust and efficient automated pre-selection methods to
robotically reduce the number of targets to be inspected.

In order to perform such a search for gravitational arcs on wide-field surveys with 
a significantly reduced number of spurious detections 
and minimal human intervention, 
in \cite{Stapelberg_2017_EasyCriticsI} we present 
our fully automated algorithm ``\textit{EasyCritics}".
Based on the assumption that light traces mass \citep[LTM;][]{Broadhurst05,UmetsuBroadhurst08, Zitrin09a},
 our algorithm constructs a simple model of the lensing potential for the
total mass projected along each line of sight (LOS) on wide field surveys  by using the 
flux and position of the brightest elliptical galaxies, 
in combination with their photometric redshift information and angular size.
\textit{EasyCritics} then finds the most likely super-critical regions on the sky, 
i.e. where the total surface mass density integrated along the 
LOS is sufficient to produce SL events.

The development of \textit{EasyCritics} has been motivated by successful results
of several previous mass reconstruction studies \citep[e.g.][]{Zitrin09a, Limousin07,Limousin09_SL_Ggroup, Limousin2012_M0717,
Richard2010locuss20,Jauzac2015, Caminha2017_MACSJ1206, Caminha2017_MACSJ0416};
in particular, by the lens modeling analysis of several massive galaxy clusters
presented in \cite{Zitrin11MACS} and by the arc-free approach
described in \cite{Zitrin2012UniversalRE}. 
In those studies, they have pointed out that, by constructing a simple \textit{preliminary} mass model 
based only on the light distribution of the  cluster members (mostly elliptical galaxies),
  critical curves can be correctly predicted. 
Additionally, we have also included into our approach the contribution of extra but uncorrelated structures projected along the LOS,
since it has been proved to significantly affect the total lensing cross section  
\citep[][]{Wambsganss2005_LOS, Hilbert07,PuchweinHilbert2009, Wong2012_LOS, Ammons2014_LOS, Bayliss2014_LOS}. 
\cite{Wong2012_LOS} have shown that multiple small cluster-scale halos along the LOS
can enhance the lensing cross section, reaching, in some cases, values comparable to
those of single massive halos. 
Therefore, \textit{EasyCritics} not only searches for single massive SL galaxy clusters
but also for alignments of multiple group- and small cluster-scale halos, which
may also lead to successful detections of SL events on the sky.

In this paper we present the results of \textit{EasyCritics}
applied to the  Canada--France--Hawaii Telescope 
Lensing Survey \citep[CFHTLenS;][]{Heymans2012_CFHTLenS},
which is a based on the same observations as the CFHTLS-Wide survey. 
We test the efficiency of our automated approach by systematically exploring
the parameter space and comparing the results with known lenses in the survey.
We show that \textit{EasyCritics} is able to identify more than 70\% of the known lenses by pre-selecting for inspection only 
$\sim0.4\%$ of the total survey area, which drastically reduces
the total post-processing time.
Among the pre-selected SL regions, we find 9 new promising lens candidates 
 and several regions having a low or medium probability of containing a lens.
Moreover we present statistics regarding the Einstein radius of the lens candidates
and mass-scaling relations.

This paper is organized as follows: 
In \S\ref{sec:EasyCritics} we give an overview of \textit{EasyCritics}, while  
in \S \ref{sec:data} we summarize the main properties 
of  CFHTLenS and describe the selection criteria for the sub-sample of the elliptical galaxies used to construct the lensing potential maps.
Then, in \S\ref{sec:parameter_config} we introduce the reference sample of known lenses used in this work
and explain the procedure of our systematic exploration of the parameter space. 
In \S \ref{sec:results_analysis} we present the results obtained from the parameter exploration analysis 
 and report on the efficiency of our approach based on the detection rate and total number of pre-selected 
 SL regions. In this section we also show the first characterizations of the SL region candidates
 and introduce the new promising lens candidates discovered by \textit{EasyCritics}. 
In \S \ref{sec:discussion} we analyze possible systematics that might affect our results.
Lastly, we summarize the main results and present the final conclusions in \S \ref{sec:conclusion}.
Throughout the paper we assume a flat \LCDM\ cosmology with  $\Omega_m = 0.27$, 
$\Omega_\Lambda = 0.73$, and  $H_0 = 70$ $h_{70}$ km s$^{-1}$ Mpc$^{-1}$.

\section{EasyCritics} \label{sec:EasyCritics}

Motivated by the successful results of the studies mentioned above,
we have developed \textit{EasyCritics} \citep{Stapelberg_2017_EasyCriticsI}, 
a new LTM approach based on a more efficient mathematical 
derivation of the critical curves, which allows us to perform a blind analysis of the data
on wide field surveys, and does not require any list of pre-selected targets.
Among other features,  \textit{EasyCritics} is built with the purpose of obtaining
direct estimates of the lensing quantities in the field of interest by taking into account all the contribution of massive structures
along the LOS.
In brief, once the galaxies are selected that best trace the DM distribution (described in \S \ref{sec:selecting_galaxies}), 
they are sliced in several lens planes, 
 distributed in the redshift range $0.2 \lesssim z \lesssim 0.9$
and located at the redshift bin center  $z^{(k)}$ (\mbox{$k \in \mathbb N$}). 
We then construct an individual lensing potential model $\smash{\psi^{(k)}}$ for each of these lens planes. 
These models are constructed by assuming that the
surface density at every lens plane 
can be idealized by a superposition of 
embedded galaxy-scale subhalos, $\smash{\Sigma^{(k)}_{\mathrm{gal}}}$, and 
a smooth group- or cluster-scale component $\smash{\Sigma^{(k)}_{\mathrm{clus}}}$.
The latter  component is derived from the galaxy distribution assuming that the DM component
approximately follows the observed galaxy distribution of the %automatically 
selected  galaxies. 
Neglecting any non-linear coupling between the lens planes \citep[\eg][]{Schneider2014}, we then compute
the total lensing potential of the field under study  
as  
\begin{align}
        \bar \psi \equiv \sum_k \psi^{(k)},  \label{total_potential}
\end{align}
where every $\smash{\psi^{(k)}}$ is evaluated at its respective lens-plane redshift.

We start by creating the galaxy component for a given lens plane at $z^{(k)}$, assuming 
a common axially symmetric power-law density profile, with slope $q$, for all 
selected galaxies. This profile is then scaled linearly in amplitude by the observed luminosity $L$
\citep{Brainerd96}. For a given galaxy $i$, we thus have 
\begin{align}
        \Sigma^{(k)}_{\mathrm{gal},i}(\theta) = K_{\mathrm{gal}} L_i \cdot \big( D^{(k)}_l \theta \big)^{-q}, \label{sigma_gal}
\end{align}
where $K_{\mathrm{gal}}$ represents the $M/L$ ratio for galaxy-scale halos
and corresponds to one of the free parameters of our approach,
$\theta \equiv \| \boldsymbol \theta \|$ refers to the angular impact parameter
and $D^{(k)}_l$ denotes the angular-diameter distance to the respective redshift bin $z^{(k)}$.
Note that the slope\footnote{We restrict $q$ to the interval \mbox{$q \in (0, 2)$} to ensure a well-defined 
lensing potential on the whole domain $\smash{\mathbb R^2}$.} $q$
is also a free parameter of \textit{EasyCritics}.

The Poisson equation relates the surface mass density profile 
to the lensing potential as 
\begin{align}
        \Delta \psi^{(k)}_{\mathrm{gal,}i}(\theta) = 2 \kappa^{(k)}_{\mathrm{gal},i}( \theta)  
                               \equiv 2 \frac{\Sigma^{(k)}_{\mathrm{gal},i}(\theta) }{\Sigma_{\mathrm{crit}}(z^{(k)}) }, \label{poisson}
\end{align}
where we introduced the  convergence $\kappa$, a dimensionless surface mass density.
The convergence is normalized by the critical surface density for lensing
$\Sigma_{\mathrm{crit}}$, which depends on the angular-diameter distances
to the lens plane $\smash{z^{(k)}}$,  $\smash{D^{(k)}_l}$, to the source plane, $\smash{D_s}$, 
and between the lens plane and source plane, $\smash{D^{(k)}_{ls}}$. 

Then, the lensing potential due to the galaxy component of
a lens plane at $z^{(k)}$ can be obtained by applying the 
Poisson equation (\ref{poisson}) and the superposition principle:
\begin{align}
        \psi^{(k)}_{\mathrm{gal}}(\boldsymbol{\theta}) 
		= \frac {2 \left(D^{(k)}_l\right)^{-q} K_{\mathrm{gal}} }
		{\Sigma_{\mathrm{crit}}(z^{(k)}) (2-q)^2} \sum_{i = 1}^N L_i
		\| \boldsymbol \theta - \boldsymbol \theta_i \|^{2-q}, \label{psi_gal_comp}
\end{align}
where the index $i$ runs over all galaxies binned to the $k$-th redshift slice
at $\smash{z^{(k)}}$.

As anticipated, we find that the cluster-scale DM halos are smoother than the galaxy components  
and have a much higher overall mass scaling.
The cluster-scale component for a lens plane at $\smash{z^{(k)}}$
is modeled as a convolution of
the galaxy component with a Gaussian function\footnote{
The choice of this function is empirically motivated \citep[\eg][]{Zitrin2013_gauss}.}, 
\begin{align}
S(\boldsymbol \theta) \equiv 
                   \frac 1{2 \pi \sigma^2_{\mathrm{clus}}} \mathrm{exp}
                   \left( - \frac{\boldsymbol \theta^2}{2 \sigma^2_{\mathrm{clus}}} \right), \label{gaussian}
\end{align}
of smoothing window $\smash{\sigma_{\mathrm{clus}}}$\footnote{ 
For simplicity, we neglect the weak redshift dependence of  $\smash{\sigma_{\mathrm{clus}}}$, 
which may arise due to the mass-concentration relation,
and we thus use a unique smoothing window for all redshift slices.}. 
The kernel $\smash{\sigma_{\mathrm{clus}}}$  not only defines
the smoothing scale but also affects the steepness 
of the profile of the cluster-scale halos and corresponds to another
free parameter of our approach.

It is expected that not all selected galaxies are living in cluster environments, since they
may also trace smaller group-scale structures along the LOS or just be isolated 
field galaxies. We thus introduce a conditional probability for our selected 
galaxies to trace a massive cluster-scale halo, $\smash{w^{(k)}(n^{(k)}|n_c)}$, 
which depends on the local number density $\smash{n^{(k)}}$ of selected galaxies
of a given  lens plane at $\smash{z^{(k)}}$ and $n_c$, which specifies the `critical' number 
density of galaxies to satisfy the condition of being in a super-critical 
massive environment, which gives  $\smash{w^{(k)} = 1}$.  The value of  $n_c$
is given by the number of elliptical cluster members located in the lens plane of the main deflector.

In analogy to the galaxy component, by applying the
Poisson equation and the superposition principle,
the lensing potential due to the cluster-scale component of
a lens plane at $z^{(k)}$ can be computed as
\begin{align}
   \psi^{(k)}_{\mathrm{clus}}(\boldsymbol{\theta}) 
      =  S(\sigma_{\mathrm{clus}}) * \left( w^{(k)}(n^{(k)}|n_c) 
           \frac{ K_{\mathrm{clus}} }{K_{\mathrm{gal}} }  \psi^{(k)}_{\mathrm{gal}}(\boldsymbol{\theta})  \right), \label{clus_potential}
\end{align}
where we introduced an additional free parameter into our approach, 
$\smash{K_{\mathrm{clus}}}$, which represents the $M/L$ ratio for  cluster-scale halos.
 
The lensing potential of a given lens plane at $\smash{z^{(k)}}$ is then just the sum 
of both components, 
\begin{align}
   \psi^{(k)} =    \psi^{(k)}_{\mathrm{gal}} +    \psi^{(k)}_{\mathrm{clus}}.  \label{potential_lens_plane}
\end{align}
Finally, the total lensing potential of the field under study is then computed via Eq. (\ref{total_potential}),
 adding all the lens-plane potentials $\smash{ \psi^{(k)}}$, evaluating them at their respective lens-plane redshifts
$\smash{z^{(k)}}$ and neglecting any correlation between the lens-planes.

This new LTM approach has only five free parameters, which are calibrated 
by fitting critical curves to known gravitational arcs (explained in \S \ref{sec:parameter_config}).
These free parameters are listed in Table~\ref{table:params}.
All the details of \textit{EasyCritics} and its practical application can be found in  
\cite{Stapelberg_2017_EasyCriticsI}. 
It should be noted that we have simplified some equations of our algorithm in
the text and we have also interchanged the order of the two-dimensional  convolutions 
in order to give a better understanding to the reader.

%%%%%%%%%%%%%%%%%%%%%%%%%%%%%%%%%%%%%%%%%%%%%%%%%%%%%%5
%%%%%%%%%%%%%%%%%%%%%%%%%%%%%%%%%%%%%%%%%%%%%%%%%%%%%%%%
% Table of summary of parameters, description
\begin{table}
\centering
\caption{Description of \textit{EasyCritics}' parameters.  \label{table:params}}
\hspace{0.0cm}
\begin{tabular}{l l}
\toprule \toprule
 Symbol & Description\\ 
\midrule
$q$                                   & Slope of the power-law density profile\\
$K_{\mathrm{gal}}$          & $M/L$ ratio for galaxy-scale halos\\
$K_{\mathrm{clus}}$         & $M/L$ ratio for group- or cluster-scale halos \\
$\sigma_{\mathrm{clus}}$ & Smoothing window for the Gaussian function\\
$n_c$                                & Critical number density of LRGs\\
\bottomrule
\end{tabular}
%}                                        
{\footnotesize\flushleft
}
\end{table}
%%%%%%%%%%%%%%%%%%%%%%%%%%%%%%%%%%%%%%%%%%%%%%%%%%%%%%%%
%%%%%%%%%%%%%%%%%%%%%%%%%%%%%%%%%%%%%%%%%%%%%%%%%%%%%%%%

\section{The LRG catalogs} \label{sec:data}   

\subsection{The CFHTLenS data} \label{sec:CFHTLenS_data}
CFHTLenS is a wide field optical survey designed to accurately measure weak gravitational lensing  
from deep multicolour images. The data have been extracted from the 
CFHTLS.
All imaging data of this survey have been obtained with the MegaPrime 
instrument\footnote{http://www.cfht.hawaii.edu/Instruments/Imaging/Megacam/},
between the semesters 2003A and 2008B inclusive.
MegaPrime is an optical multichip instrument with 
a field of view of $\sim 1\deg \times 1 \deg$ and a pixel scale resolution
of $0.187$ arcsec \citep{Boulade2003_MEGACAM}.
CFHTLenS spans 154 sq. deg in the five optical bands $u^* g' r' i' z'$,
with a $5\sigma$ point source limiting magnitude in the $i'$ band of
$i'_{AB} \sim 25.5$.
Given the survey strategy, the median seeing is 
$\lesssim 0.8''$ for the primary lensing $i'$-band, 
making CFHTLenS ideal for SL studies as well. 
Furthermore, its deep multicolour data result in
 accurate photometric redshifts (photo$-z$), with
a photo$-z$ scatter in the range $ 0.03 < \sigma_z/(1+z)  < 0.06$ and 
an average catastrophic outlier rate smaller than  10$\%$, 
when limited to the photo$-z$ range $0.1 <  z < 1.3$ and to objects 
with magnitudes in $i'_{AB} \lesssim 24.5$ \citep{Hildebrandt2012_CFHTLenS_photz}.
The data reduction has been carried out with  the {\footnotesize THELI} pipeline \citep{Schirmer2003_theli,Erben2005_theli,Schirmer2013_theli}
following the procedures described in \cite{Erben2009_CARS}. 
For a complete description of CFHTLenS data, see \cite{Erben2013_CFHTLenS}.

\subsection{Selection of the tracing galaxies} \label{sec:selecting_galaxies}
Our procedure is based on the well-tested underlying assumption that DM approximately follows 
 light; hence, the key point is to select the best galaxies to be used as tracers.
Several former studies have shown that bright 
elliptical galaxies (or luminous red galaxies; LRGs), are biased probes of the underlying
matter distribution \citep[e.g.][]{Zehavi2005_LRG,Li2006_LRG, Ho2009_LRG,
White2011_clustering_massive_gal, Wong2013_LRG},   
and are observable up to relatively high redshift 
\citep{GladdersYee00,GladdersYee05,Gilbank11}.
In particular, \cite{Wong2013_LRG} have shown that by using the light of LRGs 
projected on the sky over a wide redshift range ($0.1 \le z \le 0.7$),
one can predict the LOSs with the highest integrated mass densities.
Furthermore, the observed elliptical galaxy distribution has been successfully 
used as the starting point of several lens--mass reconstruction studies of 
SL galaxy clusters and groups \citep[e.g.][]{Broadhurst2005a,Broadhurst2005b,Zitrin09a,Zitrin11MACS,Zitrin2012UniversalRE}.
For these reasons, we use these galaxies to trace the DM distribution.

%%%%%%%%%%%%%%%%%%%%%%%%%%%%%%%%
% zphot vs zspec comparison for elliptical galaxies
%%%%%%%%%%%%%%%%%%%%%%%%%%%%%%%%
\begin{figure}
\begin{center}
\includegraphics[width=0.48\textwidth,trim= 0mm 0mm 0mm 0mm,clip]{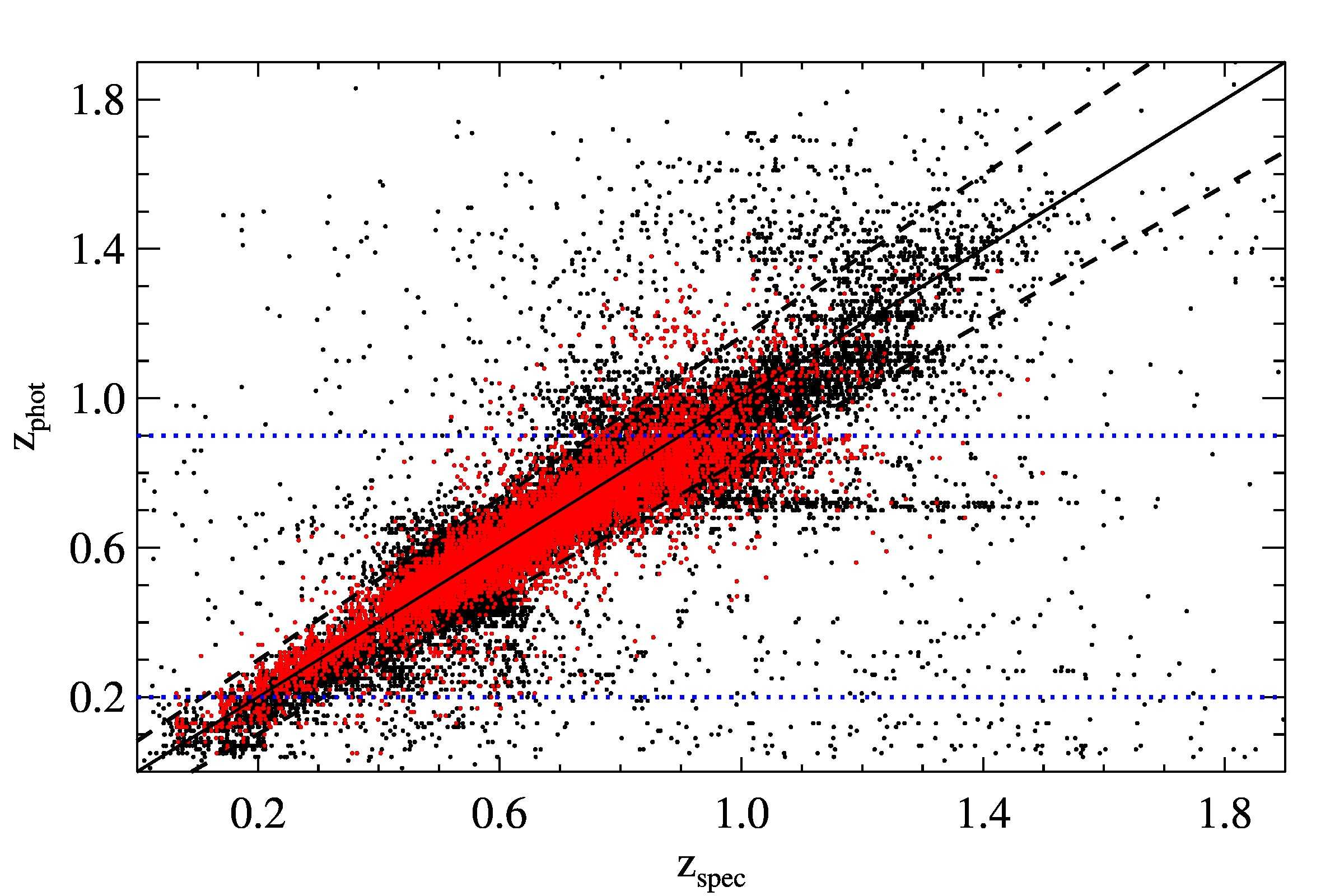}
\caption{\label{Fig:phot_z}
Comparison of photo$-z$ with spec$-z$. 
The black dots correspond to $\sim 50000$ CFHTLenS galaxies 
with spec$-z$ measurements used in \citet[private communication]{Hildebrandt2012_CFHTLenS_photz}.
Among these objects, approximately $\sim 10000$ elliptical galaxies (red dots) fall
in our pre-selection criteria; $T_B \leq 1.7$,  $17 \leq i'_{AB} \leq 24$, and $0.2 \lesssim z \lesssim 0.9$.
The solid black line represents the one-to-one relation, while dashed black lines
the  $2 \times \bar{\sigma}_z / (1+z)$ deviations.  
The dotted blue lines show the redshift range considered in this work.} 
\end{center}
\end{figure}
%%%%%%%%%%%%%%%%%%%%%%%%%%%%%%%% 

The selection procedure is based on: 
magnitude in the $i'$ band,  photo$-z$,
spectral index $\smash{T_B}$, and size (semi-major axis).
In order to include all possible elliptical galaxies, we start selecting all objects
with $\smash{T_B \leq 1.7}$\footnote{As stated in \cite{CWW80}, objects
with spectral index $\smash{T_B = 1}$ 
corresponds to 'early-type' E/S0 galaxies, 
$\smash{T_B = 2}$ to SBc barred spirals and $\smash{T_B \geq 2}$ to 'late-type' spiral and irregular galaxies.}. 
 To ensure the precision of the photo$-z$ of our galaxy candidates and 
to decrease the number of outliers,  
we limit the selection to objects falling within the magnitude and photo$-z$ ranges
$\smash{17 \leq i'_{AB} \leq 24}$ and $\smash{0.2 \lesssim z \lesssim 0.9}$, respectively. 
This selection procedure results in a sample of elliptical galaxies with 
photo$-z$ in excellent agreement with the available 
spectroscopic redshifts \citep[spec$-z$;][private communication]{Hildebrandt2012_CFHTLenS_photz}, %\footnote{Private com}
with
an average photo$-z$ scatter\footnote{The 
photo$-z$ scatter, $\smash{\sigma_z}$, is calculated as the standard deviation around the mean of
  $\Delta z =  (z_{phot} - z_{spec}) /(1+ z_{spec})$. Objects with $|\Delta z| > 0.15$ are 
  considered as outliers.}
 $\bar{\sigma}_z  <0.04$  and an outlier rate $<5\%$
(Fig. \ref{Fig:phot_z}).
Then, we apply a $k$-correction to the fluxes and magnitudes of the selected LRGs
using the template spectra described by \cite{Capak04} and the final transmittance
curves for the MegaPrime filters.

To improve the redshift accuracy of our sample, 
we replace the photo$-z$ of the galaxy candidates 
with the spec$-z$ where available. 
Therefore, the given outlier rate and scatter are just upper limits.
Hereafter, when referring to redshift in our galaxy sample,
we mean the spectroscopic redshift when available and the photo$-z$ otherwise.

%%%%%%%%%%%%%%%%%%%%%%%%%%%%%%%%
% Redshift slices and selection routine
%%%%%%%%%%%%%%%%%%%%%%%%%%%%%%%%
\begin{figure}
\begin{center}
\includegraphics[width=0.48\textwidth,trim= 0mm 0mm 0mm 0mm,clip]{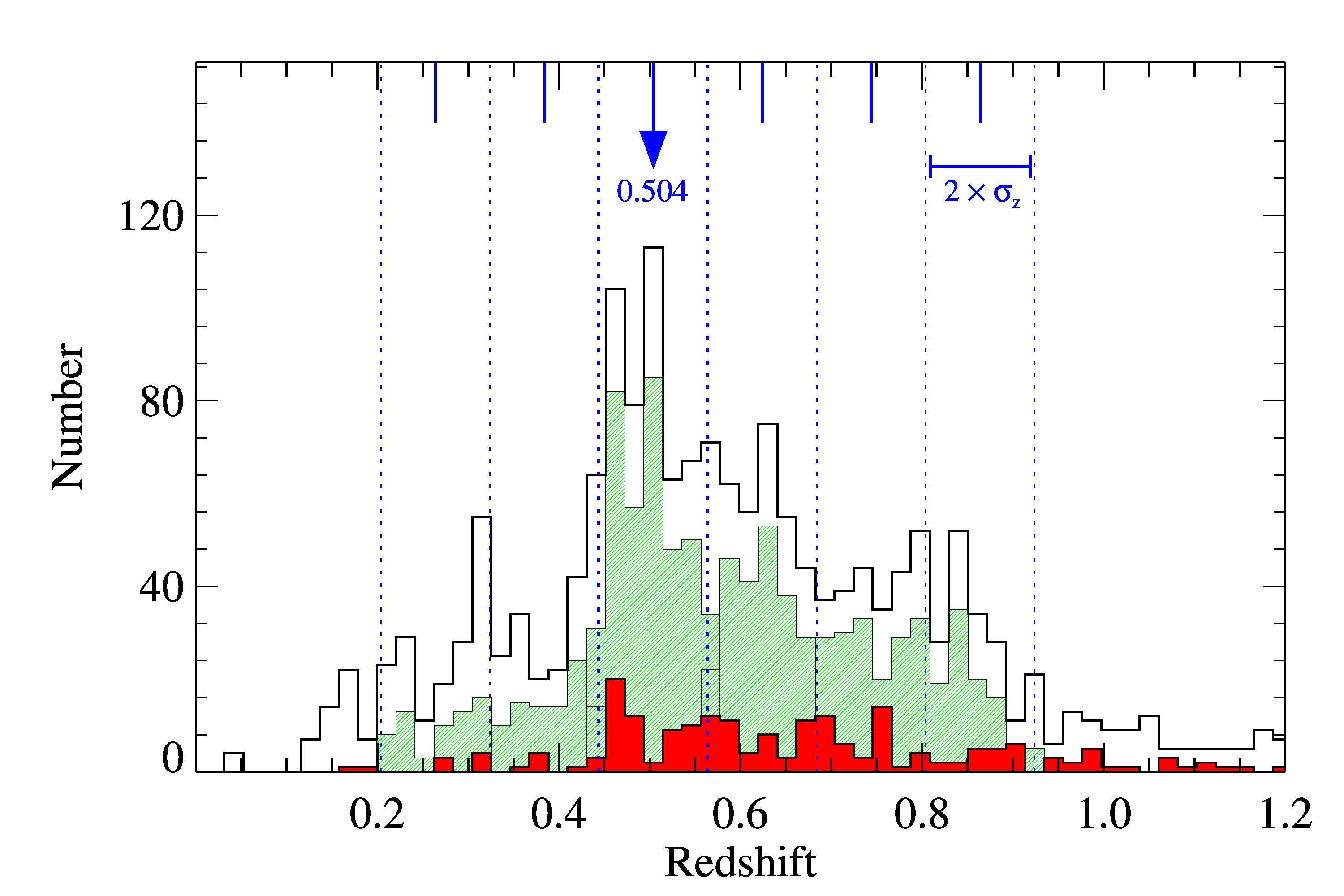} 
\caption{\label{Fig:z_dist}
Redshift distributions of elliptical galaxies within a $15'\times15'$ region, centered
on the SL galaxy cluster SA125 ($\alpha = $22:14:18.82 and $\delta = $+01:10:33.85).
The white histogram corresponds to all elliptical galaxies ($T_B \leq 1.7$) 
distributed along the LOS in the magnitude range $17 \leq i'_{AB} \leq 24$. 
The blue arrow indicates the starting center for the binning 
which corresponds to the peak of the redshift distribution, 
while the blue vertical lines correspond to the center of the remaining redshift slices. 
The width of each redshift bin is set to $2\times\bar{\sigma}_z$
and delimited by the dotted blue lines. 
The green histograms of each redshift bin correspond 
to the photo$-z$ distribution of the LRGs selected by 
our procedure as described in the text.
While the red histogram corresponds to the spec$-z$ distribution 
of the elliptical galaxies in this region.
As one can see, our binning choice helps to keep the most massive structures in the LOS.} 
\end{center}
\end{figure}
%%%%%%%%%%%%%%%%%%%%%%%%%%%%%%%% 
%%%%%%%%%%%%%%%%%%%%%%%%%%%%%%%% 

In order to create a smooth transition between the lensing potential maps,
we divide the LRG sample into galaxy catalogs of $15'\times15'$ each, 
with a conservative overlap of $5'$ per side.
As mentioned in \S \ref{sec:EasyCritics},  the survey is sliced in redshift bins 
 of $\smash{2\times\bar{\sigma}_z}$\footnote{We conservatively
decide to use the upper limit of the photo$-z$ scatter but taking into 
account the whole CFHTLenS data, i.e. $\sigma_z = 0.06$ \citep{Hildebrandt2012_CFHTLenS_photz},
to ensure that all massive structures are included.} in thickness to cope with the photometric redshift uncertainties 
but still preserve correlated structures. 
Once the slices are created, the redshift of each elliptical galaxy is updated to the redshift center  $z^{(k)}$ of its corresponding bin.
In this step, we search for possible outliers by comparing galaxy sizes;
if the size of a galaxy is $\smash{5 \sigma}$ larger than the average size 
of the brightest galaxies of the corresponding bin, 
then this galaxy is shifted to another redshift slice by applying 
an empirical $z-$size relation. % (Appendix A).
We finally derive the  Schechter luminosity function for each angular and redshift portion of the survey \citep[LF;][]{Schechter76}; 
all galaxies fainter than $\smash{M_{\star} + n_{cut}}$ are removed from the catalogs. 
The characteristic magnitude, $M_{\star}$, corresponds to the slope change of the LF,
from an exponential to a power-law form, dividing the bright tail from the faint galaxies.
In order to ensure the inclusion of all LRGs and, at the same time, 
to remain in the magnitude range $\smash{17 \leq i'_{AB} \leq 24}$, we limit the values of $n_{cut}$ to the range $\smash{1 \leq n_{cut} \leq 2}$.
For illustration, we show in Fig. \ref{Fig:z_dist} the redshift distribution of 
the LOS selected galaxies in the $15'\times15'$ cutout centered on the SL galaxy cluster SA125 \citep{More_2012_arcfinder}. 
The final selected LRGs 
are shown by the green histograms.
By comparing with the spec$-z$ distribution  
(red histogram), 
one can see that our selection procedure  properly selects galaxies that belong to
the most massive objects in the LOS, 
associated to the galaxy cluster SA125 and other small groups.

\section{The parameter set configuration} \label{sec:parameter_config}
 
In this section we explore several parameter sets in order to 
test the efficiency of \textit{EasyCritics} to identify SL regions in wide field surveys.
For this purpose, we define a reference sample of known SL systems previously found by other studies in CFHTLenS.
From the reference sample, 
we select the most representative lenses in three different mass intervals, in order to 
 perform three independent parameter calibrations, one for each mass range. 
Based on these parameter calibrations and aiming at identifying the majority of the known lenses, 
we then systematically explore the parameter space.

%%%%%%%%%%%%%%%%%%%%%%%%%%%%%%%%
% Selected KL systems for calibration in 3 mass regimes
%%%%%%%%%%%%%%%%%%%%%%%%%%%%%%%%
\begin{figure*}
\begin{center}
\begin{tabular}{c c c}
\includegraphics[width=55mm, height=55mm]{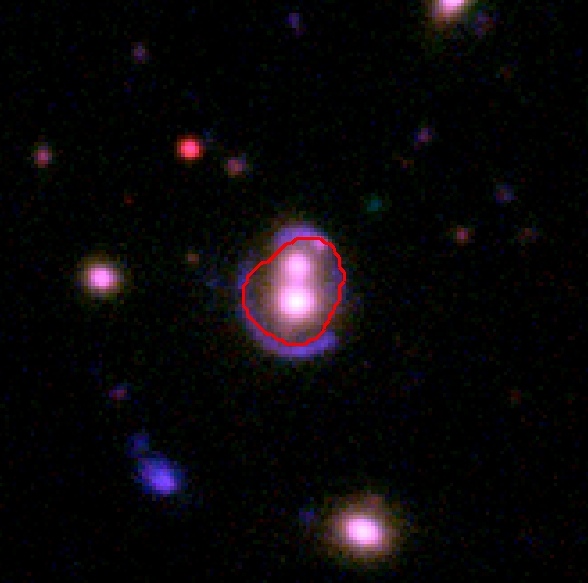} &
\includegraphics[width=55mm, height=55mm]{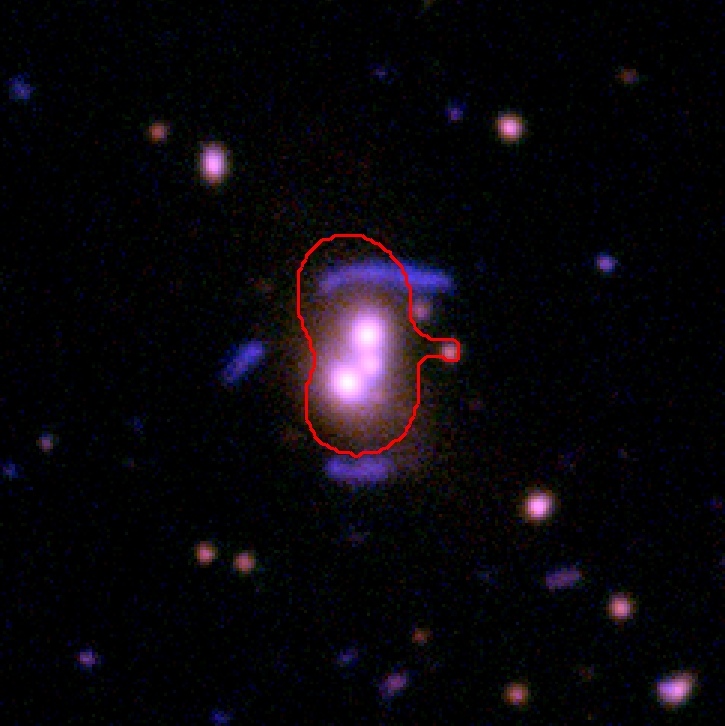} &
\includegraphics[width=55mm, height=55mm]{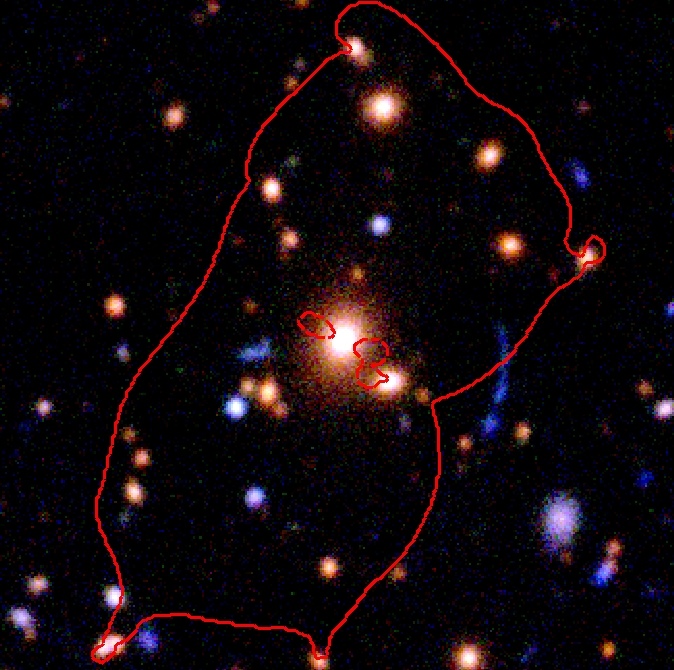} \\
\end{tabular}            
\put(-490, 60){\bf \huge \color{green} $A$} 
\put(-435, 70){\bf \large \color{cyan} SA14}  
\put(-320, 60){\bf \huge \color{green} $B$}     
\put(-265, 70){\bf \large \color{cyan} SA22}
\put(-150, 60){\bf \huge \color{green} $C$}                                                                                          
\put(-100, 70){\bf \large \color{cyan} SA100}                  
\caption{\label{Fig:KL_calib_3mass}
RGB ($i'$  $r'$ $g'$) compose images of the known SL galaxy groups SA14, SA22, and the known SL galaxy cluster 
SA100 (coordinates and references are listed in Table \ref{table:app:reference_sample}, 
Appendix \ref{app:reference_sample}), with Einstein radii of $\smash{\sim 3.2'', 7.1''}$,
and $\smash{14.7''}$, respectively. 
Given the different angular scales, these lenses are probing three mass intervals;
interval $A$: small galaxy groups; interval $B$: systems in the range of groups and small clusters;
interval $C$: massive galaxy clusters. 
The cut-outs are centered on the candidate centers and
 covering an area of $30''\times30''$, $45''\times45''$, and $60''\times60''$, respectively. 
The red closed lines correspond to the critical curves derived from 
our calibration procedure, described in \S\ref{sec:sub:sub:3_mass_regime},
and computed by using the parameter sets listed on the Table \ref{table:calibrated_par}. 
} 
\end{center}
\end{figure*}
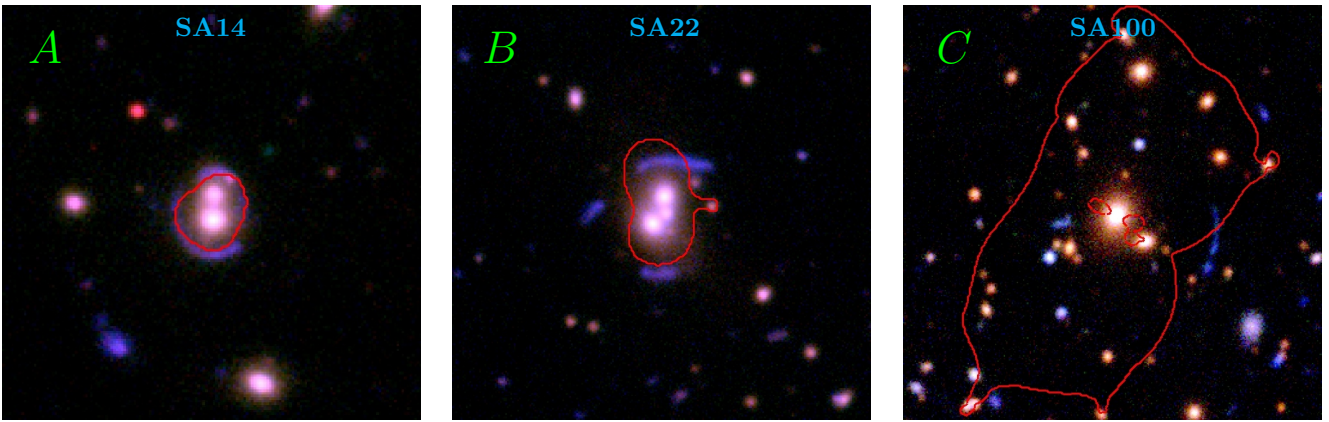
%%%%%%%%%%%%%%%%%%%%%%%%%%%%%%%%   
%%%%%%%%%%%%%%%%%%%%%%%%%%%%%%%%  

\subsection{Reference samples: known confirmed and candidate lenses} \label{sec:reference_samples}

In order to find the most suitable parameter sets,
we compare our results with confirmed lenses previously found on CFHTLenS
and its predecessor CFHTLS\footnote{As mention earlier, CFHTLenS is a sub-sample of CFHTLS.}. 
These surveys have been extensively used for the search
of SL objects since their earliest releases, yielding to date of the order 
of 500 candidates, of which approximately 150 are confirmed lenses \footnote{Note that 
confirmed lenses correspond to a subset of the lens candidates sample.}. 
Note that in the previous studies, 
the term ``confirmed'' lens not necessarily refers to
a spectroscopic confirmation of the system; instead this term usually
corresponds to objects that have been classified as real lens systems by several experts,
making them the most promising candidates.

The known SL systems discovered in these surveys cover a very wide mass range, 
from $10^{12}$ to $10^{14}$ M$_{\odot}$,
encompassing galaxy-scale DM halos with Einstein radii $\lesssim 3''$, 
group-scale DM halos with Einstein radii in the range of $\sim 3-8''$\footnote{The upper limit of $8''$ is 
somewhat arbitrary and may be considered by others as a characteristic size of
a poor cluster.}, and massive galaxy clusters with Einstein radii $\gtrsim 8''$ 
\citep{Oguri06_image_sep}.
Since our approach is designed to detect massive objects,
we restrict the comparison to known lenses on group- and cluster-scales.
For this purpose, we create our reference sample of known SL systems by selecting
confirmed lenses with an Einstein radius $> 3.5''$. 
Since some galaxy-scale lenses are placed in group- or cluster-scale DM halos, 
there is a possibility that these SL systems have also been enhanced by the smooth lensing potential 
of the massive host halo. 
Therefore, we visually inspect all objects with Einstein radii $\lesssim 3.5''$,
 in order to include in our reference sample SL systems where the creation of arcs or 
 multiple images is boosted by the external shear and convergence from the smooth group or cluster component 
\citep{Limousin07,Fassnacht06, Oguri05}.

Our reference sample of known SL systems, namely the `known lenses' (KLs) sample, lists
44 confirmed lenses.
Furthermore, by applying the same criteria as before but now also including objects that have been flagged as candidate lenses,
we create a sample of `known lens candidates' (KCs),
which is composed of 98 objects.
Note that the  KLs sample is a subset of 
the KCs sample, \eg the 44 KLs are included in the KCs list. 
These objects are listed in Table \ref{table:app:reference_sample}, in Appendix \ref{app:reference_sample}, together with their main characteristics; ID, Ra, Dec,
redshift, Einstein radius (or distance to the arc), and reference publication.

\subsection{Exploring the parameter space in three mass intervals} \label{sec:sub:par_SL_pred}

The mass range covered by the systems of the KL and KC samples,
 given by the Einstein radius or image separation,  extends from  $\smash{\sim 10^{13}}$
to $\smash{10^{15}}$ M$\smash{_{\odot}}$ \citep{Oguri06_image_sep}.
As expected, %given the nature of our approach, 
different mass regimes require different mass scalings to initialize our approach.  
Therefore, we perform three independent 
parameter calibrations for three different mass intervals
in order to find the initial parameter sets for our systematic parameter exploration.
Then, aiming at identifying most of the KLs, we explore several 
parameter setups based on these initial parameter sets.

\subsubsection{Parameter calibration: initial parameter sets for three different mass intervals} \label{sec:sub:sub:3_mass_regime}

We  select the most representative lenses in each mass interval; 
the SL galaxy groups SA14, SA22, and the SL galaxy cluster
SA100, with Einstein radii of $\smash{\sim 3.2'', 7.1''}$,
and $\smash{14.7''}$, respectively. 
For simplicity, we call these mass intervals $A$, $B$, and $C$, correspondingly. 
These systems are shown in Fig. \ref{Fig:KL_calib_3mass}, while their 
coordinates and references are listed in Table \ref{table:app:reference_sample}
(Appendix \ref{app:reference_sample}). 
Note that for those systems, the Einstein radius measurement corresponds
to the distance from the BCG to the average position where the arc is
located \citep{More_2012_arcfinder}.

For each of these lenses, we carry out an independent parameter calibration 
by fitting critical curves to the position of KL arcs;
where critical curves are derived from the total lensing potential, described in \S \ref{sec:EasyCritics} by Eq. (\ref{total_potential}), 
and by using
the selected LRG catalogs (described in \S \ref{sec:selecting_galaxies}), 
centered on the BCG of the SL systems and covering a region of $\smash{15'\times15'}$
on the sky.
Our calibration routine uses a parallelized generalization 
of the Metropolis-Hastings algorithm \citep{Metropolis1953, Hastings1970}, 
combining a Markov Chain Monte-Carlo (MCMC) sampling 
with an adaptive grid technique, introduced in \cite{Stapelberg_2017_EasyCriticsI}. This new method is 
in general much faster than similar algorithms, while preserving the resolution necessary for accurate results. 

The limited number of constraints results in highly degenerate parameters,
which translates into several different combinations of equivalent free parameters yielding 
similar results, in terms of orders of magnitude for the $\smash{\chi^2}$.
To cope with this degeneracy we fix three of our parameters to some physical values; $q$, $\smash{n_c}$, and
$\smash{K_{\mathrm{gal}}}$ (previously described in \S \ref{sec:EasyCritics}). 
Neglecting any misalignment between the BCG light profile 
and the galaxy-scale DM profile, we use the light distribution of the
brightest galaxies of the selected KLs to find the most suitable values for 
the power-law index $q$.
Although this assumption is not always accurate, the estimates of $q$
are fully consistent with those reported in previous works \citep[\eg][ and references therein]{Zitrin2012UniversalRE},
falling within the range of $\smash{1.1 < q < 1.2}$.
Hence, we set this parameter to $q = 1.14$. 
It is worth mentioning that in the studies cited above, it is shown 
that the locations of critical curves are very well predicted independently of the
 value for $q$,  as long as it is in the range $\smash{1.0 < q < 1.5}$.
Since the parameter $n_c$ is defined as the `critical' number density of selected galaxies
for having a complete DM component (\ie a SL system), it therefore can be directly extracted from the
number of LRGs located in the redshift slice of the SL system, i.e. the galaxy members 
photometrically selected as described in \S \ref{sec:selecting_galaxies}.
For a $\smash{15' \times 15'}$ region,  $n_c$
corresponds to a number of bright elliptical galaxies of $95$, $163$, and $252$,
for the SL galaxy groups SA14, SA22, and the SL galaxy cluster
SA100, respectively.  
Finally,  given the well-known abundance of baryons in clusters \citep[\eg][]{Dai_2010_baryons,Semboloni_2011_baryons,Lagana_2013_baryons,Ge_2016_baryons}, 
the $\smash{K_{\mathrm{gal}}}$ parameter can be 
directly related to the $\smash{K_{\mathrm{clus}}}$ parameter.
In \cite{Stapelberg_2017_EasyCriticsI}, we showed that 
the ratio between $\smash{K_{\mathrm{clus}}}$ and 
$\smash{ K_{\mathrm{tot}} \equiv  K_{\mathrm{clus}} +  K_{\mathrm{gal}} }$
 is limited to the range $\smash{0.86 \lesssim K_{\mathrm{clus}} / K_{\mathrm{tot}} \lesssim 0.94}$, 
which is consistent with the expectations. 
Therefore, we set the galaxy-scale component $\smash{K_{\mathrm{gal}}}$ to $10 \%$
of the total mass scaling $\smash{K_{\mathrm{tot}}}$.

Consequently, the only remaining free parameters in our model are $\smash{K_{\mathrm{clus}}} $ and $\smash{\sigma_{\mathrm{clus}}}$.
With these constraints we finally perform an independent calibration for each of the selected 
KLs  by following the procedure described in our first paper.
Throughout the calibration process,  
we assume the source plane to be at redshift $z_s = 2$, 
which is a representative source redshift for the observed giant arc 
population according to broadband photometric \citep[\eg][]{Bayliss12} 
and spectroscopic studies \citep[\eg][]{Bayliss11b, Carrasco17_spec}. 
The critical curves resulting from this minimization approach are shown in Fig. \ref{Fig:KL_calib_3mass},
while the calibrated parameters are listed in Table \ref{table:calibrated_par}.
As expected, the critical curves nicely match the locations of the arcs; 
there are relatively small deviations, 
which can be expected within the limits of the LTM
approximation and the assumptions made in this analysis. 
Furthermore, by comparing our results for the SL galaxy group SA22 
with the models from \cite{Verdugo11, Verdugo16}, 
we can conclude that the approach used in this work reaches the expectations,
\eg consistent critical line estimates well suited to our goal of finding regions potentially containing strong lensing effects.

%%%%%%%%%%%%%%%%%%%%%%%%%%%%%%%%%%%%%%%%%%%%%%%%%%%%%%5
%%%%%%%%%%%%%%%%%%%%%%%%%%%%%%%%%%%%%%%%%%%%%%%%%%%%%%%%
% Table of summary of the calibrated parameters
\begin{table}
\centering
\caption{Summary of the calibration routine for the three mass intervals $A$, $B$, and $C$. \label{table:calibrated_par}}
\hspace{0.0cm}
\resizebox{\columnwidth}{!}{
\begin{tabular}{c l c c c c c}
\toprule \toprule
M. R.$^a$ &  KL$^b$ & $q$ & $n_c$ & $K_{\mathrm{gal}}^c$  & $K_{\mathrm{clus}}^d$ & $\sigma_{\mathrm{clus}}^d$\\ 
 &  & & & [\%]& [$\times 10^{-8}$ a.u.] & [$''$] \\ 
\midrule
$A$  &  SA14    &  1.14  &  95    &  0.10  &  $1.3 \pm 1.1$  & $6.2 \pm 1.1$  \\    
$B$  &  SA22    &  1.14  &  163  &  0.10  &  $2.7 \pm 1.6$  & $12.4 \pm 1.6$  \\  
$C$  &  SA100  &  1.14  &  252  &  0.10  &  $3.9 \pm 1.9$  & $17.1 \pm 1.6$  \\
\bottomrule
\end{tabular}
}                                        
{\footnotesize\flushleft
{}$^a$ Mass interval. \\
{}$^b$ ID from \cite{More_2012_arcfinder}. \\
{}$^c$ Relative to $\smash{ K_{\mathrm{tot}} \equiv  K_{\mathrm{clus}} +  K_{\mathrm{gal}} }$. \\
{}$^d$ The errors correspond to $1\sigma$ of the parameter distribution. \\ %, as explained in \cite{Stapelberg_2017_EasyCriticsI}. \\
}
\end{table}
%%%%%%%%%%%%%%%%%%%%%%%%%%%%%%%%%%%%%%%%%%%%%%%%%%%%%%%%
%%%%%%%%%%%%%%%%%%%%%%%%%%%%%%%%%%%%%%%%%%%%%%%%%%%%%%%%

%%%%%%%%%%%%%%%%%%%%%%%%%%%%%%%%
% Convergence profile for selected KLs
%%%%%%%%%%%%%%%%%%%%%%%%%%%%%%%%
\begin{figure}
\begin{center}
\includegraphics[width=0.52\textwidth,trim= 10mm 0mm 0mm 0mm,clip]{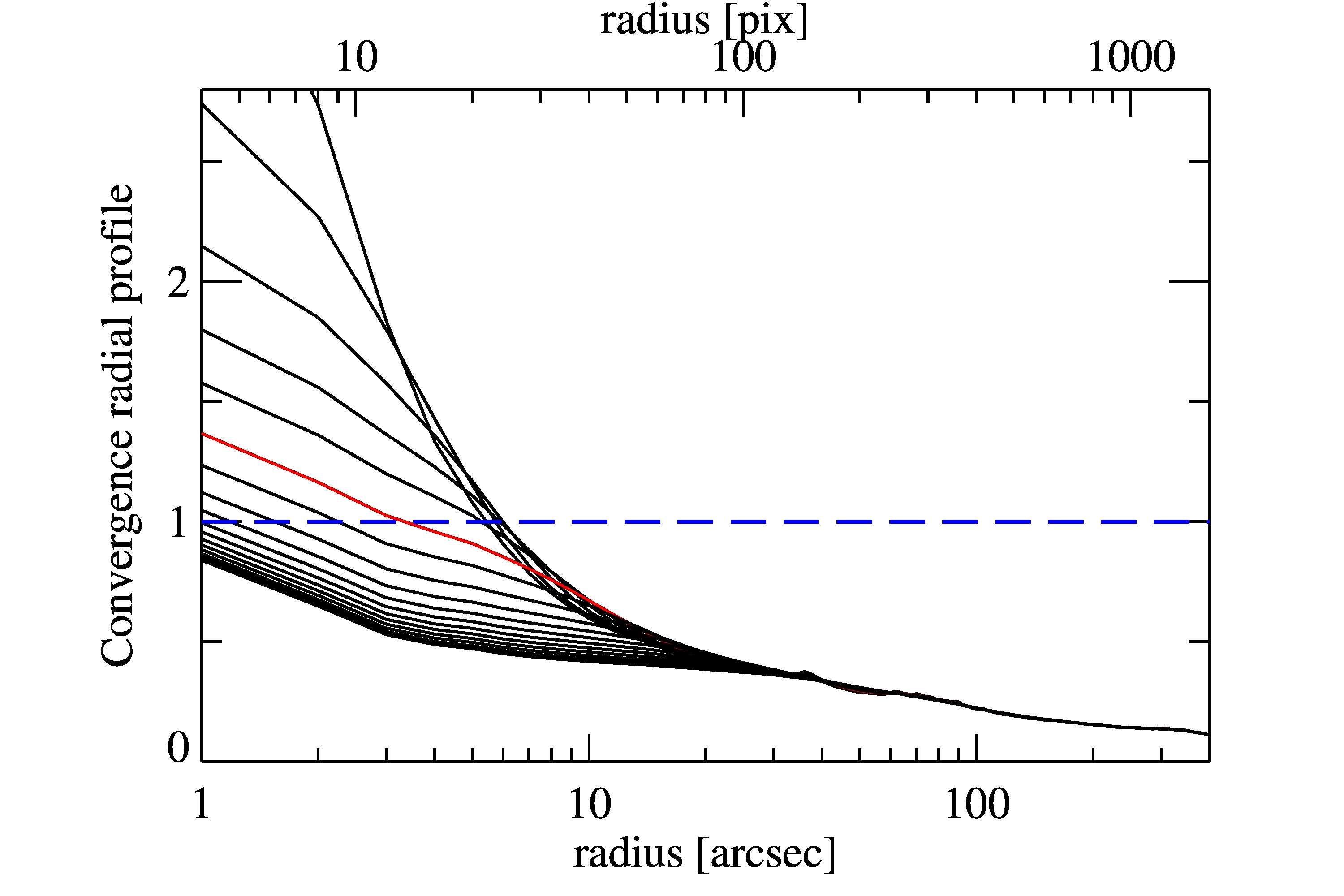}  
\caption{\label{Fig:kappa_profile}
Radial convergence profiles of the SL galaxy group SA22, for different parameter sets.  
The continuous red line corresponds to the profile derived from the calibrated 
parameters for the mass interval $B$.
Profiles derived from different values of the smoothing window ($\smash{\sigma_{\mathrm{clus}} = 2'', 4'',..., 30''}$) 
are shown by the black curves. Smaller values of $\smash{\sigma_{\mathrm{clus}}}$
correspond to steeper profiles, while larger values correspond to flatter profiles.
Note that in the computation of the black-curve profiles, the other four parameters were kept constant 
to the values listed on Table \ref{table:calibrated_par}.
The dashed blue line represents a hypothetical threshold for super-critical profiles.
} 
\end{center}
\end{figure}
%%%%%%%%%%%%%%%%%%%%%%%%%%%%%%%% 
%%%%%%%%%%%%%%%%%%%%%%%%%%%%%%%% 

\subsubsection{Parameter exploration: systematic probing of $\smash{\sigma_{\mathrm{clus}}}$ in each mass interval} \label{sec:sub:sub:3_mass_regime}

Having chosen the initial parameter sets for the three mass intervals, 
 we can systematically explore the parameters based on those initial values, aiming at identifying most of the KLs
 of the reference sample. 
Since the parameter $\smash{K_{\mathrm{clus}}}$ represents the total mass scaling of each SL system 
and since its value depends on the mass interval,
we can independently explore each mass interval
 by setting this parameter to the values in Table \ref{table:calibrated_par} for the corresponding interval,
and by varying only $\smash{\sigma_{\mathrm{clus}}}$.

Furthermore, the radial convergence profiles of our selected KLs 
can turn from flat to super-critical matter distributions
 by changing $\smash{\sigma_{\mathrm{clus}}}$ only.
This is because  $\smash{\sigma_{\mathrm{clus}}}$  does not only determine the smoothness of the cluster-scale 
component but it also affects the steepness of its profile.
In other words, in the particular case when $q$ and the 
overall mass normalization $\smash{ K_{\mathrm{tot}} }$ 
 are fixed, the smoothness and steepness of the profile play the decisive role in the determination of the critical curves.
 For illustration, we show in Fig. \ref{Fig:kappa_profile} 
 the dependence of the radial convergence profile as a function of $\smash{\sigma_{\mathrm{clus}}}$
 for the SL galaxy group SA22. 
The continuous red line corresponds to the profile derived from the calibrated 
parameters for the mass interval $B$, while the black lines correspond to 
profiles obtained from different values of $\smash{\sigma_{\mathrm{clus}}}$. 
Note that $\smash{K_{\mathrm{clus}}} $ is kept fixed through all the computations. 
From this figure,  one can see the impact of $\smash{\sigma_{\mathrm{clus}}}$
on the concentration of the projected matter, which directly affects the ability to produce critical curves. %lensing efficiency (REF). 
Once all the parameters are set, we explore the parameter space by varying $\smash{\sigma_{\mathrm{clus}}}$ only. 

For each of the three mass intervals, we create new parameter sets 
by fixing $\smash{K_{\mathrm{clus}}}$, $q$, $\smash{n_c}$, and
$\smash{K_{\mathrm{gal}}}$ to the values obtained
from the corresponding calibration based on the three reference lenses $A$, $B$ and $C$. 
We vary $\smash{\sigma_{\mathrm{clus}}}$ 
from $2''$ till $20''$, in steps of $\smash{\Delta \sigma_{\mathrm{clus}}} = 2''$. 
This exploration results in  a total of 30 different parameter sets, \ie
10 different values of $\smash{\sigma_{\mathrm{clus}}}$ for each mass interval. 
Finally, we apply \textit{EasyCritics} to the whole area of CFHTLenS by using each of these parameter sets, \ie
we create lensing potential models, convergence maps, and critical curve catalogs for more than $\smash{\sim 150}$ sq. deg, 30 times.
This exhaustive task can be carried out thanks to
the highly parallelized and efficient performance of \textit{EasyCritics},
 which is able to analyze a region of 1 sq. deg in less than 42
 seconds \citep{Stapelberg_2017_EasyCriticsI}. 
These results are analyzed in the next section, where the SL region candidates generated by \textit{EasyCritics} 
 are compared with the KLs and KCs of the reference sample.

\section{Testing EasyCritics: results, analysis, and new SL candidates} \label{sec:results_analysis}

In this section we present the results of \textit{EasyCritics}, 
after processing the whole area of CFHTLenS using 30 different parameter sets.
These outcomes correspond to lensing potential models,
 convergence maps, and critical curve catalogs
of the entire survey. 
We define the detection rate of our approach by comparing the pre-selected SL regions by \textit{EasyCritics} with  the KLs of the reference sample. 
We also determine the most suitable parameter set, as the one yielding the highest detection rate but still generating a low number of SL region candidates.
We then analyze and correlate these results with Einstein radii, mass estimates, 
and luminosity density maps.%, which have been obtained  in previous studies.}
Finally, we present a catalog of new  SL candidates, provided by our approach and  found in the completely automated
way.

\subsection{The detection rate of  \textit{EasyCritics}}

Having processed the survey, we study the performance of our algorithm  
by comparing the SL region candidates generated by \textit{EasyCritics},
 with each of the 30 parameter sets, with the KLs of the reference sample.

We  define the detection rate of our approach as the ratio between
the number of pre-selected SL regions\footnote{The center of our super-critical regions is defined as in \cite{Stapelberg_2017_EasyCriticsI}.}
 that coincide with the center of  KLs, within a certain radius $\smash{r_{\mathrm{m}}}$, and the total number
of KLs in our reference sample.
We choose a conservative value for the matching radius, given by $\smash{r_{\mathrm{m}} = 40''}$, 
in order to take into account two independent effects:  
1) the intrinsic offset between the peak of the light
distribution and the cluster-scale DM halos
\citep{Becker2007,Johnston2007, Rozo2009MassRichSDSS, Rozo2010CosmoConstraintsSDSS,OguriTakada2011};
2) and an extra misalignment due to the nature of our approach, 
where uncorrelated structures along the LOS are stacked 
to produce the total lensing potential. 
The choice of $\smash{r_{\mathrm{m}}}$ is also motivated 
by the image size used in the SW project \citep{More_2016_SWI, Marshall_2016_SWII},
 where they visually inspect several thousand tiles of  $\smash{82'' \times 82''}$ size.

The parameter sets that predict most of the KLs 
are placed in the mass range $B$, which corresponds to group and small galaxy cluster scales.
As explained in \S \ref{sec:results:Er_distribution} and \S \ref{conf_B_sigma4}, the mass scaling parameter of $B$
and concentrated DM halos are the most favorable configurations for 
the discovery of group and cluster lenses.
In configuration $\smash{B({\sigma_{\mathrm{clus}}} = 2'')}$, 
36 of the $\sim3300$ pre-selected SL regions by \textit{EasyCritics} coincide with the lenses
from our reference sample. Counting the 44 KLs in the sample, this corresponds to a detection rate of $\sim82\%$.   
These results are followed by those from configuration $\smash{B({\sigma_{\mathrm{clus}}} = 4'')}$, 
where 32 of the KLs are identified.
In this setup \textit{EasyCritics} reaches a detection rate of $\sim73\%$ by pre-selecting $\sim1200$ regions for inspection.
The detection rates in the other two mass intervals are poor, and therefore, they are not considered in the next analyses.

%%%%%%%%%%%%%%%%%%%%%%%%%%%%%%%%
% K_clus vs Sigma_clus: its impact on the SL predictions
%%%%%%%%%%%%%%%%%%%%%%%%%%%%%%%%
\begin{figure*}
\begin{center}
 \hspace{-0.4cm}
\begin{tabular}{c c c}
\includegraphics[height=48mm,trim=  70mm 100mm 40mm 30mm,clip]{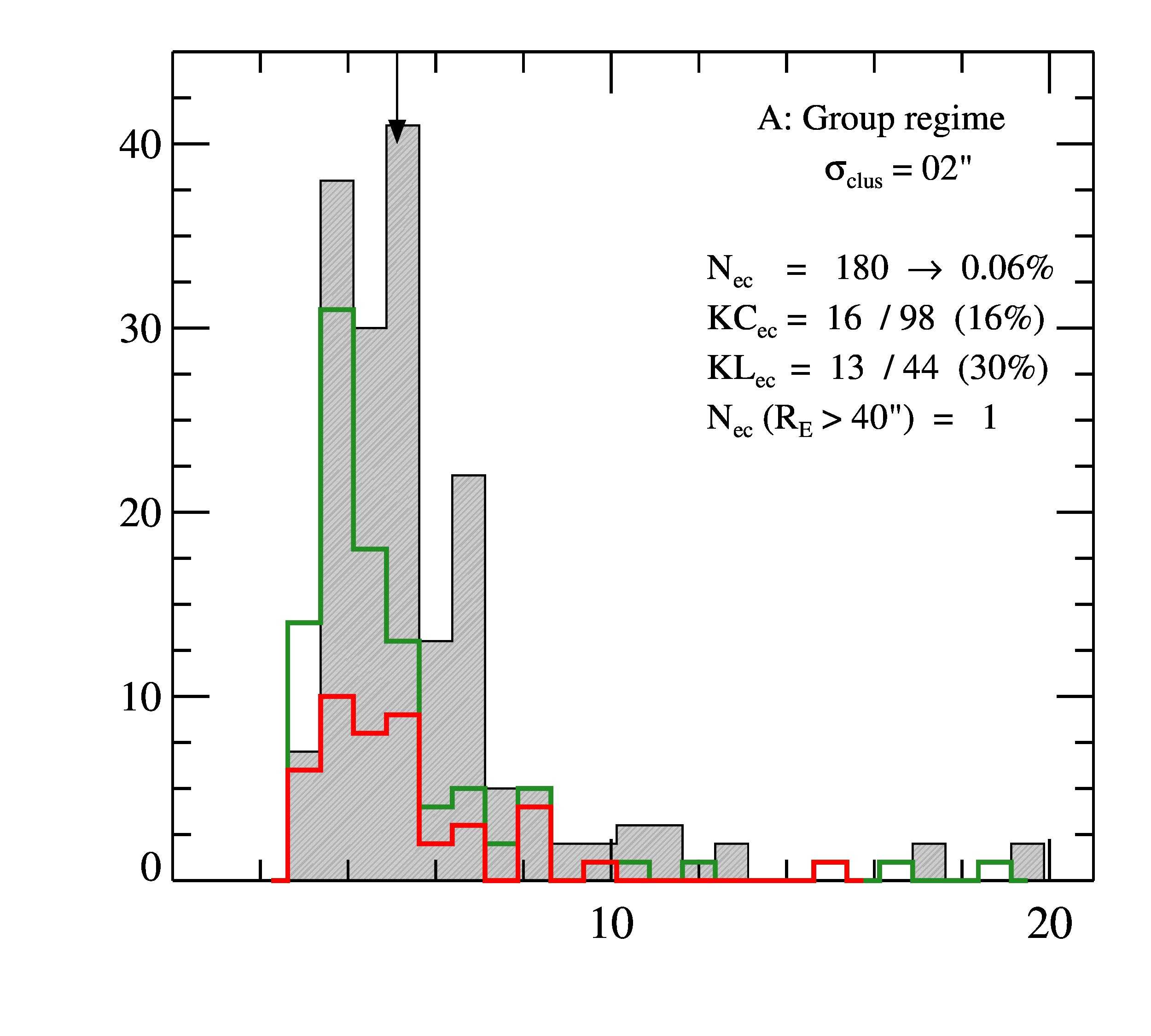} &
\includegraphics[height=48mm,trim=  70mm 100mm 40mm 30mm,clip]{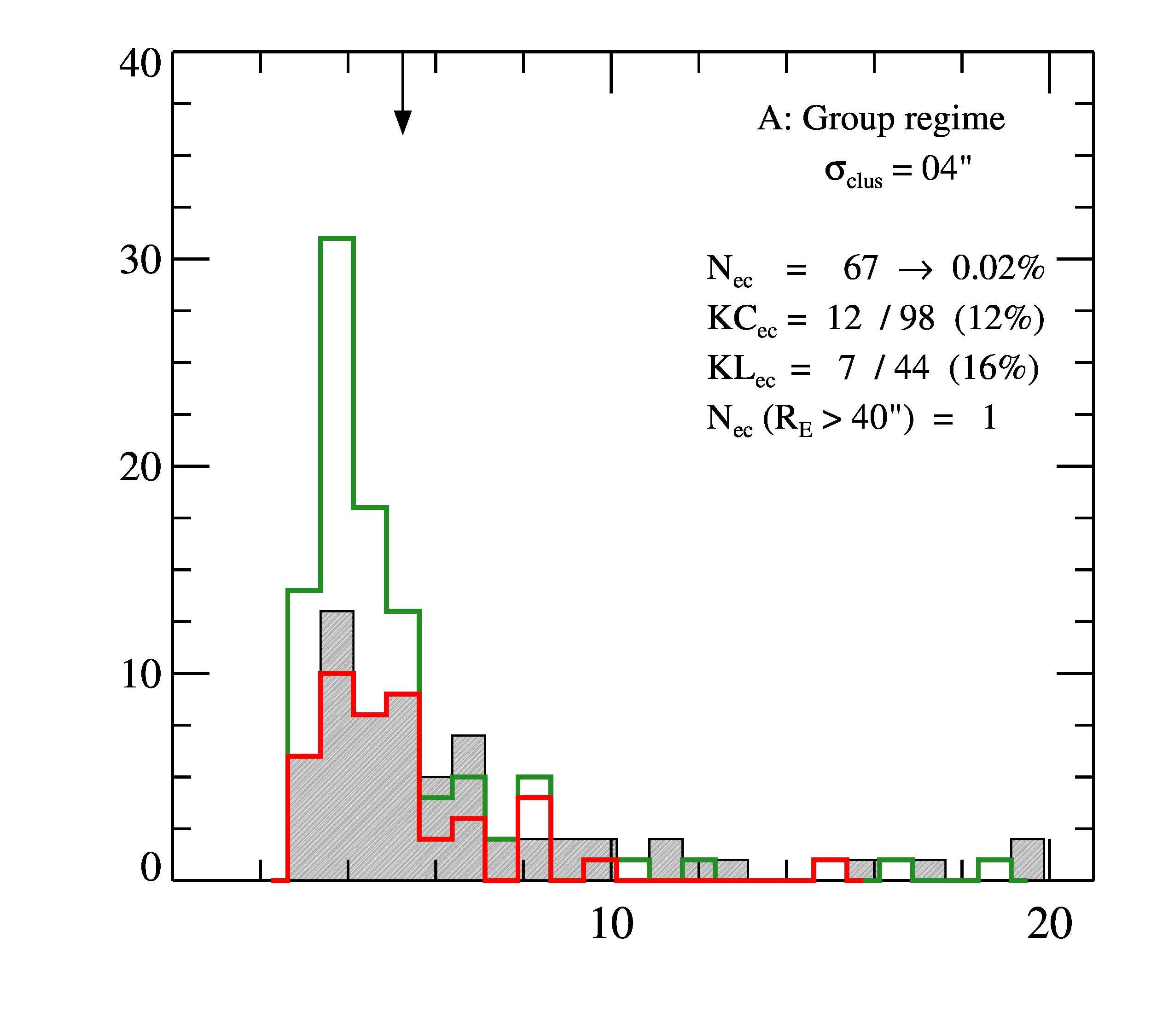} &
\includegraphics[height=48mm,trim=  70mm 100mm 40mm 30mm,clip]{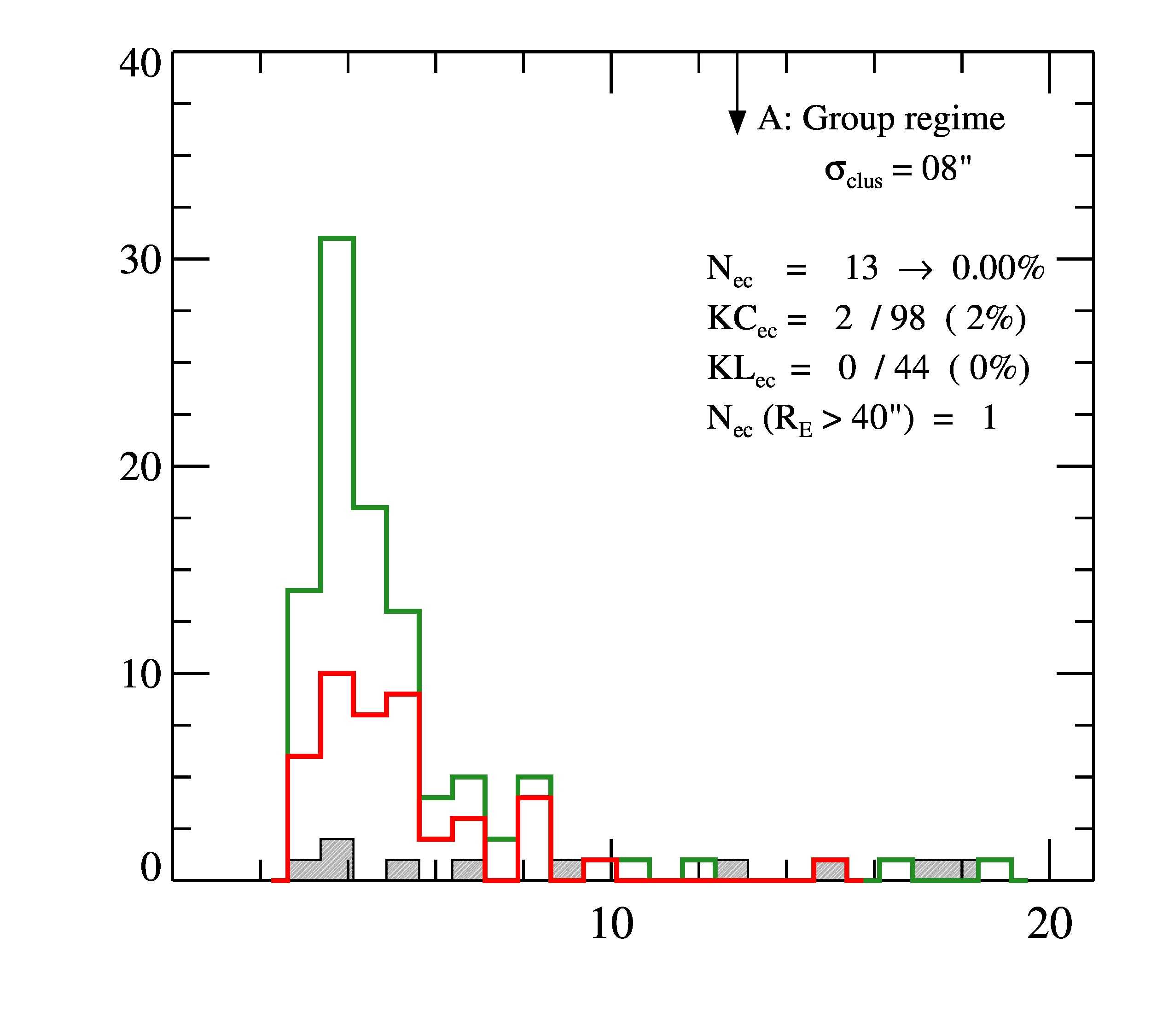} \\
\includegraphics[height=48mm,trim=  70mm 100mm 40mm 30mm,clip]{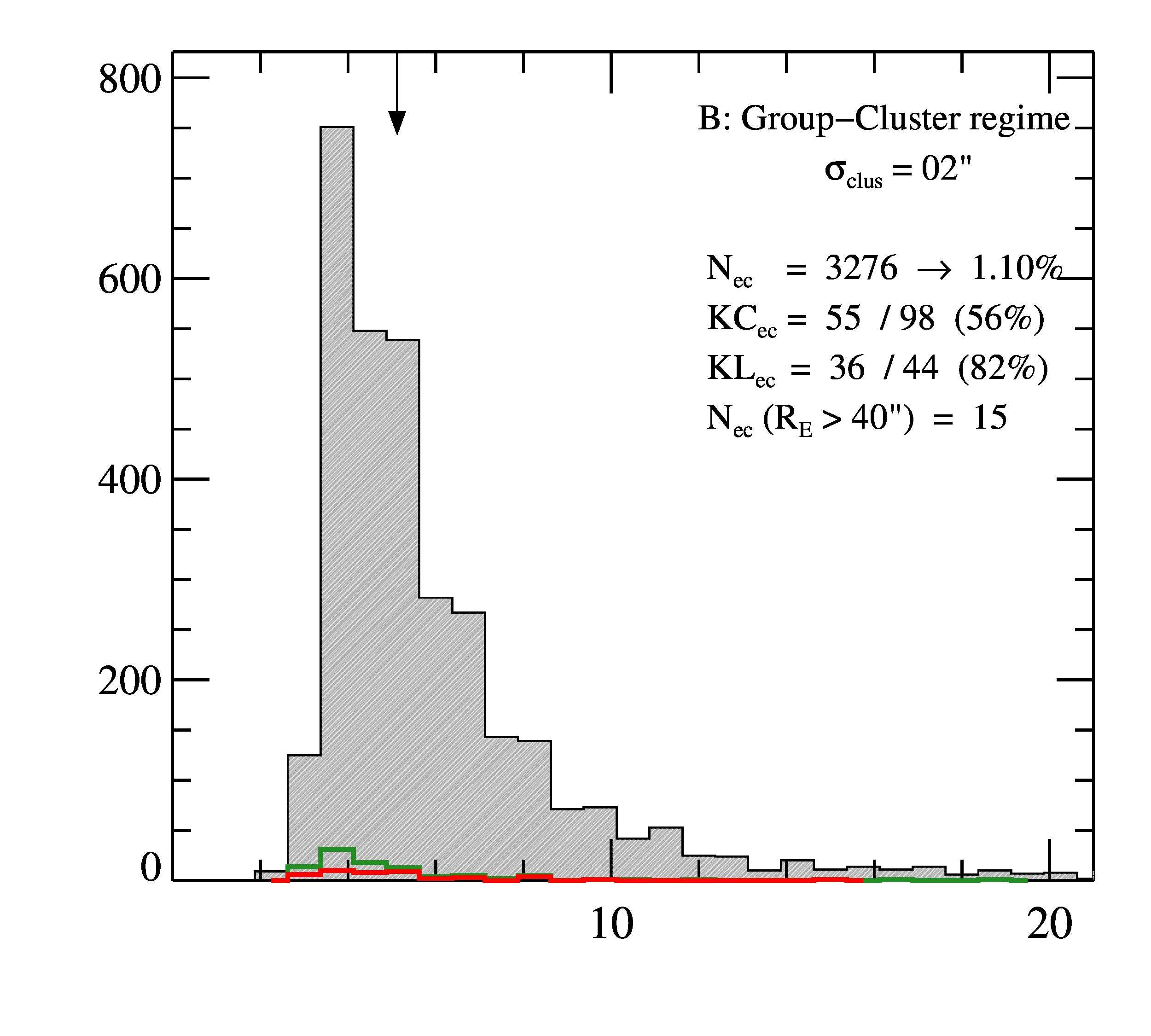} &
\includegraphics[height=48mm,trim=  70mm 100mm 40mm 30mm,clip]{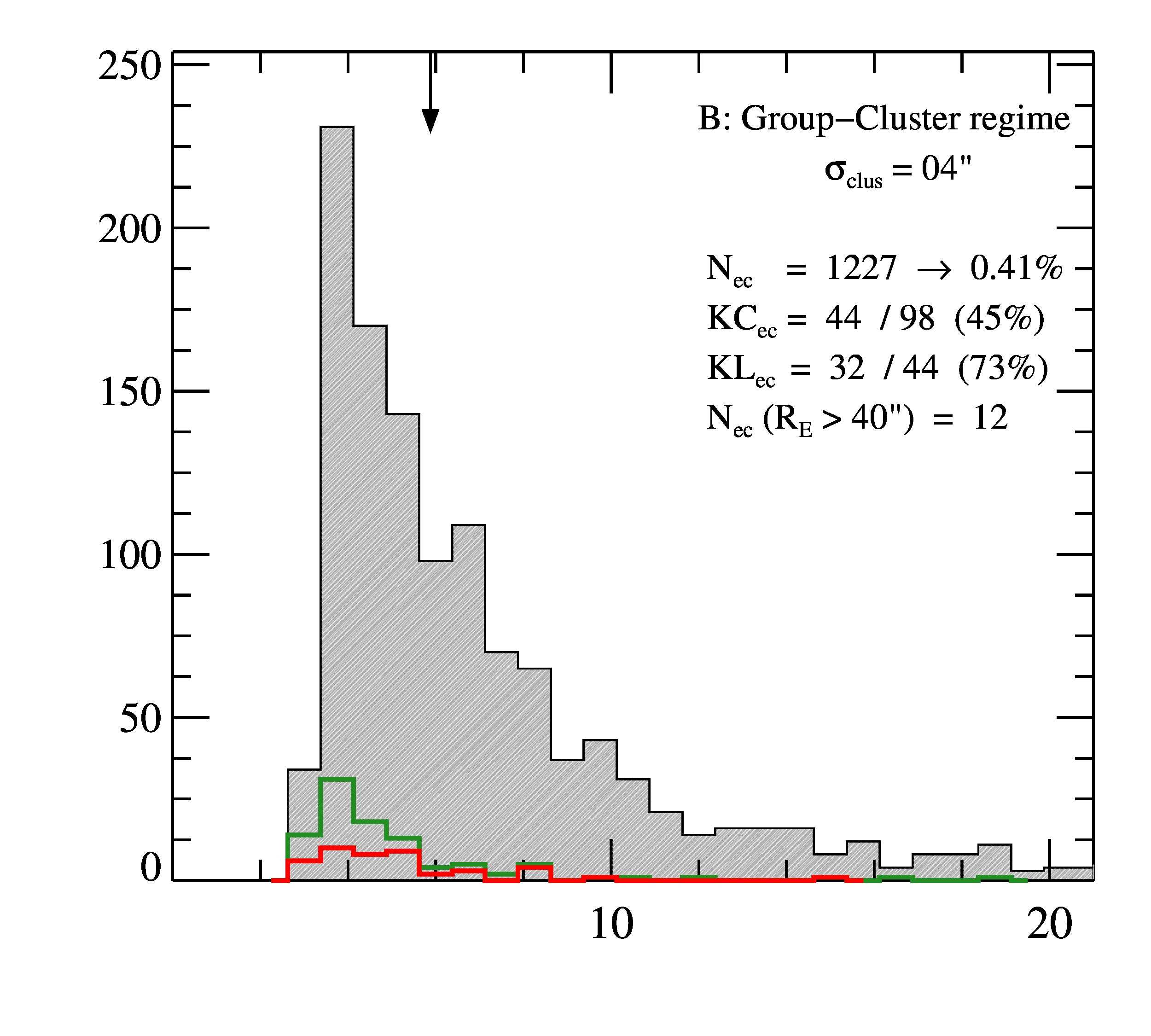} &
\includegraphics[height=48mm,trim=  70mm 100mm 40mm 30mm,clip]{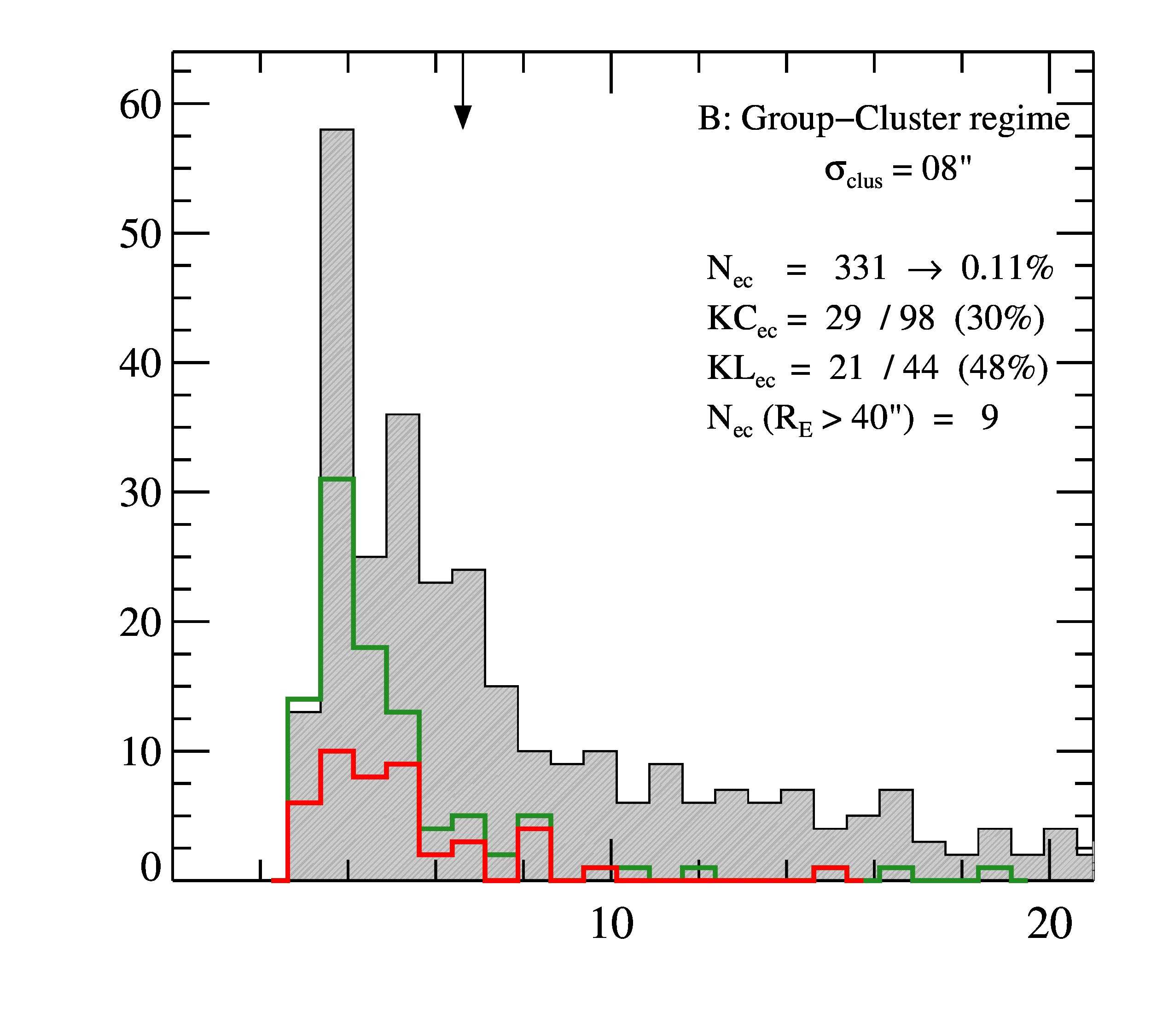} \\
\includegraphics[height=53.2mm,trim=  70mm  28mm 40mm 30mm,clip]{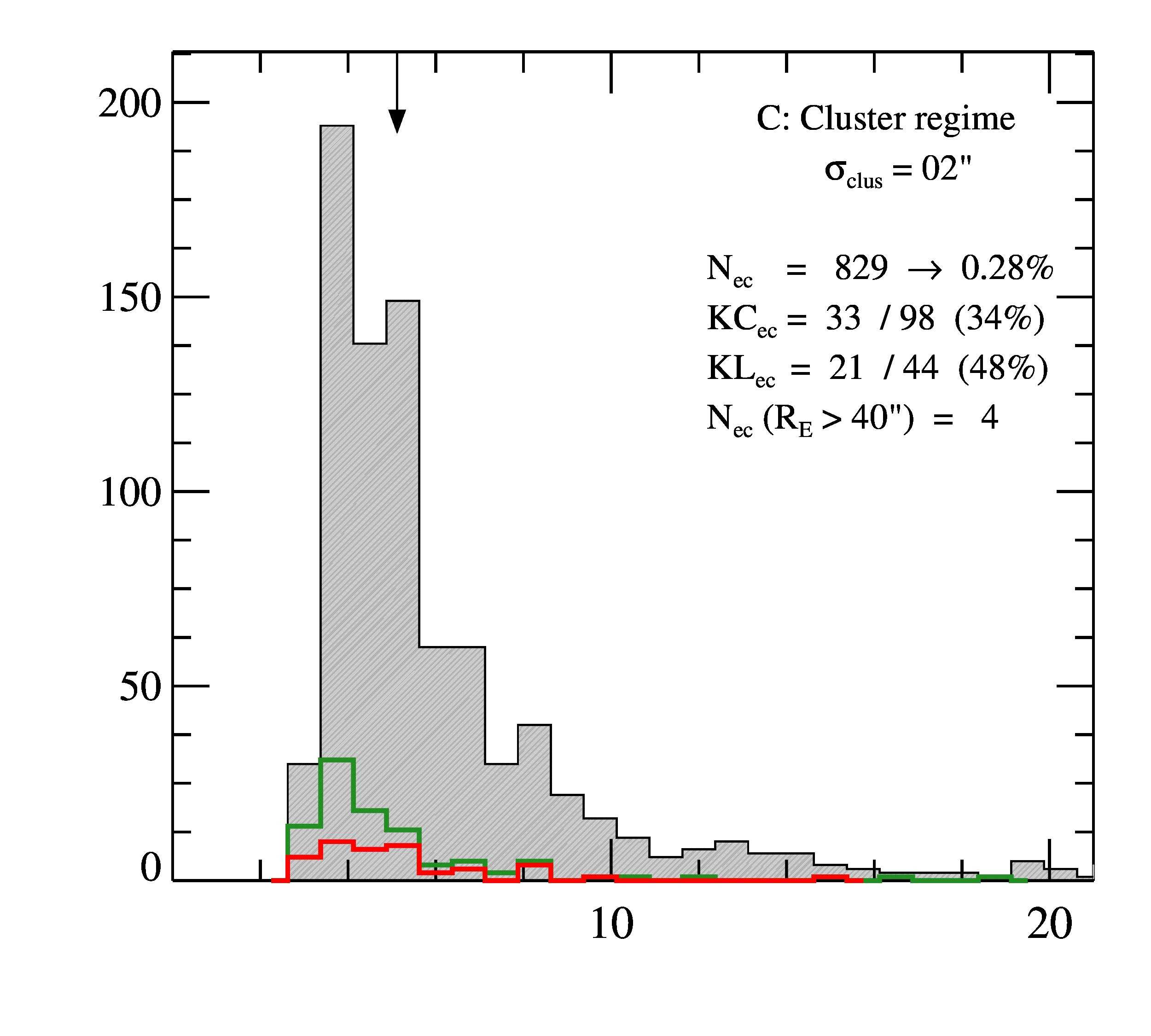} &
\includegraphics[height=53.2mm,trim= 70mm  28mm 40mm 30mm,clip]{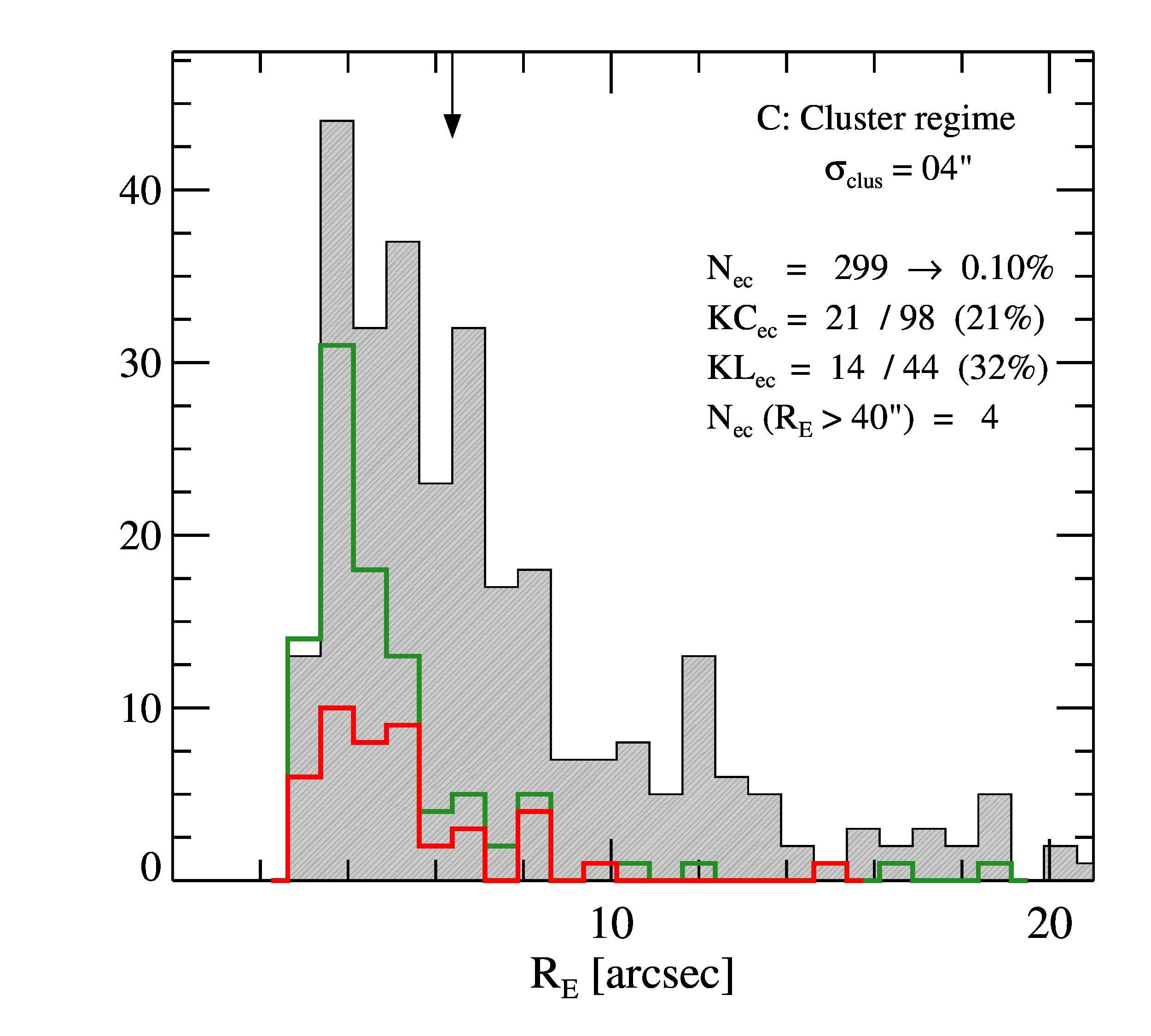} &
\includegraphics[height=53.2mm,trim= 70mm  28mm 40mm 30mm,clip]{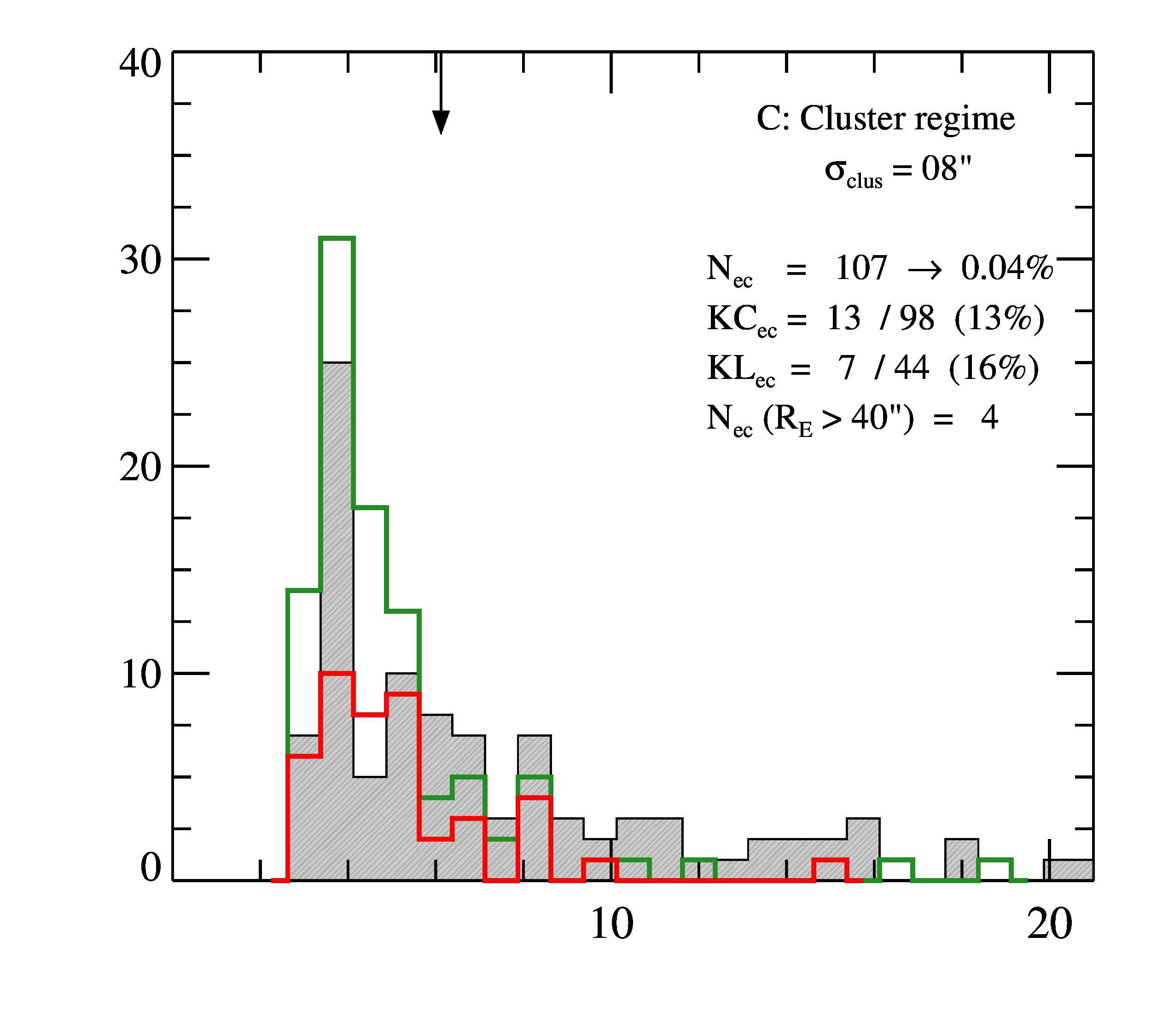} \\
\end{tabular}
\caption{\label{Fig:3mass_regime_3sigma}
Einstein-radius distribution for different parameter set configurations. The gray histograms correspond to 
the Einstein-radius distributions of the SL region candidates generated by \textit{EasyCritics} 
in the three mass intervals described in the text ($A$, $B$, and $C$: from the upper panels to the lower panels), 
and for different values of $\smash{\sigma_{\mathrm{clus}}}$. 
The green histograms correspond to the Einstein-radius distributions of the KCs from the reference sample
(Table \ref{table:app:reference_sample}). Among these 98 KCs, 44 of them have been already confirmed as secure lenses, which
are represented by the red histograms and constitute the KLs sample.
In each panel is given the total number of candidates, $\smash{{\mathrm{N_{ec}}}}$, generated by \textit{EasyCritics}
in the corresponding configuration, 
and the effective area for inspection (in percentage) 
when compared with the total area of CFHTLenS (details in \S\ref{sec:results:new_candidates}). 
The number of predicted KCs and KLs by  \textit{EasyCritics} ($\smash{{\mathrm{KC_{ec}}}}$ 
and $\smash{{\mathrm{KL_{ec}}}}$, respectively) are also presented in each configuration, 
as well as the number of SL regions that have extremely large  Einstein radii ($\smash{{\mathrm{R_{E}}} > 40''}$).
The black arrows indicate the median of each Einstein-radius distribution.
Note that the scale on the `$y$' axis differs between plots.}
\end{center}
\end{figure*}
%%%%%%%%%%%%%%%%%%%%%%%%%%%%%%%% 
%%%%%%%%%%%%%%%%%%%%%%%%%%%%%%%% 

\subsection{Einstein-radius distributions}   \label{sec:results:Er_distribution}

It has been shown that the Einstein radius, $\smash{\mathrm{R_{E}}}$, is the most direct measurement 
of the total inner mass of SL systems \citep[e.g.][]{BroadhurstBarkana08, Zitrin11MACS};
and therefore, by investigating its distribution  
one can obtain a direct understanding of the
halo mass distribution of the lens population under study.
In Fig. \ref{Fig:3mass_regime_3sigma} we show the distribution of the Einstein radii returned 
by \textit{EasyCritics}\footnote{The derivation of $\smash{\mathrm{R_{E}}}$ is given in \cite{Stapelberg_2017_EasyCriticsI}.} 
obtained from 9 different parameter sets, including the most significant
 cases we considered.

For the three mass intervals ($A$, $B$, and $C$), the Einstein-radius distributions
of the pre-selected SL regions show a qualitatively similar behavior.
There is a strong increment in the number 
of super-critical regions generated by \textit{EasyCritics}, together with 
a slight displacement of their distribution peak 
(from $\sim6''$ till $5''$; black arrows in Fig. \ref{Fig:3mass_regime_3sigma}),
when the smoothing window decreases from $\smash{\sigma_{\mathrm{clus}} = 8''}$
till $\smash{\sigma_{\mathrm{clus}} = 2''}$.
This correlation arises because the size of $\smash{\sigma_{\mathrm{clus}} }$
determines the capability of the 2D Gaussian function of boosting  the cluster-scale component. 
Then, a small kernel produces a convolution function that concentrates most of the light and hence, 
most of the mass, in a very small region. 
This enhances the surface mass density 
for poor LRG environments when having high spatial densities,
i.e. groups composed of few galaxies ($\sim3-6$) that are highly concentrated 
 can also reach the required amount of surface mass density to be critical.
This effect is better seen by comparing the middle and left panels of Fig. \ref{Fig:3mass_regime_3sigma}, 
 where the total number of super-critical regions increases by a factor of $\sim3$
  when $\smash{\sigma_{\mathrm{clus}}}$ decreases from $4''$ to $2''$.

It should be mentioned that we observe, in all the explored parameter sets,
a maximum of only 15 candidates having Einstein radii larger than $40''$; of which 8 actually correspond 
to spurious detections given by photometric problems in the original catalogs of CFHTLenS,
as explained in the following section. The other 7 remaining regions correspond to 
different group-scale objects aligned on the LOS but with  relatively high offsets of the order of $\sim20''$.  %($\sim30''$).
One of these SL regions actually contains (at $\sim20''$ from its center)
 the known candidate SA8 (Table \ref{table:kl_kc_by_EasyCritics}),
  located in a rich environment of LRGs.
Despite this, we could not identify any SL signature on the other 6 regions, 
and therefore, they are not selected as possible SL candidates.

%%%%%%%%%%%%%%%%%%%%%%%%%%%%%%%%%%%%%%%%%%%%%%%%%%%%%%5
%%%%%%%%%%%%%%%%%%%%%%%%%%%%%%%%%%%%%%%%%%%%%%%%%%%%%%%%
% Table of KLs and KCs found by EasyCritics
\begin{table*}
\centering
\caption{SL regions pre-selected by \textit{EasyCritics} in configuration $\smash{B({\sigma_{\mathrm{clus}}} = 4'')}$: 32 KLs and 12 KCs.   \label{table:kl_kc_by_EasyCritics}}
\hspace{0.0cm}
\resizebox{\textwidth}{!}{
\begin{tabular}{l l r r l c r c c c r}
\toprule \toprule
ID$^a$ & R.A. & Dec. & $R_\mathrm{E}^b$ & Lens ID$^c$ & $z_\mathrm{phot}$ & $R_\mathrm{arc}^d$ & $\sigma_{\mathrm{SIS}}^e$ &$ L^{\mathrm{T}}(\mathrm{< 1 \hspace{0.1cm} Mpc})^e$    &   Rank$^f$ &   Offset$^g$  \\
   & [J2000] & [J2000] & [$''$]  &      &     & [$''$] &  [km s$^{-1}$]  &  [$\times 10^{12} L_{\odot}$]  &    &   [$''$]  \\
\midrule
SLEC-J0202-1109 & 02:02:10.34  & -11:09:09.8  & 18.3  &  SA2            &  0.48  &  5.0  &  758$^{+88}_{-153}$   &   2.72$\pm$0.21  &   KL &   3.0  \\
SLEC-J0203-0734 & 02:03:20.40  & -07:34:43.3  & 13.5  &  SA6            &  0.59  &  5.0  &  914$^{+89 }_{-123}$  &   1.92$\pm$0.12  &   KL &   7.5  \\
SLEC-J0203-0942 & 02:03:49.88  & -09:42:55.8  & 4.8   &  SA7            &  0.25  &  5.0  &  -- --                &   -- --          &   KL &   2.7  \\
SLEC-J0205-1105 & 02:05:03.22  & -11:05:47.0  & 5.5   &  SA9            &  0.62  &  3.3  &  543$^{+139}_{-191}$  &   1.20$\pm$0.12  &   KL &   1.0  \\
SLEC-J0206-0657 & 02:06:47.92  & -06:57:01.3  & 35.4  &  SA10           &  0.49  &  3.2  &  856$^{+114}_{-100}$  &   3.65$\pm$0.36  &   KL &   8.2  \\
SLEC-J0209-0643 & 02:09:29.66  & -06:43:08.5  & 17.5  &  SA14           &  0.45  &  3.2  &  -- --                &   -- --          &   KL &   5.6  \\
SLEC-J0214-0535 & 02:14:08.24  & -05:35:22.5  & 23.5  &  SA22           &  0.44  &  7.1  &  638$^{+101}_{-152}$  &   2.21$\pm$0.19  &   KL &  10.1  \\
SLEC-J0215-0440 & 02:15:28.95  & -04:40:47.0  & 16.1  &  arc68c         &  0.31  &  8.0  &  -- --                &   -- --          &   KL &   9.7  \\
SLEC-J0219-0528 & 02:19:55.97  & -05:28:05.3  & 3.5   &  SA36           &  0.35  &  4.0  &  724$^{+65 }_{-107}$  &   3.18$\pm$0.28  &   KL &   9.0  \\
SLEC-J0224-1058 & 02:24:09.40  & -10:58:08.0  & 8.6   &  SW1            &  0.50  &  4.8  &  -- --                &   -- --          &   KL &   2.1  \\
SLEC-J0225-0737 & 02:25:46.11  & -07:37:41.5  & 9.4   &  SA50           &  0.51  &  5.8  &  540$^{+130}_{-172}$  &   1.32$\pm$0.15  &   KL &   3.0  \\
SLEC-J0227-1056 & 02:27:16.45  & -10:56:02.5  & 4.0   &  SW22           &  0.40  &  4.8  &  -- --                &   -- --          &   KL &   0.2  \\
SLEC-J0227-0451 & 02:27:40.34  & -04:51:32.0  & 8.5   &  XLSSC022       &  0.29  &  5.0  &  -- --                &   -- --          &   KL &   1.4  \\
SLEC-J0229-0554 & 02:29:17.34  & -05:54:05.8  & 6.5   &  SA55           &  0.38  &  3.2  &  701$^{+69 }_{-109}$  &   2.58$\pm$0.47  &   KL &   0.4  \\
SLEC-J0852-0343 & 08:52:07.20  & -03:43:15.8  & 4.8   &  SA63           &  0.48  &  5.0  &  561$^{+116}_{-155}$  &   1.51$\pm$0.16  &   KL &   0.6  \\
SLEC-J0854-0121 & 08:54:46.60  & -01:21:37.3  & 10.3  &  SA66           &  0.35  &  4.8  &  644$^{+69 }_{-102}$  &   2.42$\pm$0.17  &   KL &   0.7  \\
SLEC-J0858-0240 & 08:58:49.95  & -02:40:00.3  & 25.8  &  SA71           &  0.36  &  3.7  &  -- --                &   -- --          &   KL &  38.3  \\
SLEC-J0859-0345 & 08:59:14.54  & -03:45:14.7  & 6.0   &  SA72           &  0.64  &  4.5  &  466$^{+150}_{-160}$  &   1.44$\pm$0.16  &   KL &   0.1  \\
SLEC-J0901-0158 & 09:01:39.23  & -01:58:56.3  & 15.0  &  SL2SJ0901-0158 &  0.29  &  6.8  &  -- --                &   -- --          &   KL &   5.3  \\
SLEC-J1357+5317 & 13:57:25.69  & +53:17:42.7  & 15.8  &  SA87           &  0.54  &  3.5  &  425$^{+192}_{-124}$  &   2.35$\pm$0.28  &   KL &   2.2  \\
SLEC-J1401+5654 & 14:01:10.37  & +56:54:20.9  & 4.9   &  SA90           &  0.53  &  3.7  & 1015$^{+70 }_{-79 }$  &   2.82$\pm$0.27  &   KL &   0.8  \\
SLEC-J1405+5445 & 14:05:54.47  & +54:45:48.4  & 10.8  &  SA96           &  0.41  &  3.9  &  449$^{+123}_{-138}$  &   2.74$\pm$0.42  &   KL &   1.1  \\
SLEC-J1408+5429 & 14:08:13.70  & +54:29:03.7  & 10.0  &  SA97           &  0.42  &  8.0  &  384$^{+162}_{-109}$  &   1.09$\pm$0.23  &   KL &   4.5  \\
SLEC-J1414+5447 & 14:14:47.24  & +54:47:04.4  & 28.5  &  SA100          &  0.63  &  14.7 &  969$^{+100}_{-130}$  &   5.49$\pm$0.35  &   KL &   0.9  \\
SLEC-J1419+5326 & 14:19:11.92  & +53:26:13.7  & 9.3   &  SA102          &  0.69  &  9.9  & 1028$^{+140}_{-272}$  &   3.40$\pm$0.27  &   KL &   3.1  \\
SLEC-J1429+5625 & 14:29:34.06  & +56:25:40.2  & 14.6  &  SW4            &  0.50  &  5.9  &  -- --                &   -- --          &   KL &   1.6  \\
SLEC-J1431+5533 & 14:31:39.72  & +55:33:24.7  & 3.4   &  SA113          &  0.67  &  4.0  &  745$^{+139}_{-210}$  &   2.21$\pm$0.23  &   KL &   1.9  \\
SLEC-J2202+0234 & 22:02:57.01  & +02:34:33.2  & 21.6  &  SW7            &  0.50  &  6.8  &  -- --                &   -- --          &   KL &   2.3  \\
SLEC-J2206+0411 & 22:06:41.94  & +04:11:30.7  & 4.1   &  SA121          &  0.62  &  3.7  &  443$^{+170}_{-137}$  &   1.76$\pm$0.26  &   KL &   1.3  \\
SLEC-J2213-0018 & 22:13:06.30  & -00:18:28.8  & 10.5  &  arc20a         &  0.49  &  5.0  &  -- --                &   -- --          &   KL &   9.0  \\
SLEC-J2214+0110 & 22:14:18.98  & +01:10:31.9  & 20.4  &  SA125          &  0.74  &  8.0  &  695$^{+176}_{-235}$  &   3.66$\pm$0.34  &   KL &   3.0  \\
SLEC-J2220+0058 & 22:20:52.24  & +00:58:14.9  & 13.5  &  arc54c         &  0.41  &  8.0  &  -- --                &   -- --          &   KL &  11.0  \\
SLEC-J0204-1024 & 02:04:55.36  & -10:24:18.3  & 40.8  &  SA8            &  0.33  &  10.8 &  654$^{+61 }_{-96 }$  &   2.85$\pm$0.44  &   KC &  20.1  \\
SLEC-J0209-0354 & 02:09:57.61  & -03:54:58.0  & 3.6   &  SA15           &  0.44  &  3.9  &  534$^{+129}_{-152}$  &   0.78$\pm$0.06  &   KC &   1.3  \\
SLEC-J0216-0935 & 02:16:04.63  & -09:35:06.5  & 7.8   &  SA26           &  0.69  &  16.4 &  853$^{+145}_{-153}$  &   5.60$\pm$0.31  &   KC &   0.4  \\
SLEC-J0228-0949 & 02:28:32.70  & -09:49:27.5  & 33.5  &  SA54           &  0.45  &  6.3  &  793$^{+72 }_{-96 }$  &   3.73$\pm$0.38  &   KC &  20.3  \\
SLEC-J0848-0407 & 08:48:23.51  & -04:07:26.5  & 10.5  &  SA61           &  0.51  &  7.4  &  677$^{+108}_{-133}$  &   3.30$\pm$0.35  &   KC &  11.4  \\
SLEC-J0900-0230 & 09:00:49.33  & -02:30:53.5  & 28.0  &  SA74           &  0.36  &  3.2  &  672$^{+89 }_{-94 }$  &   2.83$\pm$0.31  &   KC &  11.5  \\
SLEC-J1356+5527 & 13:56:49.31  & +55:27:07.2  & 8.9   &  SA86           &  0.46  &  3.7  &  600$^{+82 }_{-151}$  &   2.88$\pm$0.25  &   KC &   0.2  \\
SLEC-J1402+5257 & 14:02:06.35  & +52:57:07.2  & 4.5   &  arc81c         &  0.51  &  5.0  &  -- --                &   -- --          &   KC &   0.4  \\
SLEC-J1411+5212 & 14:11:20.91  & +52:12:09.7  & 29.6  &  SA98           &  0.52  &  18.4 &  932$^{+107}_{-133}$  &   3.37$\pm$0.46  &   KC &   3.5  \\
SLEC-J1428+5213 & 14:28:34.84  & +52:13:13.2  & 18.4  &  SA111          &  0.52  &  5.0  &  889$^{+67 }_{-88 }$  &   3.73$\pm$0.26  &   KC &   6.7  \\
SLEC-J2210+0023 & 22:10:33.03  & +00:23:51.4  & 3.5   &  arc53c         &  0.58  &  4.0  &  -- --                &   -- --          &   KC &   1.1  \\
SLEC-J2215+0102 & 22:15:13.77  & +01:02:39.9  & 9.6   &  arc23a         &  0.69  &  7.0  &  -- --                &   -- --          &   KC &   5.6  \\
                                                                                                          
\bottomrule                                                                                               
\end{tabular}                                                                                             
}             
{\footnotesize\flushleft
{}$^a$ ID of the SL region candidates generated by \textit{EasyCritics}. The name is given by the acronym 
\textbf{S}trong \textbf{L}ensing regions by \textit{\textbf{E}asy\textbf{C}ritcs} (SLEC) 
and the corresponding location in sexagesimal coordinates.\\
{}$^b$ Einstein radius derived from the critical curves of the SL regions.\\
{}$^c$ IDs from \cite{Cabanac07}, \cite{Limousin09_SL_Ggroup},
\cite{More_2012_arcfinder}, \cite{Maturi14}, and \cite{More_2016_SWI}. The corresponding publications
of each lens are listed in Table \ref{table:app:reference_sample}, Appendix \ref{app:reference_sample}.\\
{}$^d$ Corresponds to the distance from the BCG of the lens system
till the average location of the arc. \\
{}$^e$ Measurements from \cite{Foex13_SARCS_WL}. \\
{}$^f$ Ranking from previous studies. KL stands for known lenses, while
KC stands for known candidates. More details in Table \ref{table:app:reference_sample}, Appendix \ref{app:reference_sample}.\\
{}$^g$ Distance between the location of the KLs or KCs and the center of the predicted critical curves,
defined in \cite{Stapelberg_2017_EasyCriticsI}. \\
}                                                                                           
\end{table*}
%%%%%%%%%%%%%%%%%%%%%%%%%%%%%%%%%%%%%%%%%%%%%%%%%%%%%%%%
%%%%%%%%%%%%%%%%%%%%%%%%%%%%%%%%%%%%%%%%%%%%%%%%%%%%%%%%

\subsection{The most suitable parameter set: $\smash{B({\sigma_{\mathrm{clus}}} = 4'')}$} \label{conf_B_sigma4}

As mentioned above, the parameter sets that yield the highest efficiency to correctly
identify KLs belong to the intermediate mass interval $B$, 
which maximizes the probability of generating lenses at the group- and small cluster-scales.
These results are indeed expected, since the majority of the  galaxies
of the Universe are living in groups \citep[][]{Limousin09_SL_Ggroup}. 
In fact, extrapolating the results from \cite{Oguri06} to the total area of the CFHTLenS
($\smash{\sim154}$ sq. deg) and assuming an equivalent 
magnification factor in all cases, we anticipate $\sim68$ lenses in the group-scale 
and $\sim18$ SL massive galaxy clusters, \ie we expect to discover about 4 times more 
SL galaxy groups than clusters \citep{Limousin09_SL_Ggroup, Cabanac07}.

In configuration $\smash{B({\sigma_{\mathrm{clus}}} = 4'')}$, \textit{EasyCritics} generates only $\sim1200$ possibly super-critical regions.
Among these pre-selected SL regions, we identify 32 of the 44 previously KLs 
corresponding to a detection rate of $\sim73\%$.
We adopt this specific setup as the `best' parameter set because of 
the high number of identified KLs and the relatively low amount of SL region candidates,
which helps to decrease both the total post-processing time and the size in the final sample of tentative SL candidates.
However, determining the parameter set representing the best compromise between purity and completeness 
can only be done with extensive simulations or exhaustive observational follow-up campaigns.
The  coordinates of the pre-selected SL regions generated by our approach in configuration $\smash{B({\sigma_{\mathrm{clus}}} = 4'')}$ 
that correctly match the KLs of the reference sample,
 together with their main properties, are listed in Table \ref{table:kl_kc_by_EasyCritics}.
Note that when compared with the KCs, the efficiency drops to $\sim 45\%$; however, it is expected
to find few previously known low ranked candidates, since their nature is not clear yet.

The Einstein-radius distribution (central panel in Fig. \ref{Fig:3mass_regime_3sigma})
 derived from configuration $\smash{B({\sigma_{\mathrm{clus}}} = 4'')}$
 peaks at  $\smash{\mathrm{R_{E}}^{peak}  \sim 5.9''}$, %{}^{+3.5}_{-1.6}}$, 
 which is in agreement, within the errors, with the results of the Einstein-radius distribution 
derived from 10 000 SDSS clusters in \cite{Zitrin2012UniversalRE}. 
Note that even if most of the SL regions are in the range $\smash{\sim3'' - 15''}$,  %($\smash{> 70\%}$ ), 
our approach also generates super-critical regions with larger Einstein radii.
In fact, several of these massive regions are those 
matching the lenses from the reference samples. These
objects are studied below, where their characteristics are compared with 
the properties of the KLs and KCs.

%%%%%%%%%%%%%%%%%%%%%%%%%%%%%%%%
% Er from Easycritics vs R_arc,  Dispersion vs Er, Luminosity vs Er
%%%%%%%%%%%%%%%%%%%%%%%%%%%%%%%%
\begin{figure}
\begin{center}
\begin{tabular}{c}
\hspace{-0.8cm}
\includegraphics[width=0.52\textwidth,trim= 0mm 0mm 0mm 0mm,clip]{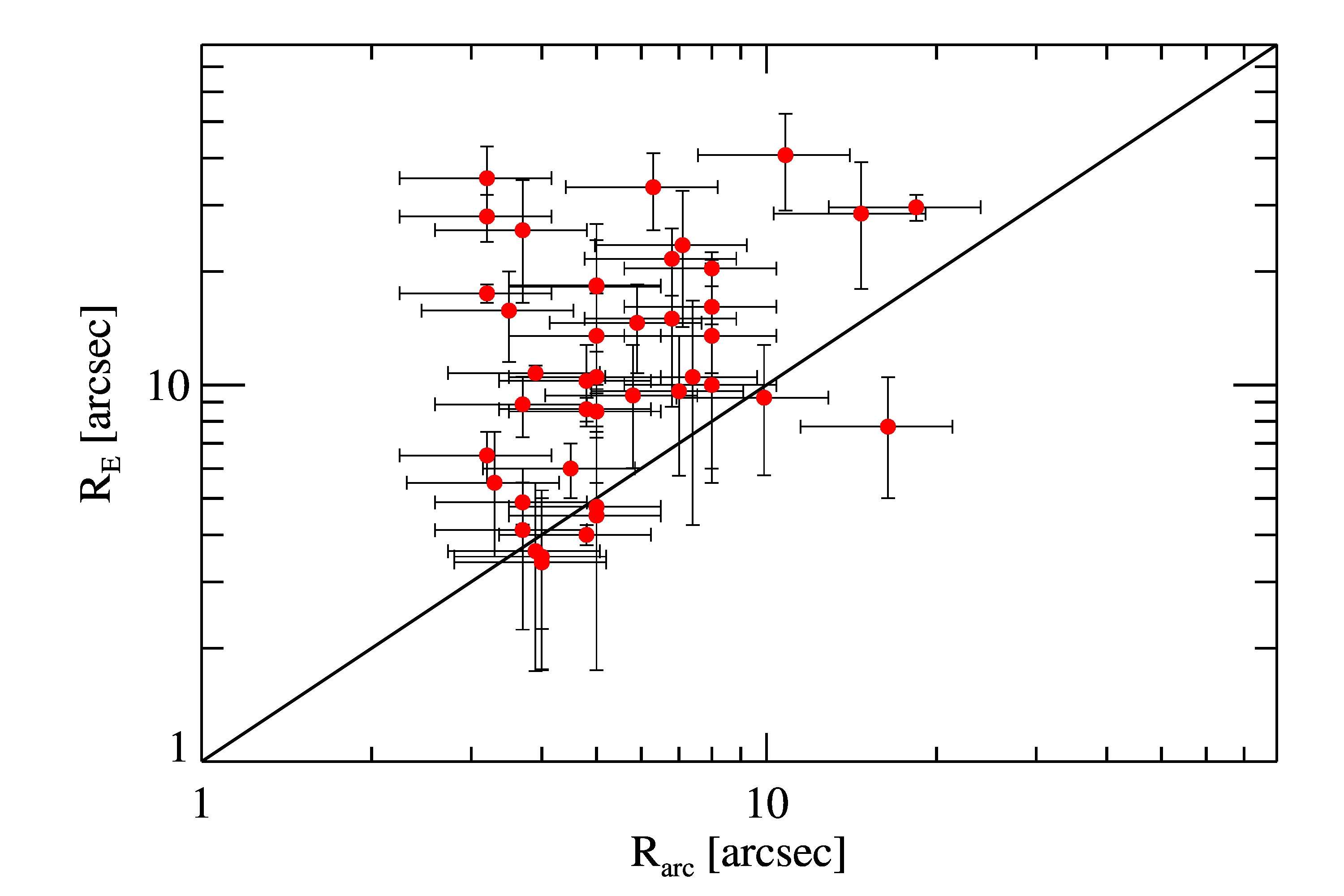}  \\
\hspace{-0.8cm}
\includegraphics[width=0.52\textwidth,trim= 0mm 0mm 0mm 0mm,clip]{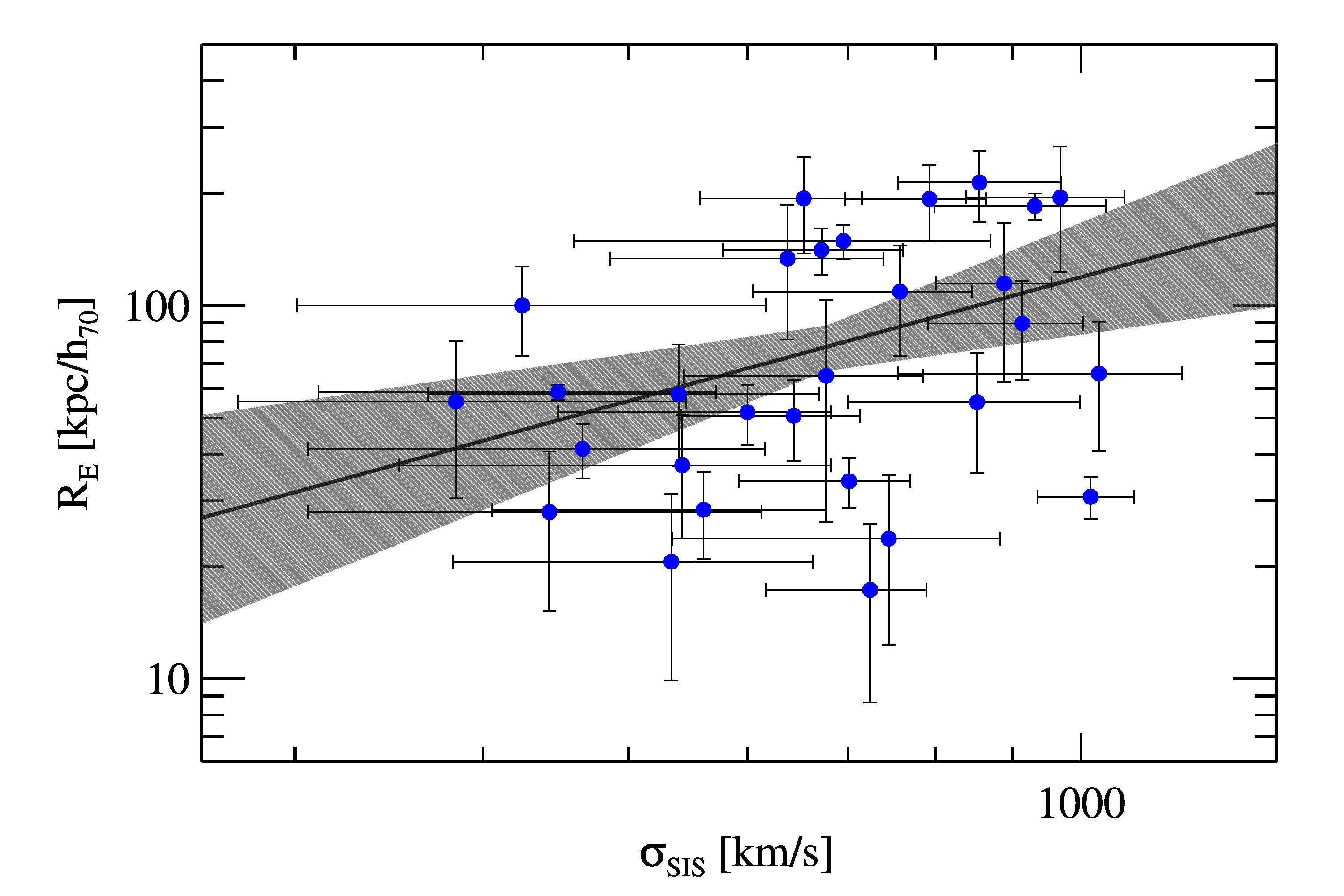}  \\
\hspace{-0.8cm}
\includegraphics[width=0.52\textwidth,trim= 0mm 0mm 0mm 0mm,clip]{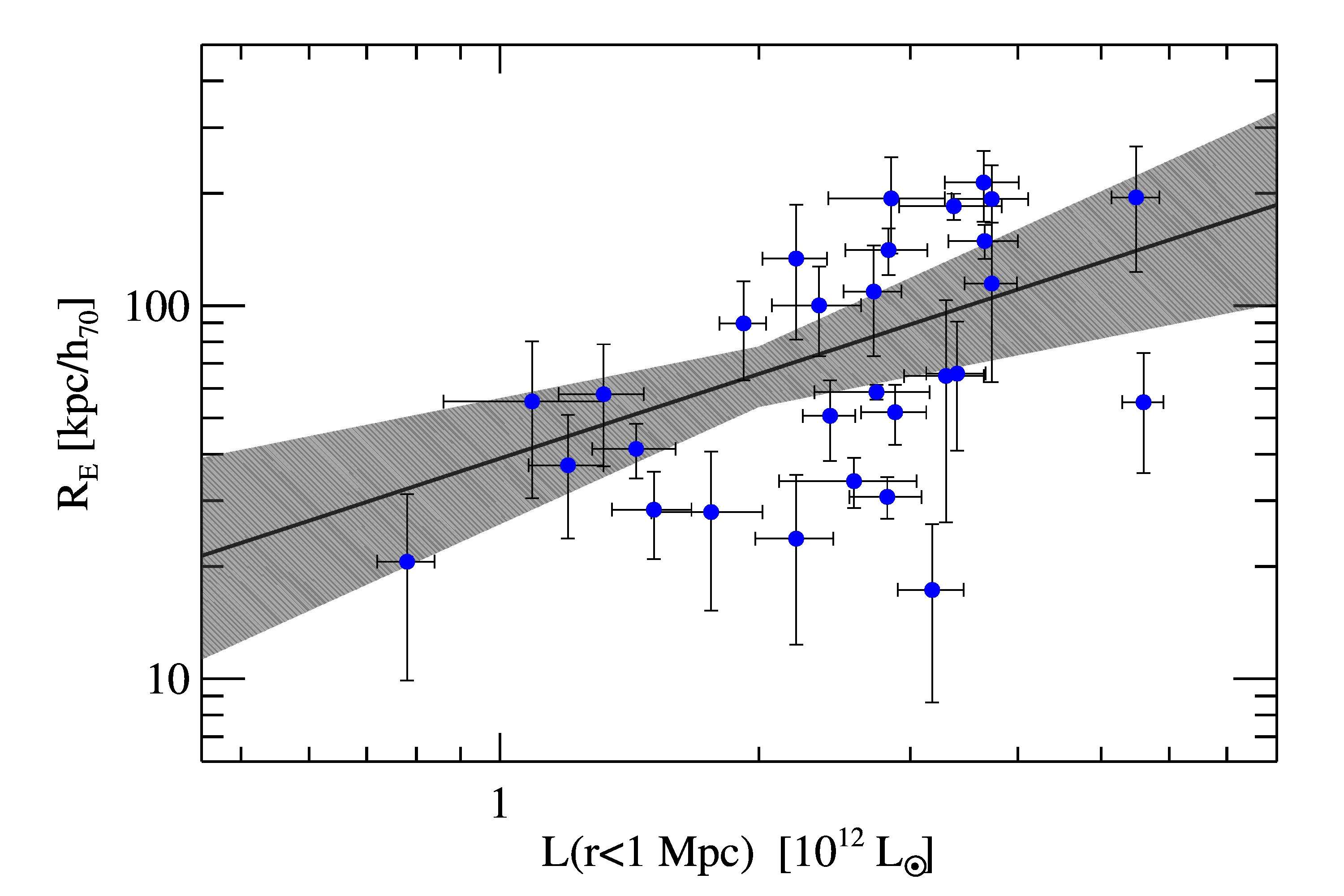}  
\end{tabular}
\put(-220,  240){\bf \huge \color{blue} $a$} 
\put(-220, 60){\bf \huge \color{blue} $b$}     
\put(-220, -120){\bf \huge \color{blue} $c$}   
\caption{\label{Fig:Er_Rarc}
Panel $a$: Einstein radius versus $R_{arc}$. The red points correspond to 44 pre-selected SL regions
 by \textit{EasyCritics} in configuration $\smash{B({\sigma_{\mathrm{clus}}} = 4'')}$, including both the KLs and KCs.
The black solid line correspond to the one-to-one relation.
Panel $b$: Einstein radius vs $\smash{\sigma_{\mathrm{SIS}}}$ .
The blue points correspond to 29 objects matching the SL region
candidates generated by \textit{EasyCritics}, and having velocity dispersion
measurements. The best-fitting power-law scaling relation, given by Eq. (\ref{eq:power_law_fit}),
is represented by the black continuous line, while the errors of its 
parameters are illustrated by the gray region. 
Panel $c$: Einstein radius vs $\smash{L^{\mathrm{T}}(\mathrm{< 1 \hspace{0.1cm} Mpc})}$. 
This figure is displayed in the same fashion as in panel $b$). 
The best-fitting parameters and their errors are listed in Table \ref{table:fit_results}.
} 
\end{center}
\end{figure}
%%%%%%%%%%%%%%%%%%%%%%%%%%%%%%%% 
%%%%%%%%%%%%%%%%%%%%%%%%%%%%%%%% 

\subsection{Einstein radius vs lens properties} \label{sec:correction}

In order to verify whether the features of the pre-selected SL regions by our approach agree
with the properties of the cluster lensing population, we compare our results with  
mass and luminosity estimates of the reference lenses, which have been previously computed in former studies.
In doing so, it is important to remember that here we are not fitting lens by lens as done in the literature,
instead we are predicting the lens properties by using fixed values of the parameters for the whole population.
In configuration $\smash{B({\sigma_{\mathrm{clus}}} = 4'')}$, there are 32 super-critical 
regions matching with KLs and 12 with  KCs. 
We then have a total of 44 pre-selected SL regions that coincide with formerly studied lenses.
These regions and their main properties derived from \textit{EasyCritics}
are listed in Table \ref{table:kl_kc_by_EasyCritics}, together with all the 
relevant information from previous studies. 

We start by comparing the  Einstein radius, $\smash{\mathrm{R_{E}}}$, of our pre-selected SL regions with the arc radius, 
$\smash{\mathrm{R_{arc}}}$, which is defined as the distance from the BCG of the lens system
till the average location of the visible arc. 
A one-to-one relation is not expected due to the fixed value for the source redshift used 
in our calculations and the different redshifts where the lensed galaxies are placed.
Furthermore, we are in general overestimating the inner mass of the regions by decreasing the kernel size
 as explained in \S\ref{sec:results:Er_distribution}. 
Thus, a large scatter is expected as well as an 
excess
in the size of our critical curves as shown in Fig. \ref{Fig:Er_Rarc} (panel $a$).

Then, by assuming that most of the mass is concentrated in the lens plane of the main
deflector, we compute the physical Einstein radius of each region by using the
redshift of the associated lens. 
This quantity is less biased when comparing with the intrinsic properties of lenses, 
since it is corrected by angular scale factors and then represents a physical size in the lens plane.
We use the weak lensing (WL) measurements from \cite{Limousin09_SL_Ggroup} and 
\cite{Foex13_SARCS_WL} to compare the total mass 
 of the reference lenses with our results. 
In these WL studies, they assume a singular isothermal sphere (SIS)
 profile to compute their mass estimates, which is parameterized by the velocity dispersion 
 parameter $\smash{\sigma_{\mathrm{SIS}}}$.
 Then, we focus on estimates of the total mass via the velocity dispersion parameter $\smash{\sigma_{\mathrm{SIS}}}$
 and core masses through the Einstein radius $\smash{\mathrm{R_{E}}}$.
Due to the low mass of some small groups, not all lenses have WL measurements and
not all of them coincide with our 44 pre-selected SL regions. This
results in only 29 objects matching our SL candidates and having WL data. The range of this SL--WL sample goes from 
$\smash{\sigma_{\mathrm{SIS}} \sim 350 }$ km s$^{-1}$ to $\sim 1000$  km s$^{-1}$,
with an average value of $\smash{\langle \sigma_{\mathrm{SIS}} \rangle \sim 677}$ km s$^{-1}$, 
which is  used as a pivot value for the next step of the analysis.
Fig. \ref{Fig:Er_Rarc} (panel $b$) shows the $\smash{\mathrm{R_{E}} - \sigma_{\mathrm{SIS}}}$ plane for 
these  29 SL--WL systems. 
The correlation is expected as well as a large dispersion in the relation.
For a more quantitative assessment, we fit the data
by performing 
a Levenberg-Marquardt least-squares fit and assuming a
power-law scaling relation given by
\begin{align}
   {\mathrm{R_E}}(\sigma_{\mathrm{SIS}}) 
      =   {\mathrm{R^{norm}_E}} \times \left(
           \frac{ \sigma_{\mathrm{SIS}} }{\sigma^{piv}_{\mathrm{SIS}}} \right)^{\alpha}, \label{eq:power_law_fit}
\end{align}
where $\smash{\sigma^{piv}_{\mathrm{SIS}}}$ is chosen to be the mean value $\smash{\langle \sigma_{\mathrm{SIS}}\rangle}$ of the SL--WL sample,
which is a representative normalization of these objects and optimizes the fit of the scaling relation, 
reducing the correlation between the logarithmic slope and normalization $\smash{\mathrm{R^{norm}_{E}}}$,
as pointed out in \cite{Foex13_SARCS_WL}.
The best-fitting results are listed in Table \ref{table:fit_results} and shown in Fig. \ref{Fig:Er_Rarc} (panel $b$)
by the continuous black line and the gray region.
The average scatter of this relation is $\smash{ 37\%}$.
This large dispersion can be explained by the combination of the large uncertainties in the measurements of the velocity dispersion parameter, 
due to the intrinsic low WL signal of groups, and the construction of our method itself.
Nonetheless, the fact that these two independent estimates correlate with an accuracy 
of better than $\sim37\%$ is by itself remarkable, since the WL approach is based on the statistical 
analysis of the background galaxies, without using any photometric information of the galaxy members, 
 while our estimates are derived directly from the luminosity data
of the LRGs. 

%%%%%%%%%%%%%%%%%%%%%%%%%%%%%%%%%%%%%%%%%%%%%%%%%%%%%%5
%%%%%%%%%%%%%%%%%%%%%%%%%%%%%%%%%%%%%%%%%%%%%%%%%%%%%%%%
% Table of Summary of fitting results mass estimates and luminosities
\begin{table}
\centering
\caption{Summary of fitting results of Einstein radius vs lens properties: mass estimates and total luminosities. \label{table:fit_results}}
\hspace{0.0cm} %-1.0cm
\resizebox{\columnwidth}{!}{
\begin{tabular}{l c c c c}
\toprule \toprule
Scaling law &  Pivot value$^a$ & $\mathrm{R^{norm}_E}$ &  $\alpha$ & $\langle \sigma^s \rangle^{b}$ \\
            &        &  [kpc$/h_{70}$]              &           &  \\
\midrule
$\smash{\mathrm{R_{E}} - \sigma_{\mathrm{SIS}}}$ &  677 [km s$^{-1}$] & 79$\pm$11 & 1.05$\pm$0.52  & 0.37$\pm$0.12 \\
$\smash{\mathrm{R_{E}} - L^{\mathrm{T}}(\mathrm{< 1 \hspace{0.1cm} Mpc})}$  & $2.7\times10^{12}[L_{\odot}]$ & 67$\pm$12& 0.73$\pm$0.29 & 0.31$\pm$0.09  \\ 
\bottomrule
\end{tabular}
}                                        
{\footnotesize\flushleft
{}$^a$ These normalizations correspond to the average values of our SL--WL sample, 
$\smash{\langle\sigma_{\mathrm{SIS}}\rangle}$ and $\smash{\langle L^{\mathrm{T}}(\mathrm{< 1 \hspace{0.1cm} Mpc}) \rangle}$,
respectively.  \\
{}$^b$ Average scatter of the corresponding correlation.\\
}
\end{table}
%%%%%%%%%%%%%%%%%%%%%%%%%%%%%%%%%%%%%%%%%%%%%%%%%%%%%%%%
%%%%%%%%%%%%%%%%%%%%%%%%%%%%%%%%%%%%%%%%%%%%%%%%%%%%%%%%

A better correlation is expected between the Einstein radii of the SL regions
 and the total luminosities of the systems, since the luminosity is the observable on which \textit{EasyCritics} is based. 
 The luminosity density maps of the 29 SL--WL systems are also obtained from 
 \cite{Foex13_SARCS_WL}. They have derived the total luminosity of each lens by 
 collapsing the individual luminosities of the galaxy members (within a radius of 1 Mpc), which have been
previously selected by using a red-sequence technique \citep[e.g.][]{GladdersYee00, GladdersYee05} and
 applying a luminosity cut ($\smash{M_{i'} < -21}$). 
 This process results in total luminosities ranging from 
 $\smash{L^{\mathrm{T}}(\mathrm{< 1 \hspace{0.1cm} Mpc}) \sim 0.6 \times 10^{12} L_{\odot} }$ till 
 $\smash{\sim 6 \times 10^{12} L_{\odot} }$.
The total luminosities of our SL--WL sample are in the same range above and have an 
average total luminosity of $\smash{\sim 2.7 \times 10^{12} L_{\odot} }$, 
which is consistent with the previous results as well.
The $\smash{\mathrm{R_{E}} - L^{\mathrm{T}}(\mathrm{< 1 \hspace{0.1cm} Mpc})}$ plane for 
the 29 SL systems is presented in Fig. \ref{Fig:Er_Rarc} (panel $c$).
By fitting the same power-law scaling relation given by Eq. (\ref{eq:power_law_fit}), but using 
the average total luminosity as normalization, we obtain an average scatter of $\sim31\%$, 
which is slightly smaller than the dispersion obtained in the
 $\smash{\mathrm{R_{E}} - \sigma_{\mathrm{SIS}}}$ relation.
 The best-fitting parameters are listed in Table \ref{table:fit_results} and its
 relation is presented in panel $c$ in analogy to the previous panel.

In interpreting these results, note that the purpose 
of \textit{EasyCritics} is not to estimate the lens features but to find super-critical regions; 
thus, this analysis serves just as a consistency check.

\subsection{New SL candidates and inspection efficiency}  \label{sec:results:new_candidates}

Given the expected number of SL galaxy clusters and groups in numerical simulations
\citep{Oguri06} and the number of KLs,
one can see that it is still possible to find missing lenses in the survey.
Thus, we systematically search for new  SL candidates. 
In configuration $\smash{B({\sigma_{\mathrm{clus}}} = 4'')}$, \textit{EasyCritics} produces of the order of $\sim1200$
super-critical regions, which are sorted by their Einstein radii in order to start 
the search by looking first at the more promising regions.
We use the $g'$-, $r'$-, and $i'$-band imaging data from CFHTLenS to generate color composite images 
and RGB FITS files of all pre-selected SL regions. These files are created by using our own IDL routines and by choosing an adequate color scale,
in order to maximize the contrast between faint extended objects and LRGs. 
The RGB FITS cut-outs are centered on the candidate centers and
 cover an area of $82''\times82''$. 
This choice is motivated by the average angular size of the 
Einstein radius of known massive gravitational lenses, and in order to compare with the results of SW 
as well. 
Note that there are only 12 regions with Einstein radii $>40''$, where the
analysis is carried out in a larger area of $210''\times210''$.

Aiming at obtaining a sample minimally biased by subjective decisions, all candidates in this configuration
are visually inspected by three authors of this work (MC, SS, MM)\footnote{
MC: Mauricio Carrasco; SS: Sebastian Stapelberg; MM: Matteo Maturi.}. 
Then, in order to standardize our results, the pre-selected SL regions are ranked by applying the
same scale used in the SW project, from $0 - 3$ (with a step size of 0.5). Where $rank=0$ corresponds to
regions unlikely to contain a lens; $rank=1$ to regions possibly containing a lens; 
$rank=2$ to regions probably containing a lens; 
and $rank=3$ to regions almost certainly containing a lens \citep{More_2016_SWI}.
The fact that CFHTLenS has been analyzed several times for many years 
significantly decreases the probability of discovering new spectacular SL systems,
 \eg giant arcs around clusters or groups.
Thus, we do not expect to find any new candidates with $rank = 3$.
However, we have found 9 systems showing  prominent SL features, which are
classified with a final $rank \geq 2$. They are presented in Fig. \ref{Fig:new_SL_candidates}, Appendix \ref{app:new_SL_candidates}.
Our visual inspection has also yielded several
regions having a low or medium probability of containing a lens ($1\leq rank < 2$).
The new SL candidates identified by \textit{EasyCritics} (with $rank \geq 2$)
are listed in Table \ref{table:new_candidates}.
We have named our SL regions as SLEC-Jhhmm$+$ddmm, which stands for 
\textbf{S}trong \textbf{L}ensing regions by \textit{\textbf{E}asy\textbf{C}ritics}
 and the corresponding location in sexagesimal coordinates.

It is worth mentioning that,
given the low number of SL region candidates produced by \textit{EasyCritics},
the effective area for inspection can be dramatically reduced 
by assuming that the chosen size of  $82''\times82''$ is enough to detect SL signatures around the 
center of group- and cluster-scale lenses, which has been already confirmed by the SW project \citep{More_2016_SWI, Marshall_2016_SWII}. 
The area of one tile ($82''\times82''$) times $1200$ pre-selected SL region candidates to analyze,
results in an effective area of $\sim0.623$ sq. deg, which
corresponds to only $\sim0.4\%$  of the total area of CFHTLenS ($\sim 154$ sq. deg).
Therefore, the final area where the search is performed is more than two orders of magnitude smaller than
the original field.
In other words, by using \textit{EasyCritics}, we effectively reduce the area 
which has to be visually inspected
or where any arcfinder algorithm is executed, which  is
translated into a large decrease of the total achievement time, from years to weeks. 
Consequently,  \textit{EasyCritics} is ideal for the upcoming $\smash{10^{4}}$ sq. deg surveys where 
pre-selection of  candidates for inspection is mandatory due to the enormous amount of data
that these missions are going to yield. 

%%%%%%%%%%%%%%%%%%%%%%%%%%%%%%%%%%%%%%%%%%%%%%%%%%%%%%5
%%%%%%%%%%%%%%%%%%%%%%%%%%%%%%%%%%%%%%%%%%%%%%%%%%%%%%%%
% Table of new tentative SL candidate
\begin{table}
\centering
\caption{New  SL candidate pre-selected by \textit{EasyCritics}. \label{table:new_candidates}}
\hspace{0.0cm}
\resizebox{\columnwidth}{!}{
\begin{tabular}{c c c c c}
\toprule \toprule
ID$^a$ &   R.A.  &   Dec.   & $R_\mathrm{E}$ & Rank  \\ 
       & [J2000] & [J2000]  &   [$''$]       &  \\ 
\midrule
SLEC-J0211--0609 & 02:11:13.90 & -06:09:48.8 &  19  &  2.5   \\
SLEC-J1405+5356 & 14:05:33.73 & +53:56:12.7 &  7   &   2.5  \\
SLEC-J0211--0422 & 02:11:22.58 & -04:22:05.2 &  7   &   2.4  \\
SLEC-J0213--0951 & 02:13:23.75 & -09:51:40.2 & 11   &   2.1  \\
SLEC-J0204--1017 & 02:04:58.29 & -10:17:31.3 & 13   &   2.0  \\
SLEC-J2220+0058 & 22:20:51.59 & +00:58:15.7 & 12   &   2.0  \\
SLEC-J0233--0530 & 02:33:39.75 & -05:30:36.7 & 11   &   2.0  \\
SLEC-J0216--0558 & 02:16:23.52 & -05:58:46.0 &  7   &   2.0  \\
SLEC-J0212--0820 & 02:12:35.22 & -08:20:45.3 & 12   &   2.0  \\
\bottomrule
\end{tabular}
}             

{\footnotesize\flushleft
{}$^a$ ID of the new  SL candidates. The name is given by the acronym:
\textbf{S}trong \textbf{L}ensing regions by \textit{\textbf{E}asy\textbf{C}ritcs} (SLEC) 
and the corresponding location in sexagesimal coordinates.\\
}
\end{table}
%%%%%%%%%%%%%%%%%%%%%%%%%%%%%%%%%%%%%%%%%%%%%%%%%%%%%%%%
%%%%%%%%%%%%%%%%%%%%%%%%%%%%%%%%%%%%%%%%%%%%%%%%%%%%%%%% 

\section{Discussion} \label{sec:discussion}

In this section we explore some aspects that may affect our results and interpretations, such
as the mass distribution of the KLs matching with the
pre-selected SL regions, problems in the photometric catalogs of CFHTLenS, and
possible biases in our calibration routine or parameter selection procedure.
% Since the latter is not completely independent from the rest, we analyze it together in all subsections.

\subsection{Consistency checks}

The correlations discussed in \S\ref{sec:correction} might suffer some bias if the super-critical regions
generated by \textit{EasyCritics} were not well distributed over the three mass intervals studied in this work.
Nevertheless, from Fig. \ref{Fig:Er_Rarc}  %\ref{Fig:Er_sigma}, and \ref{Fig:Er_luminosity},
one can see that the pre-selected super-critical regions are homogeneously spread within
$\smash{3'' \lesssim \mathrm{R_{E}} \lesssim 40''}$, % (20'')}$\footnote{If $\smash{\mathrm{R_{arc}}}$ is used instead of $\smash{\mathrm{R_{E}}}$.}, 
covering a very wide mass range of approximately 3 orders of magnitude \citep{Oguri06}. 
%After performing this simple inspection, 
Thus, we do not find any indication of possible biases in our results due to 
a preferred mass interval.  % \footnote{With respect to the three mass regime explored in this work.}.}

\subsection{Missed known lenses}

In configuration $\smash{B({\sigma_{\mathrm{clus}}} = 4'')}$, \textit{EasyCritics} is not able to
predict super-critical regions for $12$ of the $44$ KLs of the reference sample.
Due to extended halos of bright stars, several regions in the CFHTLenS imaging data 
are suffering from photometric imperfections, and therefore have been masked \citep{Erben2013_CFHTLenS}.
This masking procedure results in a large decrease in the number of galaxies within areas of several square arcseconds.
Since our method is based on the luminosity data of LRGs only, 
a decline in the galaxy count is directly translated into a decrease in the surface mass density of the DM component,
as well as in its weight $\smash{w^{(k)}(n^{(k)}|n_c)}$.
Therefore, it is expected to create sub-critical regions in masked areas.
Among the 12 missed KLs, 3 have large masked regions around their centers,
resulting in flat sub-critical convergence profiles.
These objects correspond to the KLs SA103, SA123, and SW3.
Taking this into account, the final detection rate of our approach reaches $\sim78\%$  
(and $\sim88\%$ for $\smash{B({\sigma_{\mathrm{clus}}} = 2'')}$), since there are only 41 KLs with reliable photometric data.

The other 9 remaining  missed KLs might easily reach the super-critical condition by a slight modification 
in the used parameter set, 
as shown in Fig. \ref{Fig:3mass_regime_3sigma}.
For a better compromise between purity and completeness, a new calibration routine is currently being implemented, 
where specific parameter set configurations are applied to different mass intervals, 
given by the density of LRGs in the studied field.
This new procedure is explained in the next subsection and presented in a forthcoming paper. % (Stapelberg et al. 2018, in prep.).

\subsection{New calibration routine}
 
We are developing a new calibration routine for \textit{EasyCritics}, 
which is going to extend the currently implemented approach by replacing 
the calibration of parameters on individual known lenses by a method capable of
a fully-automated optimization on an entire set of (arbitrarily many)
known lenses simultaneously. This ensures that the calibration will be most sensitive to 
the statistical properties of the lens population rather than to individual
properties that may vary from case to case. 

This new strategy starts with a classification of known lenses into different 
mass intervals according to a criterion based on the richness and galaxy number density profiles. 
For each set defined in such a way, a simultaneous multi-candidate optimization is applied.
This new optimization process uses the same MCMC approach described in our first paper, 
but with a new objective function that includes a larger number of $\chi^2$ terms.
These terms could include the mean deviation between the Einstein radii and the location of arcs, 
the number of pre-selected SL regions, and the total number of selected or generated  SL regions.
In other words, the new objective function is defined such that an optimal compromise is achieved between
 completeness and purity. 
This new routine is fully described in a forthcoming paper (Stapelberg et al 2018, in prep).

\subsection{Large Einstein radii and catalog problems}

As mentioned earlier,  in configuration $\smash{B({\sigma_{\mathrm{clus}}} = 4'')}$,
only 12 SL region candidates created by \textit{EasyCritics} have Einstein radii larger than $40''$.
Among them, 5 regions are boosted by uncorrelated small group-scale objects aligned in the LOS,
but with offsets of the order of $\sim20''$.
One of these SL regions is located in a rich environment of LRGs, which
 actually encloses (at $\sim20''$ from its center)
 the known candidate SA8 (Table \ref{table:kl_kc_by_EasyCritics}).
However, the other 4 super-critical regions do not show any clear SL signature,
and are therefore not selected as possible SL candidates.

The 7 remaining region candidates actually correspond 
to spurious detections given by some photometric problems in the CFHTLenS catalogs.
In 5 of these regions, critical curves have been produced by
a wrong association of satellite tracks to LRGs. 
The long spatial alignment of these fake measurements considerably augments 
the luminosity density of the field, 
resulting in large massive regions having long but narrow critical curves.
The other 2 super-critical regions are the product of erroneous classifications
of bright stars as elliptical galaxies,
which affect the shapes and sizes of critical curves as well.
In general, these photometric problems are wrongly generating critical curves of all sizes; 
nonetheless, these artifacts are producing not more than $9\%$ of the total pre-selected SL regions. 
These objects are then classified as spurious detections or artificial candidates.
These issues can be solved by a simple modification of the SExtractor parameters \citep{SExtractor} 
and by running an automated satellite removal procedure; 
however, these tasks are beyond the scope of this work. 
Since the number of spurious detections due to these problems is very low, 
we leave these improvements for future studies.
 
\section{Summary and conclusions} \label{sec:conclusion}

In this paper we reported the results of \textit{EasyCritics} applied to CFHTLenS.
Based on the well-known LTM assumption, our algorithm  
constructs a simple lensing potential model of the total mass 
projected along the LOS, using only the photometric information of the brightest LRGs
distributed in the redshift range $\smash{0.2 \lesssim z \lesssim 0.9}$. 

We used CFHTLenS data, which have been extensively analyzed 
in the past, to construct the reference samples for finding the initial parameter sets and for testing the performance of our code.
These catalogs are composed of 44 secure or confirmed KLs, from a total of
98 KCs, spanning a mass range  from  $\smash{\sim 10^{13}}$
to $\smash{10^{15}}$ M$\smash{_{\odot}}$.
Given the nature of our approach, a different mass scaling parameter is required for a
different mass interval. 
Therefore, in order to cover the majority of this mass range and to find the initial parameter sets for our systematic parameter exploration,
we  performed three independent parameter calibrations; one for each different mass interval ($A$, $B$, and $C$).
We chose the most representative lenses in these mass scales;
 the SL galaxy groups SA14, SA22, and the SL galaxy cluster
SA100, with Einstein radii of $\smash{\sim 3.2'', 7.1''}$, and $\smash{14.7''}$, respectively. 
Once the initial parameter sets were found,
 we systematically explored the parameter space based on those initial values, aiming at identifying
most of the KLs of the reference sample.
For each of the three mass intervals, we created several new parameter sets 
by fixing $\smash{K_{\mathrm{clus}}}$, $q$, $\smash{n_c}$, and
$\smash{K_{\mathrm{gal}}}$  to the values obtained
from the corresponding calibration (Table \ref{table:calibrated_par}), and by varying $\smash{\sigma_{\mathrm{clus}}}$ 
from $2''$ till $20''$, in steps of $\smash{\Delta \sigma_{\mathrm{clus}}} = 2''$. 
This exploration results in a total of 30 different parameter sets, \ie
10 different values of $\smash{\sigma_{\mathrm{clus}}}$ for each mass interval. 
We then applied \textit{EasyCritics} to the whole area of CFHTLenS using each of these parameter sets.

Once the survey had been processed, we studied the performance of our algorithm  
by comparing the SL region candidates generated by \textit{EasyCritics},
 in each of the 30 parameter sets, with the KLs of the reference sample.
 The parameter set configurations that predicted most of the KLs are placed in the mass range $B$, 
which maximizes the probability of generating lenses in the group and small cluster scales.
These results were indeed expected since the majority of the  galaxies
of the Universe are living in groups.
In configuration $\smash{B({\sigma_{\mathrm{clus}}} = 4'')}$,  
our algorithm pre-selected 32 out of 44 KLs, corresponding to a detection rate of $\sim73\%$. 
To achieve this, \textit{EasyCritics} left only $\sim1200$  possibly super-critical regions for inspection.
The parameter set $\smash{B({\sigma_{\mathrm{clus}}} = 2'')}$ reached
a detection rate of $\sim82\%$ by generating $\sim3300$ SL region candidates.
Even though the `best' parameter set can only be found by using numerical simulations, we
chose configuration $\smash{B({\sigma_{\mathrm{clus}}} = 4'')}$ to perform our analysis and
search for new  candidates due to the high number of identified KLs and the low amount of SL region candidates left for inspection.

We presented the Einstein-radius distribution of the pre-selected SL regions of configuration $\smash{B({\sigma_{\mathrm{clus}}} = 4'')}$, 
which peaked at $\smash{\mathrm{R_{E}}\sim 5.9''}$ and agreed well with previous observations.
We found that only 12 SL regions showed excessively large  Einstein radii ($\smash{\mathrm{R_{E}} > 40''}$),
of which 7 were produced by photometric problems in the CFHTLenS catalogs.
The other 5 remaining SL regions were boosted by uncorrelated small group-scale objects aligned in the LOS,
but with large offsets. 
Among them, one  SL region is enclosing the known candidate SA8, at $\sim20''$ from its center.
In total, among the $\sim1200$ pre-selected candidates, there were 32 super-critical 
regions matching with KLs and 12 with KCs, and only $\sim9\%$ of the total SL regions 
were spurious detection caused by imperfections in the photometric catalog. 

We also compared our results with WL velocity dispersion and luminosity estimates of the reference lenses.
These quantities are direct indicators of the total mass of the clusters, and therefore, 
a  correlation with the Einstein radius is expected. 
In configuration $\smash{B({\sigma_{\mathrm{clus}}} = 4'')}$, there were 29 KLs matching with our pre-selected SL regions and having WL velocity dispersion
and luminosity measurements, whose ranges go from 
$\smash{\sigma_{\mathrm{SIS}} \sim 350 }$ km s$^{-1}$ to $\sim 1000$  km s$^{-1}$,
with an average value of $\smash{\langle \sigma_{\mathrm{SIS}} \rangle \sim 677}$ km s$^{-1}$,
and from $\smash{L^{\mathrm{T}}(\mathrm{< 1 \hspace{0.1cm} Mpc}) \sim 0.6 \times 10^{12} L_{\odot} }$ to
 $\smash{\sim 6 \times 10^{12} L_{\odot} }$, with an average total luminosity of $\smash{\sim 2.7 \times 10^{12} L_{\odot} }$,
 respectively. 
 By analyzing the $\smash{\mathrm{R_{E}} - \sigma_{\mathrm{SIS}}}$ and 
 $\smash{\mathrm{R_{E}} - L^{\mathrm{T}}(\mathrm{< 1 \hspace{0.1cm} Mpc})}$ planes, 
 we showed that the Einstein radius measurements derived from our SL regions correlate well with
 estimates of total luminosity and  WL velocity dispersion, and therefore, with the total
projected mass of the lenses.
In fact, these correlations are well characterized by a power-law scaling relation, with 
average scatter of $\smash{37\%}$ and $\smash{31\%}$, respectively.
 Despite the large scatter, the fact that these independent estimates correlated with an accuracy 
of the order of $\sim30 - 40\%$ is by itself remarkable, since these  approaches 
are based on completely independent methods.

Finally, having confirmed that \textit{EasyCritics} was able to identify the location of most
of the KLs  and reproduced their features as well,
we performed a systematic search of new  SL candidates. 
We created RGB FITS files for all the $\sim1200$ pre-selected candidates generated in configuration $\smash{B({\sigma_{\mathrm{clus}}} = 4'')}$.
Each of the cut-outs was centered on the candidate center and
 covered an area of $82''\times82''$. 
Despite the fact that CFHTLenS data have been extensively analyzed by various teams and various methods for several years,
we found 9 new systems showing  prominent SL features and several
regions having a low or medium probability of containing a lens.
We also showed that given the low number of SL region candidates pre-selected by \textit{EasyCritics},
the effective area for inspection was dramatically reduced from
$\sim 154$ sq. deg to $\sim0.623$ sq. deg, 
which corresponds to only $\sim0.4\%$  of the total area,
indicating the power of \textit{EasyCritics}  as a pre-selection method.

Summing up, \textit{EasyCritics} is a very successful and efficient method for the search of SL systems. 
It was able to identify  $\sim73\%$  of the previously KLs  by generating only  $\sim1200$ SL regions for inspection, with 
less than $\sim9\%$ of spurious detections. 
This method effectively reduced the area where the inspection took place by more than two orders of magnitude,
resulting in a large decrease of both the total achievement time and the manual post-processing. 
Furthermore, \textit{EasyCritics} correctly  provided a first characterization of the lensing properties of the region candidates,
 such as their Einstein radii and convergence profiles.
Even though the final answers regarding its efficiency 
and the implications of its results for cosmological inference
will be addressed in a forthcoming paper, in which EasyCritics will be tested against extensive numerical simulations, 
we can already conclude that \textit{EasyCritics} is a very promising algorithm to find and characterize 
SL systems in wide-field surveys. 
Consequently,  this new approach is ideal for upcoming surveys covering $\smash{10^{4}}$ sq. deg,
such as the Euclid mission and \textit{LSST}, where a pre-selection of candidates for any kind of SL
 analysis will be indispensable due to the expected enormous data volume.

\section*{Acknowledgments}

This work was supported by the Transregional  Collaborative  Research  Center  TRR  33 ``The Dark Universe''.

%%%%%%%%%%%%%%%%%%%%%%%%%%%%%%%%%%%%%%%%%%%%%%%%%%



%\bibliographystyle{mnras}
%\bibliography{papersb} % if your bibtex file is called example.bib

\begin{thebibliography}{99}
% \bibitem[\protect\citeauthoryear{Author}{2012}]{Author2012}
% Author A.~N., 2013, Journal of Improbable Astronomy, 1, 1
% \bibitem[\protect\citeauthoryear{Others}{2013}]{Others2013}
% Others S., 2012, Journal of Interesting Stuff, 17, 198
\makeatletter
\relax
\def\mn@urlcharsother{\let\do\@makeother \do\$\do\&\do\#\do\^\do\_\do\%\do\~}
\def\mn@doi{\begingroup\mn@urlcharsother \@ifnextchar [ {\mn@doi@}
  {\mn@doi@[]}}
\def\mn@doi@[#1]#2{\def\@tempa{#1}\ifx\@tempa\@empty \href
  {http://dx.doi.org/#2} {doi:#2}\else \href {http://dx.doi.org/#2} {#1}\fi
  \endgroup}
\def\mn@eprint#1#2{\mn@eprint@#1:#2::\@nil}
\def\mn@eprint@arXiv#1{\href {http://arxiv.org/abs/#1} {{\tt arXiv:#1}}}
\def\mn@eprint@dblp#1{\href {http://dblp.uni-trier.de/rec/bibtex/#1.xml}
  {dblp:#1}}
\def\mn@eprint@#1:#2:#3:#4\@nil{\def\@tempa {#1}\def\@tempb {#2}\def\@tempc
  {#3}\ifx \@tempc \@empty \let \@tempc \@tempb \let \@tempb \@tempa \fi \ifx
  \@tempb \@empty \def\@tempb {arXiv}\fi \@ifundefined
  {mn@eprint@\@tempb}{\@tempb:\@tempc}{\expandafter \expandafter \csname
  mn@eprint@\@tempb\endcsname \expandafter{\@tempc}}}
  
\bibitem[\protect\citeauthoryear{{Alard}}{{Alard}}{2006}]{Alard2006_arcfinder}
{Alard} C.,  2006, ArXiv Astrophysics e-prints, \href
  {http://adsabs.harvard.edu/abs/2006astro.ph..6757A} {}

\bibitem[\protect\citeauthoryear{{Ammons}, {Wong}, {Zabludoff}  \&
  {Keeton}}{{Ammons} et~al.}{2014}]{Ammons2014_LOS}
{Ammons} S.~M.,  {Wong} K.~C.,  {Zabludoff} A.~I.,   {Keeton} C.~R.,  2014,
  \mn@doi [\apj] {10.1088/0004-637X/781/1/2}, \href
  {http://adsabs.harvard.edu/abs/2014ApJ...781....2A} {781, 2}

\bibitem[\protect\citeauthoryear{{Bartelmann}, {Huss}, {Colberg}, {Jenkins}  \&
  {Pearce}}{{Bartelmann} et~al.}{1998}]{Bartelmann98}
{Bartelmann} M.,  {Huss} A.,  {Colberg} J.~M.,  {Jenkins} A.,   {Pearce} F.~R.,
   1998, \aap, \href {http://adsabs.harvard.edu/abs/1998A%26A...330....1B}
  {330, 1}

\bibitem[\protect\citeauthoryear{{Bartelmann}, {Meneghetti}, {Perrotta},
  {Baccigalupi}  \& {Moscardini}}{{Bartelmann}
  et~al.}{2003}]{Bartelmann2003_arc_st}
{Bartelmann} M.,  {Meneghetti} M.,  {Perrotta} F.,  {Baccigalupi} C.,
  {Moscardini} L.,  2003, \mn@doi [\aap] {10.1051/0004-6361:20031158}, \href
  {http://adsabs.harvard.edu/abs/2003A%26A...409..449B} {409, 449}

\bibitem[\protect\citeauthoryear{{Bayliss}}{{Bayliss}}{2012}]{Bayliss12}
{Bayliss} M.~B.,  2012, \mn@doi [\apj] {10.1088/0004-637X/744/2/156}, 744, 156

\bibitem[\protect\citeauthoryear{{Bayliss}, {Hennawi}, {Gladders}, {Koester},
  {Sharon}, {Dahle}  \& {Oguri}}{{Bayliss} et~al.}{2011}]{Bayliss11b}
{Bayliss} M.~B.,  {Hennawi} J.~F.,  {Gladders} M.~D.,  {Koester} B.~P.,
  {Sharon} K.,  {Dahle} H.,   {Oguri} M.,  2011, \mn@doi [\apjs]
  {10.1088/0067-0049/193/1/8}, \href
  {http://adsabs.harvard.edu/abs/2011ApJS..193....8B} {193, 8}

\bibitem[\protect\citeauthoryear{{Bayliss}, {Johnson}, {Gladders}, {Sharon}  \&
  {Oguri}}{{Bayliss} et~al.}{2014}]{Bayliss2014_LOS}
{Bayliss} M.~B.,  {Johnson} T.,  {Gladders} M.~D.,  {Sharon} K.,   {Oguri} M.,
  2014, \mn@doi [\apj] {10.1088/0004-637X/783/1/41}, \href
  {http://adsabs.harvard.edu/abs/2014ApJ...783...41B} {783, 41}

\bibitem[\protect\citeauthoryear{{Becker} et~al.,}{{Becker}
  et~al.}{2007}]{Becker2007}
{Becker} M.~R.,  et~al., 2007, \mn@doi [\apj] {10.1086/521920}, \href
  {http://adsabs.harvard.edu/abs/2007ApJ...669..905B} {669, 905}

\bibitem[\protect\citeauthoryear{{Bertin} \& {Arnouts}}{{Bertin} \&
  {Arnouts}}{1996}]{SExtractor}
{Bertin} E.,  {Arnouts} S.,  1996, \aaps, 117, 393

\bibitem[\protect\citeauthoryear{{Boulade} et~al.,}{{Boulade}
  et~al.}{2003}]{Boulade2003_MEGACAM}
{Boulade} O.,  et~al., 2003, in {Iye} M.,  {Moorwood} A.~F.~M.,  eds,
  \procspie Vol. 4841, Instrument Design and Performance for Optical/Infrared
  Ground-based Telescopes. pp 72--81, \mn@doi{10.1117/12.459890}

\bibitem[\protect\citeauthoryear{{Bouwens} et~al.,}{{Bouwens}
  et~al.}{2009}]{Bouwens09}
{Bouwens} R.~J.,  et~al., 2009, \mn@doi [\apj] {10.1088/0004-637X/690/2/1764},
  690, 1764

\bibitem[\protect\citeauthoryear{{Brada{\v c}} et~al.,}{{Brada{\v c}}
  et~al.}{2009}]{Bradac09}
{Brada{\v c}} M.,  et~al., 2009, \mn@doi [\apj] {10.1088/0004-637X/706/2/1201},
  706, 1201

\bibitem[\protect\citeauthoryear{{Brainerd}, {Blandford}  \&
  {Smail}}{{Brainerd} et~al.}{1996}]{Brainerd96}
{Brainerd} T.~G.,  {Blandford} R.~D.,   {Smail} I.,  1996, \mn@doi [\apj]
  {10.1086/177537}, \href {http://adsabs.harvard.edu/abs/1996ApJ...466..623B}
  {466, 623}

\bibitem[\protect\citeauthoryear{{Broadhurst} \& {Barkana}}{{Broadhurst} \&
  {Barkana}}{2008}]{BroadhurstBarkana08}
{Broadhurst} T.~J.,  {Barkana} R.,  2008, \mn@doi [\mnras]
  {10.1111/j.1365-2966.2008.13852.x}, 390, 1647

\bibitem[\protect\citeauthoryear{{Broadhurst}, {Takada}, {Umetsu}, {Kong},
  {Arimoto}, {Chiba}  \& {Futamase}}{{Broadhurst}
  et~al.}{2005a}]{Broadhurst2005b}
{Broadhurst} T.,  {Takada} M.,  {Umetsu} K.,  {Kong} X.,  {Arimoto} N.,
  {Chiba} M.,   {Futamase} T.,  2005a, \mn@doi [\apjl] {10.1086/428122}, \href
  {http://adsabs.harvard.edu/abs/2005ApJ...619L.143B} {619, L143}

\bibitem[\protect\citeauthoryear{{Broadhurst} et~al.,}{{Broadhurst}
  et~al.}{2005b}]{Broadhurst2005a}
{Broadhurst} T.,  et~al., 2005b, \mn@doi [\apj] {10.1086/426494}, \href
  {http://adsabs.harvard.edu/abs/2005ApJ...621...53B} {621, 53}

\bibitem[\protect\citeauthoryear{{Broadhurst} et~al.,}{{Broadhurst}
  et~al.}{2005c}]{Broadhurst05}
{Broadhurst} T.,  et~al., 2005c, \mn@doi [\apj] {10.1086/426494}, \href
  {http://adsabs.harvard.edu/abs/2005ApJ...621...53B} {621, 53}

\bibitem[\protect\citeauthoryear{{Cabanac} et~al.,}{{Cabanac}
  et~al.}{2007}]{Cabanac07}
{Cabanac} R.~A.,  et~al., 2007, \mn@doi [\aap] {10.1051/0004-6361:20065810},
  \href {http://adsabs.harvard.edu/abs/2007A%26A...461..813C} {461, 813}

\bibitem[\protect\citeauthoryear{{Caminha} et~al.,}{{Caminha}
  et~al.}{2017a}]{Caminha2017_MACSJ0416}
{Caminha} G.~B.,  et~al., 2017a, \mn@doi [\aap] {10.1051/0004-6361/201629297},
  \href {http://adsabs.harvard.edu/abs/2017A%26A...600A..90C} {600, A90}

\bibitem[\protect\citeauthoryear{{Caminha} et~al.,}{{Caminha}
  et~al.}{2017b}]{Caminha2017_MACSJ1206}
{Caminha} G.~B.,  et~al., 2017b, \mn@doi [\aap] {10.1051/0004-6361/201731498},
  \href {http://adsabs.harvard.edu/abs/2017A%26A...607A..93C} {607, A93}

\bibitem[\protect\citeauthoryear{{Capak}}{{Capak}}{2004}]{Capak04}
{Capak} P.~L.,  2004, PhD thesis, UNIVERSITY OF HAWAI'I

\bibitem[\protect\citeauthoryear{{Carrasco} et~al.,}{{Carrasco}
  et~al.}{2017}]{Carrasco17_spec}
{Carrasco} M.,  et~al., 2017, \mn@doi [\apj] {10.3847/1538-4357/834/2/210},
  \href {http://adsabs.harvard.edu/abs/2017ApJ...834..210C} {834, 210}

\bibitem[\protect\citeauthoryear{{Coe} et~al.,}{{Coe} et~al.}{2012}]{Coe12}
{Coe} D.,  et~al., 2012, arXiv, 1201.1616

\bibitem[\protect\citeauthoryear{{Coe} et~al.,}{{Coe} et~al.}{2013}]{Coe13}
{Coe} D.,  et~al., 2013, \mn@doi [\apj] {10.1088/0004-637X/762/1/32}, \href
  {http://adsabs.harvard.edu/abs/2013ApJ...762...32C} {762, 32}

\bibitem[\protect\citeauthoryear{{Coleman}, {Wu}  \& {Weedman}}{{Coleman}
  et~al.}{1980}]{CWW80}
{Coleman} G.~D.,  {Wu} C.-C.,   {Weedman} D.~W.,  1980, \mn@doi [\apjs]
  {10.1086/190674}, 43, 393

\bibitem[\protect\citeauthoryear{{Dai}, {Bregman}, {Kochanek}  \&
  {Rasia}}{{Dai} et~al.}{2010}]{Dai_2010_baryons}
{Dai} X.,  {Bregman} J.~N.,  {Kochanek} C.~S.,   {Rasia} E.,  2010, \mn@doi
  [\apj] {10.1088/0004-637X/719/1/119}, \href
  {http://adsabs.harvard.edu/abs/2010ApJ...719..119D} {719, 119}

\bibitem[\protect\citeauthoryear{{Dalal}, {Holder}  \& {Hennawi}}{{Dalal}
  et~al.}{2004}]{Dalal+2004arcs}
{Dalal} N.,  {Holder} G.,   {Hennawi} J.~F.,  2004, \mn@doi [\apj]
  {10.1086/420960}, \href {http://adsabs.harvard.edu/abs/2004ApJ...609...50D}
  {609, 50}

\bibitem[\protect\citeauthoryear{{Erben} et~al.,}{{Erben}
  et~al.}{2005}]{Erben2005_theli}
{Erben} T.,  et~al., 2005, \mn@doi [Astronomische Nachrichten]
  {10.1002/asna.200510396}, \href
  {http://adsabs.harvard.edu/abs/2005AN....326..432E} {326, 432}

\bibitem[\protect\citeauthoryear{{Erben} et~al.,}{{Erben}
  et~al.}{2009}]{Erben2009_CARS}
{Erben} T.,  et~al., 2009, \mn@doi [\aap] {10.1051/0004-6361:200810426}, \href
  {http://adsabs.harvard.edu/abs/2009A%26A...493.1197E} {493, 1197}

\bibitem[\protect\citeauthoryear{{Erben} et~al.,}{{Erben}
  et~al.}{2013}]{Erben2013_CFHTLenS}
{Erben} T.,  et~al., 2013, \mn@doi [\mnras] {10.1093/mnras/stt928}, \href
  {http://adsabs.harvard.edu/abs/2013MNRAS.433.2545E} {433, 2545}

\bibitem[\protect\citeauthoryear{{Fassnacht}, {Gal}, {Lubin}, {McKean},
  {Squires}  \& {Readhead}}{{Fassnacht} et~al.}{2006}]{Fassnacht06}
{Fassnacht} C.~D.,  {Gal} R.~R.,  {Lubin} L.~M.,  {McKean} J.~P.,  {Squires}
  G.~K.,   {Readhead} A.~C.~S.,  2006, \mn@doi [\apj] {10.1086/500927}, 642, 30

\bibitem[\protect\citeauthoryear{{Fo{\"e}x}, {Soucail}, {Pointecouteau},
  {Arnaud}, {Limousin}  \& {Pratt}}{{Fo{\"e}x} et~al.}{2012}]{Foex13_SARCS_WL}
{Fo{\"e}x} G.,  {Soucail} G.,  {Pointecouteau} E.,  {Arnaud} M.,  {Limousin}
  M.,   {Pratt} G.~W.,  2012, \mn@doi [\aap] {10.1051/0004-6361/201218973},
  \href {http://adsabs.harvard.edu/abs/2012A%26A...546A.106F} {546, A106}

\bibitem[\protect\citeauthoryear{{Gavazzi}, {Marshall}, {Treu}  \&
  {Sonnenfeld}}{{Gavazzi} et~al.}{2014}]{Gavazzi_2014_ringfinder}
{Gavazzi} R.,  {Marshall} P.~J.,  {Treu} T.,   {Sonnenfeld} A.,  2014, \mn@doi
  [\apj] {10.1088/0004-637X/785/2/144}, \href
  {http://adsabs.harvard.edu/abs/2014ApJ...785..144G} {785, 144}

\bibitem[\protect\citeauthoryear{{Ge}, {Wang}, {Tripp}, {Li}, {Gu}  \&
  {Ji}}{{Ge} et~al.}{2016}]{Ge_2016_baryons}
{Ge} C.,  {Wang} Q.~D.,  {Tripp} T.~M.,  {Li} Z.,  {Gu} Q.,   {Ji} L.,  2016,
  \mn@doi [\mnras] {10.1093/mnras/stw599}, \href
  {http://adsabs.harvard.edu/abs/2016MNRAS.459..366G} {459, 366}

\bibitem[\protect\citeauthoryear{{Gilbank}, {Gladders}, {Yee}  \&
  {Hsieh}}{{Gilbank} et~al.}{2011}]{Gilbank11}
{Gilbank} D.~G.,  {Gladders} M.~D.,  {Yee} H.~K.~C.,   {Hsieh} B.~C.,  2011,
  \mn@doi [\aj] {10.1088/0004-6256/141/3/94}, \href
  {http://adsabs.harvard.edu/abs/2011AJ....141...94G} {141, 94}

\bibitem[\protect\citeauthoryear{{Gladders} \& {Yee}}{{Gladders} \&
  {Yee}}{2000}]{GladdersYee00}
{Gladders} M.~D.,  {Yee} H.~K.~C.,  2000, \mn@doi [\aj] {10.1086/301557}, \href
  {http://adsabs.harvard.edu/abs/2000AJ....120.2148G} {120, 2148}

\bibitem[\protect\citeauthoryear{{Gladders} \& {Yee}}{{Gladders} \&
  {Yee}}{2005}]{GladdersYee05}
{Gladders} M.~D.,  {Yee} H.~K.~C.,  2005, VizieR Online Data Catalog, \href
  {http://adsabs.harvard.edu/abs/2005yCat..21570001G} {215, 70001}

\bibitem[\protect\citeauthoryear{{Gladders}, {Hoekstra}, {Yee}, {Hall}  \&
  {Barrientos}}{{Gladders} et~al.}{2003}]{Gladders03}
{Gladders} M.~D.,  {Hoekstra} H.,  {Yee} H.~K.~C.,  {Hall} P.~B.,
  {Barrientos} L.~F.,  2003, \mn@doi [\apj] {10.1086/376518}, \href
  {http://adsabs.harvard.edu/abs/2003ApJ...593...48G} {593, 48}

\bibitem[\protect\citeauthoryear{{Hall} et~al.,}{{Hall} et~al.}{2012}]{Hall12}
{Hall} N.,  et~al., 2012, \mn@doi [\apj] {10.1088/0004-637X/745/2/155}, 745,
  155

\bibitem[\protect\citeauthoryear{Hastings}{Hastings}{1970}]{Hastings1970}
Hastings W.~K.,  1970, Biometrika, 57, 97

\bibitem[\protect\citeauthoryear{{Heymans} et~al.,}{{Heymans}
  et~al.}{2012}]{Heymans2012_CFHTLenS}
{Heymans} C.,  et~al., 2012, \mn@doi [\mnras]
  {10.1111/j.1365-2966.2012.21952.x}, \href
  {http://adsabs.harvard.edu/abs/2012MNRAS.427..146H} {427, 146}

\bibitem[\protect\citeauthoryear{{Hilbert}, {White}, {Hartlap}  \&
  {Schneider}}{{Hilbert} et~al.}{2007}]{Hilbert07}
{Hilbert} S.,  {White} S.~D.~M.,  {Hartlap} J.,   {Schneider} P.,  2007,
  \mn@doi [\mnras] {10.1111/j.1365-2966.2007.12391.x}, 382, 121

\bibitem[\protect\citeauthoryear{{Hildebrandt} et~al.,}{{Hildebrandt}
  et~al.}{2012}]{Hildebrandt2012_CFHTLenS_photz}
{Hildebrandt} H.,  et~al., 2012, \mn@doi [\mnras]
  {10.1111/j.1365-2966.2012.20468.x}, \href
  {http://adsabs.harvard.edu/abs/2012MNRAS.421.2355H} {421, 2355}

\bibitem[\protect\citeauthoryear{{Ho}, {Lin}, {Spergel}  \& {Hirata}}{{Ho}
  et~al.}{2009}]{Ho2009_LRG}
{Ho} S.,  {Lin} Y.-T.,  {Spergel} D.,   {Hirata} C.~M.,  2009, \mn@doi [\apj]
  {10.1088/0004-637X/697/2/1358}, \href
  {http://adsabs.harvard.edu/abs/2009ApJ...697.1358H} {697, 1358}

\bibitem[\protect\citeauthoryear{{Horesh}, {Ofek}, {Maoz}, {Bartelmann},
  {Meneghetti}  \& {Rix}}{{Horesh} et~al.}{2005}]{Horesh2005_arcfinder}
{Horesh} A.,  {Ofek} E.~O.,  {Maoz} D.,  {Bartelmann} M.,  {Meneghetti} M.,
  {Rix} H.-W.,  2005, \mn@doi [\apj] {10.1086/466519}, \href
  {http://adsabs.harvard.edu/abs/2005ApJ...633..768H} {633, 768}

\bibitem[\protect\citeauthoryear{{Jauzac} et~al.,}{{Jauzac}
  et~al.}{2015}]{Jauzac2015}
{Jauzac} M.,  et~al., 2015, \mn@doi [\mnras] {10.1093/mnras/stv1402}, \href
  {http://adsabs.harvard.edu/abs/2015MNRAS.452.1437J} {452, 1437}

\bibitem[\protect\citeauthoryear{{Johnston}, {Sheldon}, {Tasitsiomi},
  {Frieman}, {Wechsler}  \& {McKay}}{{Johnston} et~al.}{2007}]{Johnston2007}
{Johnston} D.~E.,  {Sheldon} E.~S.,  {Tasitsiomi} A.,  {Frieman} J.~A.,
  {Wechsler} R.~H.,   {McKay} T.~A.,  2007, \mn@doi [\apj] {10.1086/510060},
  \href {http://adsabs.harvard.edu/abs/2007ApJ...656...27J} {656, 27}

\bibitem[\protect\citeauthoryear{{Joseph} et~al.,}{{Joseph}
  et~al.}{2014}]{Joseph2014_arcfinder}
{Joseph} R.,  et~al., 2014, \mn@doi [\aap] {10.1051/0004-6361/201423365}, \href
  {http://adsabs.harvard.edu/abs/2014A%26A...566A..63J} {566, A63}

\bibitem[\protect\citeauthoryear{{Kneib} et~al.,}{{Kneib}
  et~al.}{2003}]{Kneib2003}
{Kneib} J.-P.,  et~al., 2003, \mn@doi [\apj] {10.1086/378633}, \href
  {http://adsabs.harvard.edu/abs/2003ApJ...598..804K} {598, 804}

\bibitem[\protect\citeauthoryear{{Lagan{\'a}}, {Martinet}, {Durret}, {Lima
  Neto}, {Maughan}  \& {Zhang}}{{Lagan{\'a}}
  et~al.}{2013}]{Lagana_2013_baryons}
{Lagan{\'a}} T.~F.,  {Martinet} N.,  {Durret} F.,  {Lima Neto} G.~B.,
  {Maughan} B.,   {Zhang} Y.-Y.,  2013, \mn@doi [\aap]
  {10.1051/0004-6361/201220423}, \href
  {http://adsabs.harvard.edu/abs/2013A%26A...555A..66L} {555, A66}

\bibitem[\protect\citeauthoryear{{Lenzen}, {Schindler}  \& {Scherzer}}{{Lenzen}
  et~al.}{2004}]{Lenzen2004_arcfinder}
{Lenzen} F.,  {Schindler} S.,   {Scherzer} O.,  2004, \mn@doi [\aap]
  {10.1051/0004-6361:20034619}, \href
  {http://adsabs.harvard.edu/abs/2004A%26A...416..391L} {416, 391}

\bibitem[\protect\citeauthoryear{{Li}, {Kauffmann}, {Jing}, {White},
  {B{\"o}rner}  \& {Cheng}}{{Li} et~al.}{2006a}]{Li2006_LRG}
{Li} C.,  {Kauffmann} G.,  {Jing} Y.~P.,  {White} S.~D.~M.,  {B{\"o}rner} G.,
  {Cheng} F.~Z.,  2006a, \mn@doi [\mnras] {10.1111/j.1365-2966.2006.10066.x},
  \href {http://adsabs.harvard.edu/abs/2006MNRAS.368...21L} {368, 21}

\bibitem[\protect\citeauthoryear{{Li}, {Mao}, {Jing}, {Mo}, {Gao}  \&
  {Lin}}{{Li} et~al.}{2006b}]{Li06}
{Li} G.~L.,  {Mao} S.,  {Jing} Y.~P.,  {Mo} H.~J.,  {Gao} L.,   {Lin} W.~P.,
  2006b, \mn@doi [\mnras] {10.1111/j.1745-3933.2006.00230.x}, \href
  {http://adsabs.harvard.edu/abs/2006MNRAS.372L..73L} {372, L73}

\bibitem[\protect\citeauthoryear{{Limousin} et~al.,}{{Limousin}
  et~al.}{2007}]{Limousin07}
{Limousin} M.,  et~al., 2007, \mn@doi [\apj] {10.1086/521293}, 668, 643

\bibitem[\protect\citeauthoryear{{Limousin} et~al.,}{{Limousin}
  et~al.}{2009}]{Limousin09_SL_Ggroup}
{Limousin} M.,  et~al., 2009, \mn@doi [\aap] {10.1051/0004-6361/200811473},
  \href {http://adsabs.harvard.edu/abs/2009A%26A...502..445L} {502, 445}

\bibitem[\protect\citeauthoryear{{Limousin} et~al.,}{{Limousin}
  et~al.}{2012}]{Limousin2012_M0717}
{Limousin} M.,  et~al., 2012, \mn@doi [\aap] {10.1051/0004-6361/201117921},
  \href {http://adsabs.harvard.edu/abs/2012A%26A...544A..71L} {544, A71}

\bibitem[\protect\citeauthoryear{{Marshall} et~al.,}{{Marshall}
  et~al.}{2016}]{Marshall_2016_SWII}
{Marshall} P.~J.,  et~al., 2016, \mn@doi [\mnras] {10.1093/mnras/stv2009},
  \href {http://adsabs.harvard.edu/abs/2016MNRAS.455.1171M} {455, 1171}

\bibitem[\protect\citeauthoryear{{Maturi}, {Mizera}  \& {Seidel}}{{Maturi}
  et~al.}{2014}]{Maturi14}
{Maturi} M.,  {Mizera} S.,   {Seidel} G.,  2014, \mn@doi [\aap]
  {10.1051/0004-6361/201321634}, \href
  {http://adsabs.harvard.edu/abs/2014A%26A...567A.111M} {567, A111}

\bibitem[\protect\citeauthoryear{{Mead}, {King}, {Sijacki}, {Leonard},
  {Puchwein}  \& {McCarthy}}{{Mead} et~al.}{2010}]{Mead10}
{Mead} J.~M.~G.,  {King} L.~J.,  {Sijacki} D.,  {Leonard} A.,  {Puchwein} E.,
  {McCarthy} I.~G.,  2010, \mn@doi [\mnras] {10.1111/j.1365-2966.2010.16674.x},
  406, 434

\bibitem[\protect\citeauthoryear{{Meneghetti}, {Bartelmann}  \&
  {Moscardini}}{{Meneghetti} et~al.}{2003}]{Meneghetti2003}
{Meneghetti} M.,  {Bartelmann} M.,   {Moscardini} L.,  2003, \mn@doi [\mnras]
  {10.1046/j.1365-2966.2003.07068.x}, \href
  {http://adsabs.harvard.edu/abs/2003MNRAS.346...67M} {346, 67}

\bibitem[\protect\citeauthoryear{{Meneghetti}, {Argazzi}, {Pace}, {Moscardini},
  {Dolag}, {Bartelmann}, {Li}  \& {Oguri}}{{Meneghetti}
  et~al.}{2007}]{meneghetti2007}
{Meneghetti} M.,  {Argazzi} R.,  {Pace} F.,  {Moscardini} L.,  {Dolag} K.,
  {Bartelmann} M.,  {Li} G.,   {Oguri} M.,  2007, \mn@doi [\aap]
  {10.1051/0004-6361:20065722}, \href
  {http://cdsads.u-strasbg.fr/abs/2007A%26A...461...25M} {461, 25}

\bibitem[\protect\citeauthoryear{{Meneghetti}, {Fedeli}, {Zitrin},
  {Bartelmann}, {Broadhurst}, {Gottloeber}, {Moscardini}  \&
  {Yepes}}{{Meneghetti} et~al.}{2011}]{Meneghetti2011}
{Meneghetti} M.,  {Fedeli} C.,  {Zitrin} A.,  {Bartelmann} M.,  {Broadhurst}
  T.,  {Gottloeber} S.,  {Moscardini} L.,   {Yepes} G.,  2011, arXiv, \href
  {http://adsabs.harvard.edu/abs/2011arXiv1103.0044M} {1103.0044}

\bibitem[\protect\citeauthoryear{{Meneghetti}, {Bartelmann}, {Dahle}  \&
  {Limousin}}{{Meneghetti} et~al.}{2013}]{Meneghetti2013_arc_st}
{Meneghetti} M.,  {Bartelmann} M.,  {Dahle} H.,   {Limousin} M.,  2013, \mn@doi
  [\ssr] {10.1007/s11214-013-9981-x}, \href
  {http://adsabs.harvard.edu/abs/2013SSRv..177...31M} {177, 31}

\bibitem[\protect\citeauthoryear{{Metropolis}, {Rosenbluth}, {Rosenbluth},
  {Teller}  \& {Teller}}{{Metropolis} et~al.}{1953}]{Metropolis1953}
{Metropolis} N.,  {Rosenbluth} A.~W.,  {Rosenbluth} M.~N.,  {Teller} A.~H.,
  {Teller} E.,  1953, \mn@doi [\jcp] {10.1063/1.1699114}, \href
  {http://adsabs.harvard.edu/abs/1953JChPh..21.1087M} {21, 1087}

\bibitem[\protect\citeauthoryear{{More}, {Cabanac}, {More}, {Alard},
  {Limousin}, {Kneib}, {Gavazzi}  \& {Motta}}{{More}
  et~al.}{2012}]{More_2012_arcfinder}
{More} A.,  {Cabanac} R.,  {More} S.,  {Alard} C.,  {Limousin} M.,  {Kneib}
  J.-P.,  {Gavazzi} R.,   {Motta} V.,  2012, \mn@doi [\apj]
  {10.1088/0004-637X/749/1/38}, \href
  {http://adsabs.harvard.edu/abs/2012ApJ...749...38M} {749, 38}

\bibitem[\protect\citeauthoryear{{More} et~al.,}{{More}
  et~al.}{2016}]{More_2016_SWI}
{More} A.,  et~al., 2016, \mn@doi [\mnras] {10.1093/mnras/stv1965}, \href
  {http://adsabs.harvard.edu/abs/2016MNRAS.455.1191M} {455, 1191}

\bibitem[\protect\citeauthoryear{{Oguri}}{{Oguri}}{2006}]{Oguri06_image_sep}
{Oguri} M.,  2006, \mn@doi [\mnras] {10.1111/j.1365-2966.2006.10043.x}, \href
  {http://adsabs.harvard.edu/abs/2006MNRAS.367.1241O} {367, 1241}

\bibitem[\protect\citeauthoryear{{Oguri} \& {Takada}}{{Oguri} \&
  {Takada}}{2011}]{OguriTakada2011}
{Oguri} M.,  {Takada} M.,  2011, \mn@doi [\prd] {10.1103/PhysRevD.83.023008},
  \href {http://adsabs.harvard.edu/abs/2011PhRvD..83b3008O} {83, 023008}

\bibitem[\protect\citeauthoryear{{Oguri}, {Takada}, {Umetsu}  \&
  {Broadhurst}}{{Oguri} et~al.}{2005}]{Oguri05}
{Oguri} M.,  {Takada} M.,  {Umetsu} K.,   {Broadhurst} T.,  2005, \mn@doi
  [\apj] {10.1086/452629}, 632, 841

\bibitem[\protect\citeauthoryear{{Oguri} et~al.,}{{Oguri}
  et~al.}{2006}]{Oguri06}
{Oguri} M.,  et~al., 2006, \mn@doi [\aj] {10.1086/506019}, 132, 999

\bibitem[\protect\citeauthoryear{{Oguri}, {Bayliss}, {Dahle}, {Sharon},
  {Gladders}, {Natarajan}, {Hennawi}  \& {Koester}}{{Oguri}
  et~al.}{2012}]{Oguri12}
{Oguri} M.,  {Bayliss} M.~B.,  {Dahle} H.,  {Sharon} K.,  {Gladders} M.~D.,
  {Natarajan} P.,  {Hennawi} J.~F.,   {Koester} B.~P.,  2012, \mn@doi [\mnras]
  {10.1111/j.1365-2966.2011.20248.x}, 420, 3213

\bibitem[\protect\citeauthoryear{{Puchwein} \& {Hilbert}}{{Puchwein} \&
  {Hilbert}}{2009}]{PuchweinHilbert2009}
{Puchwein} E.,  {Hilbert} S.,  2009, \mn@doi [\mnras]
  {10.1111/j.1365-2966.2009.15227.x}, \href
  {http://adsabs.harvard.edu/abs/2009MNRAS.398.1298P} {398, 1298}

\bibitem[\protect\citeauthoryear{{Puchwein}, {Bartelmann}, {Dolag}  \&
  {Meneghetti}}{{Puchwein} et~al.}{2005}]{Puchwein05}
{Puchwein} E.,  {Bartelmann} M.,  {Dolag} K.,   {Meneghetti} M.,  2005, \mn@doi
  [\aap] {10.1051/0004-6361:20053216}, \href
  {http://adsabs.harvard.edu/abs/2005A%26A...442..405P} {442, 405}

\bibitem[\protect\citeauthoryear{{Richard}, {Stark}, {Ellis}, {George},
  {Egami}, {Kneib}  \& {Smith}}{{Richard} et~al.}{2008}]{Richard08}
{Richard} J.,  {Stark} D.~P.,  {Ellis} R.~S.,  {George} M.~R.,  {Egami} E.,
  {Kneib} J.,   {Smith} G.~P.,  2008, \mn@doi [\apj] {10.1086/591312}, 685, 705

\bibitem[\protect\citeauthoryear{{Richard} et~al.,}{{Richard}
  et~al.}{2010}]{Richard2010locuss20}
{Richard} J.,  et~al., 2010, \mn@doi [\mnras]
  {10.1111/j.1365-2966.2009.16274.x}, \href
  {http://adsabs.harvard.edu/abs/2010MNRAS.404..325R} {404, 325}

\bibitem[\protect\citeauthoryear{{Rozo} et~al.,}{{Rozo}
  et~al.}{2009}]{Rozo2009MassRichSDSS}
{Rozo} E.,  et~al., 2009, \mn@doi [\apj] {10.1088/0004-637X/699/1/768}, \href
  {http://adsabs.harvard.edu/abs/2009ApJ...699..768R} {699, 768}

\bibitem[\protect\citeauthoryear{{Rozo} et~al.,}{{Rozo}
  et~al.}{2010}]{Rozo2010CosmoConstraintsSDSS}
{Rozo} E.,  et~al., 2010, \mn@doi [\apj] {10.1088/0004-637X/708/1/645}, \href
  {http://adsabs.harvard.edu/abs/2010ApJ...708..645R} {708, 645}

\bibitem[\protect\citeauthoryear{{Schechter}}{{Schechter}}{1976}]{Schechter76}
{Schechter} P.,  1976, \mn@doi [\apj] {10.1086/154079}, 203, 297

\bibitem[\protect\citeauthoryear{{Schirmer}}{{Schirmer}}{2013}]{Schirmer2013_theli}
{Schirmer} M.,  2013, \mn@doi [\apjs] {10.1088/0067-0049/209/2/21}, \href
  {http://adsabs.harvard.edu/abs/2013ApJS..209...21S} {209, 21}

\bibitem[\protect\citeauthoryear{{Schirmer}, {Erben}, {Schneider},
  {Pietrzynski}, {Gieren}, {Carpano}, {Micol}  \& {Pierfederici}}{{Schirmer}
  et~al.}{2003}]{Schirmer2003_theli}
{Schirmer} M.,  {Erben} T.,  {Schneider} P.,  {Pietrzynski} G.,  {Gieren} W.,
  {Carpano} S.,  {Micol} A.,   {Pierfederici} F.,  2003, \mn@doi [\aap]
  {10.1051/0004-6361:20031026}, \href
  {http://adsabs.harvard.edu/abs/2003A%26A...407..869S} {407, 869}

\bibitem[\protect\citeauthoryear{{Schneider}}{{Schneider}}{2014}]{Schneider2014}
{Schneider} P.,  2014, preprint, \href
  {http://adsabs.harvard.edu/abs/2014arXiv1409.0015S} {} (\mn@eprint {arXiv}
  {1409.0015})

\bibitem[\protect\citeauthoryear{{Seidel} \& {Bartelmann}}{{Seidel} \&
  {Bartelmann}}{2007}]{Seidel2007Arcfinder}
{Seidel} G.,  {Bartelmann} M.,  2007, \mn@doi [\aap]
  {10.1051/0004-6361:20066097}, \href
  {http://adsabs.harvard.edu/abs/2007A%26A...472..341S} {472, 341}

\bibitem[\protect\citeauthoryear{{Semboloni}, {Hoekstra}, {Schaye}, {van
  Daalen}  \& {McCarthy}}{{Semboloni} et~al.}{2011}]{Semboloni_2011_baryons}
{Semboloni} E.,  {Hoekstra} H.,  {Schaye} J.,  {van Daalen} M.~P.,   {McCarthy}
  I.~G.,  2011, \mn@doi [\mnras] {10.1111/j.1365-2966.2011.19385.x}, \href
  {http://adsabs.harvard.edu/abs/2011MNRAS.417.2020S} {417, 2020}

\bibitem[\protect\citeauthoryear{{Stapelberg}, {Carrasco}  \&
  {Maturi}}{{Stapelberg} et~al.}{2018}]{Stapelberg_2017_EasyCriticsI}
{Stapelberg} S.,  {Carrasco} M.,   {Maturi} M.,  2018, preprint, \href
  {http://adsabs.harvard.edu/abs/2017arXiv170909758S} {} (\mn@eprint {arXiv}
  {1709.09758})

\bibitem[\protect\citeauthoryear{{Stark}, {Ellis}, {Richard}, {Kneib}, {Smith}
  \& {Santos}}{{Stark} et~al.}{2007}]{Stark07}
{Stark} D.~P.,  {Ellis} R.~S.,  {Richard} J.,  {Kneib} J.-P.,  {Smith} G.~P.,
  {Santos} M.~R.,  2007, \mn@doi [\apj] {10.1086/518098}, 663, 10

\bibitem[\protect\citeauthoryear{{Thanjavur}}{{Thanjavur}}{2009}]{Thanjavur2009_SL}
{Thanjavur} K.~G.,  2009, PhD thesis, University of Victoria, Canada

\bibitem[\protect\citeauthoryear{{Torri}, {Meneghetti}, {Bartelmann},
  {Moscardini}, {Rasia}  \& {Tormen}}{{Torri} et~al.}{2004}]{Torri2004}
{Torri} E.,  {Meneghetti} M.,  {Bartelmann} M.,  {Moscardini} L.,  {Rasia} E.,
   {Tormen} G.,  2004, \mn@doi [\mnras] {10.1111/j.1365-2966.2004.07508.x},
  \href {http://adsabs.harvard.edu/abs/2004MNRAS.349..476T} {349, 476}

\bibitem[\protect\citeauthoryear{{Umetsu} \& {Broadhurst}}{{Umetsu} \&
  {Broadhurst}}{2008}]{UmetsuBroadhurst08}
{Umetsu} K.,  {Broadhurst} T.,  2008, \mn@doi [\apj] {10.1086/589683}, 684, 177

\bibitem[\protect\citeauthoryear{{Verdugo}, {Motta}, {Mu{\~n}oz}, {Limousin},
  {Cabanac}  \& {Richard}}{{Verdugo} et~al.}{2011}]{Verdugo11}
{Verdugo} T.,  {Motta} V.,  {Mu{\~n}oz} R.~P.,  {Limousin} M.,  {Cabanac} R.,
  {Richard} J.,  2011, \mn@doi [\aap] {10.1051/0004-6361/201014965}, \href
  {http://adsabs.harvard.edu/abs/2011A%26A...527A.124V} {527, A124}

\bibitem[\protect\citeauthoryear{{Verdugo} et~al.,}{{Verdugo}
  et~al.}{2016}]{Verdugo16}
{Verdugo} T.,  et~al., 2016, \mn@doi [\aap] {10.1051/0004-6361/201628629},
  \href {http://adsabs.harvard.edu/abs/2016A%26A...595A..30V} {595, A30}

\bibitem[\protect\citeauthoryear{{Wambsganss}, {Bode}  \&
  {Ostriker}}{{Wambsganss} et~al.}{2004}]{Wambsganss2004_arc_st}
{Wambsganss} J.,  {Bode} P.,   {Ostriker} J.~P.,  2004, \mn@doi [\apjl]
  {10.1086/421459}, \href {http://adsabs.harvard.edu/abs/2004ApJ...606L..93W}
  {606, L93}

\bibitem[\protect\citeauthoryear{{Wambsganss}, {Bode}  \&
  {Ostriker}}{{Wambsganss} et~al.}{2005}]{Wambsganss2005_LOS}
{Wambsganss} J.,  {Bode} P.,   {Ostriker} J.~P.,  2005, \mn@doi [\apjl]
  {10.1086/498976}, \href {http://adsabs.harvard.edu/abs/2005ApJ...635L...1W}
  {635, L1}

\bibitem[\protect\citeauthoryear{{Wambsganss}, {Ostriker}  \&
  {Bode}}{{Wambsganss} et~al.}{2008}]{Wambsganss2008_arc_st}
{Wambsganss} J.,  {Ostriker} J.~P.,   {Bode} P.,  2008, \mn@doi [\apj]
  {10.1086/527529}, \href {http://adsabs.harvard.edu/abs/2008ApJ...676..753W}
  {676, 753}

\bibitem[\protect\citeauthoryear{{White} et~al.,}{{White}
  et~al.}{2011}]{White2011_clustering_massive_gal}
{White} M.,  et~al., 2011, \mn@doi [\apj] {10.1088/0004-637X/728/2/126}, \href
  {http://adsabs.harvard.edu/abs/2011ApJ...728..126W} {728, 126}

\bibitem[\protect\citeauthoryear{{Wong}, {Ammons}, {Keeton}  \&
  {Zabludoff}}{{Wong} et~al.}{2012}]{Wong2012_LOS}
{Wong} K.~C.,  {Ammons} S.~M.,  {Keeton} C.~R.,   {Zabludoff} A.~I.,  2012,
  arXiv, 1203.2614

\bibitem[\protect\citeauthoryear{{Wong}, {Zabludoff}, {Ammons}, {Keeton},
  {Hogg}  \& {Gonzalez}}{{Wong} et~al.}{2013}]{Wong2013_LRG}
{Wong} K.~C.,  {Zabludoff} A.~I.,  {Ammons} S.~M.,  {Keeton} C.~R.,  {Hogg}
  D.~W.,   {Gonzalez} A.~H.,  2013, \mn@doi [\apj]
  {10.1088/0004-637X/769/1/52}, \href
  {http://adsabs.harvard.edu/abs/2013ApJ...769...52W} {769, 52}

\bibitem[\protect\citeauthoryear{{Zehavi} et~al.,}{{Zehavi}
  et~al.}{2005}]{Zehavi2005_LRG}
{Zehavi} I.,  et~al., 2005, \mn@doi [\apj] {10.1086/427495}, \href
  {http://adsabs.harvard.edu/abs/2005ApJ...621...22Z} {621, 22}

\bibitem[\protect\citeauthoryear{{Zheng} et~al.,}{{Zheng}
  et~al.}{2012}]{Zheng2012NaturZ}
{Zheng} W.,  et~al., 2012, \mn@doi [\nat] {10.1038/nature11446}, \href
  {http://adsabs.harvard.edu/abs/2012Natur.489..406Z} {489, 406}

\bibitem[\protect\citeauthoryear{{Zitrin} et~al.,}{{Zitrin}
  et~al.}{2009}]{Zitrin09a}
{Zitrin} A.,  et~al., 2009, \mn@doi [\mnras]
  {10.1111/j.1365-2966.2009.14899.x}, 396, 1985

\bibitem[\protect\citeauthoryear{{Zitrin}, {Broadhurst}, {Barkana}, {Rephaeli}
  \& {Ben{\'{\i}}tez}}{{Zitrin} et~al.}{2011}]{Zitrin11MACS}
{Zitrin} A.,  {Broadhurst} T.,  {Barkana} R.,  {Rephaeli} Y.,
  {Ben{\'{\i}}tez} N.,  2011, \mn@doi [\mnras]
  {10.1111/j.1365-2966.2010.17574.x}, 410, 1939

\bibitem[\protect\citeauthoryear{{Zitrin}, {Broadhurst}, {Bartelmann},
  {Rephaeli}, {Oguri}, {Ben{\'{\i}}tez}, {Hao}  \& {Umetsu}}{{Zitrin}
  et~al.}{2012a}]{Zitrin2012UniversalRE}
{Zitrin} A.,  {Broadhurst} T.,  {Bartelmann} M.,  {Rephaeli} Y.,  {Oguri} M.,
  {Ben{\'{\i}}tez} N.,  {Hao} J.,   {Umetsu} K.,  2012a, \mn@doi [\mnras]
  {10.1111/j.1365-2966.2012.21041.x}, \href
  {http://adsabs.harvard.edu/abs/2012MNRAS.423.2308Z} {423, 2308}

\bibitem[\protect\citeauthoryear{{Zitrin} et~al.,}{{Zitrin}
  et~al.}{2012b}]{Zitrin2012CLASH0329}
{Zitrin} A.,  et~al., 2012b, \mn@doi [\apjl] {10.1088/2041-8205/747/1/L9},
  \href {http://adsabs.harvard.edu/abs/2012ApJ...747L...9Z} {747, L9}

\bibitem[\protect\citeauthoryear{{Zitrin} et~al.,}{{Zitrin}
  et~al.}{2013}]{Zitrin2013_gauss}
{Zitrin} A.,  et~al., 2013, \mn@doi [\apjl] {10.1088/2041-8205/762/2/L30},
  \href {http://adsabs.harvard.edu/abs/2013ApJ...762L..30Z} {762, L30}

\end{thebibliography}

%% Alternatively you could enter them by hand, like this:
%% This method is tedious and prone to error if you have lots of references

%%%%%%%%%%%%%%%%%%%%%%%%%%%%%%%%%%%%%%%%%%%%%%%%%%

%%%%%%%%%%%%%%%%% APPENDICES %%%%%%%%%%%%%%%%%%%%%
\onecolumn
\appendix

\section{Reference samples} \label{app:reference_sample}

%%%%%%%%%%%%%%%%%%%%%%%%%%%%%%%%%%%%%%%%%%%%%%%%%%%%%%5
%%%%%%%%%%%%%%%%%%%%%%%%%%%%%%%%%%%%%%%%%%%%%%%%%%%%%%%%
% Table of the reference sample of known lenses (KLs) and known candidates (KCs)
% %%%%%%%%%%%%%%%%%%%%%%%%%%%%%%%%
\begin{center}
\begin{longtable}{l c c c c c c}
\caption{Reference sample of known lenses (KLs) and known candidates (KCs).  \label{table:app:reference_sample}} \\
\toprule \toprule
 Name$^a$              &  R.A.       &   Dec.       &$z_\mathrm{phot}$ & $R_\mathrm{arc}^b$ & Rank$^c$ & Reference$^d$    \\
                    &  [J2000]    &   [J2000]    &       & [$''$] &     \\
\midrule
\endfirsthead
\multicolumn{6}{c}{\tablename\ \thetable\ -- \textit{Continued from previous page}} \\
\toprule \toprule
 Name$^a$              &  R.A.       &   Dec.       &$z_\mathrm{phot}$ & $R_\mathrm{arc}^b$ & Rank$^c$ & Reference$^d$    \\
                    &  [J2000]    &   [J2000]    &       & [$''$] &     \\
\midrule
\endhead   
\midrule 
\multicolumn{6}{r}{\textit{Continued on next page}} \\
\endfoot
\midrule
\endlastfoot
  SA2               & 02:02:10.50 & -11:09:11.7  & 0.48  &  5.0   &  KL & Mo12   \\
  SA6               & 02:03:20.43 & -07:34:50.8  & 0.59  &  5.0   &  KL & Mo12   \\ 
  SA7               & 02:03:49.98 & -09:42:53.5  & 0.25  &  5.0   &  KL & Mo12   \\
  SA9               & 02:05:03.15 & -11:05:46.6  & 0.62  &  3.3   &  KL & Mo12   \\
  SA10              & 02:06:48.47 & -06:57:01.3  & 0.49  &  3.2   &  KL & Mo12   \\
  SA12              & 02:08:16.87 & -09:36:52.7  & 0.74  &  3.4   &  KL & Mo12   \\
  SA14              & 02:09:29.33 & -06:43:11.3  & 0.45  &  3.2   &  KL & Mo12   \\
  SA22              & 02:14:08.07 & -05:35:32.4  & 0.44  &  7.1   &  KL & Mo12   \\
  arc68c            & 02:15:29.40 & -04:40:54.0  & 0.31  &  8.0   &  KL & Ma14   \\
  SA30              & 02:16:49.25 & -07:03:23.8  & 0.43  &  5.6   &  KL & Mo12   \\
  SA33              & 02:18:07.29 & -05:15:36.2  & 0.64  &  3.1   &  KL & Mo12   \\
  SA36              & 02:19:56.42 & -05:27:59.2  & 0.35  &  4.0   &  KL & Mo12   \\
  SA39              & 02:21:51.18 & -06:47:32.7  & 0.61  &  5.2   &  KL & Mo12   \\
  SW1               & 02:24:09.55 & -10:58:07.9  & 0.50  &  4.8   &  KL & Mo16   \\
  SW21              & 02:25:33.32 & -05:32:04.6  & 0.50  &  3.6   &  KL & Mo16   \\
  SA50              & 02:25:46.13 & -07:37:38.5  & 0.51  &  5.8   &  KL & Mo12   \\
  LSSC017           & 02:26:28.18 & -04:59:48.1  & 0.38  &  5.0   &  KL & Th09   \\
  SW22              & 02:27:16.45 & -10:56:02.8  & 0.40  &  4.8   &  KL & Mo16   \\
  XLSSC022          & 02:27:40.27 & -04:51:31.0  & 0.29  &  5.0   &  KL & Th09   \\
  SA55              & 02:29:17.36 & -05:54:05.5  & 0.38  &  3.2   &  KL & Mo16   \\
  SA63              & 08:52:07.18 & -03:43:16.3  & 0.48  &  5.0   &  KL &  Mo12  \\
  SA66              & 08:54:46.55 & -01:21:37.1  & 0.35  &  4.8   &  KL &  Mo12  \\
  SA71              & 08:58:48.83 & -02:39:25.8  & 0.36  &  3.7   &  KL &  Mo12  \\
  SA72              & 08:59:14.55 & -03:45:14.9  & 0.64  &  4.5   &  KL &  Mo12  \\
  SL2SJ0901-0158    & 09:01:39.46 & -01:58:52.2  & 0.29  &  6.8   &  KL & Ca07  \\
  SA87              & 13:57:25.48 & +53:17:44.0  & 0.54  &  3.5   &  KL &  Mo12  \\
  SA90              & 14:01:10.46 & +56:54:20.5  & 0.53  &  3.7   &  KL &  Mo12  \\
  SA96              & 14:05:54.33 & +54:45:48.7  & 0.41  &  3.9   &  KL &  Mo12  \\
  SA97              & 14:08:13.82 & +54:29:08.1  & 0.42  &  8.0   &  KL &  Mo12  \\
  SA100             & 14:14:47.19 & +54:47:03.6  & 0.63  &  14.7  &  KL & Mo12   \\
  SA102             & 14:19:12.17 & +53:26:11.4  & 0.69  &  9.9   &  KL &   Mo12 \\
  SA103             & 14:19:17.25 & +51:17:28.6  & 0.47  &  4.1   &  KL &   Mo12 \\
  SW3               & 14:26:03.30 & +51:14:21.7  & 0.50  &  4.4   &  KL &   Mo16 \\
  SW4               & 14:29:34.23 & +56:25:41.1  & 0.50  &  5.9   &  KL &   Mo16 \\
  SA112             & 14:30:00.65 & +55:46:48.0  & 0.50  &  4.3   &  KL &  Mo12  \\
  SA113             & 14:31:39.77 & +55:33:22.8  & 0.67  &  4.0   &  KL &   Mo12 \\
  SW5               & 14:34:54.48 & +52:28:50.9  & 0.60  &  4.4   &  KL &    Mo16\\
  SW7               & 22:02:56.86 & +02:34:32.9  & 0.50  &  6.8   &  KL &    Mo16\\
  SA121             & 22:06:42.03 & +04:11:30.8  & 0.62  &  3.7   &  KL &   Mo12 \\
  arc20a            & 22:13:06.90 & -00:18:30.0  & 0.49  &  5.0   &  KL &    Ma14\\
  SA122             & 22:13:06.93 & -00:30:37.1  & 0.69  &  3.1   &  KL &   Mo12 \\
  SA123             & 22:13:31.85 & +00:48:36.1  & 1.00  &  4.8   &  KL &  Mo12  \\
  SA125             & 22:14:18.82 & +01:10:33.8  & 0.74  &  8.0   &  KL &  Mo12  \\
  arc54c            & 22:20:51.50 & +00:58:14.0  & 0.41  &  8.0   &  KL &   Ma14 \\
  SA8               & 02:04:54.51 & -10:24:02.5  & 0.33  &  10.8  &  KC &    Mo12\\
  SA11              & 02:08:15.66 & -07:24:57.8  & 0.62  &  4.3   &  KC &   Mo12 \\
  SA13              & 02:08:41.61 & -07:01:28.1  & 0.29  &  3.5   &  KC &   Mo12 \\
  SA15              & 02:09:57.67 & -03:54:57.1  & 0.44  &  3.9   &  KC &   Mo12 \\
  SA19              & 02:11:18.49 & -04:27:29.2  & 1.19  &  3.5   &  KC &   Mo12 \\
  arc1a             & 02:13:17.20 & -06:25:58.0  & 0.39  &  4.0   &  KC &    Ma14\\
  arc67c            & 02:13:28.40 & -05:11:45.0  & 0.49  &  8.5   &  KC &   Ma14 \\
  arc26b            & 02:14:26.40 & -05:39:39.0  & 0.55  &  5.0   &  KC &   Ma14 \\
  SA24              & 02:15:23.03 & -07:36:23.6  & 1.05  &  3.7   &  KC &   Mo12 \\
  SA26              & 02:16:04.66 & -09:35:06.6  & 0.69  &  16.4  &  KC &  Mo12  \\
  SA31              & 02:17:23.76 & -10:15:50.3  & 0.27  &  3.2   &  KC &   Mo12 \\
  SA35              & 02:19:09.86 & -04:01:43.3  & 0.45  &  4.3   &  KC &   Mo12 \\
  SA41              & 02:23:18.33 & -10:58:48.5  & 0.52  &  6.1   &  KC &   Mo12 \\
  SW32              & 02:23:59.89 & -08:36:51.8  & 0.50  &  3.1   &  KC &   Mo16 \\
  SA43              & 02:24:05.01 & -04:47:07.0  & 0.36  &  4.3   &  KC &    Mo12\\
  SA45              & 02:24:35.26 & -04:01:57.9  & 1.13  &  3.5   &  KC &   Mo12 \\
  SA46              & 02:24:39.06 & -04:00:45.2  & 0.43  &  3.2   &  KC &   Mo12 \\
  SA49              & 02:25:38.74 & -04:03:20.4  & 0.62  &  4.3   &  KC &   Mo12 \\
  SA51              & 02:26:07.15 & -04:27:26.3  & 0.17  &  3.7   &  KC &   Mo12 \\
  SA53              & 02:27:59.21 & -09:07:29.9  & 0.55  &  3.9   &  KC &   Mo12 \\
  SA54              & 02:28:32.05 & -09:49:45.4  & 0.45  &  6.3   &  KC &   Mo12 \\
  SL2SJ0230-0550    & 02:30:11.60 & -05:50:21.0  & 0.49  &  7.0   &  KC &  Li09   \\
  SA57              & 02:31:06.46 & -05:55:04.6  & 0.52  &  3.7   &  KC &   Mo12 \\
  SA60              & 02:35:01.61 & -09:58:32.8  & 0.70  &  4.7   &  KC &   Mo12 \\
  SA61              & 08:48:23.66 & -04:07:15.3  & 0.51  &  7.4   &  KC &   Mo12 \\
  SA62              & 08:50:07.72 & -01:23:53.3  & 0.37  &  3.5   &  KC &   Mo12 \\
  SA70              & 08:57:49.10 & -01:13:00.7  & 0.29  &  3.9   &  KC &   Mo12 \\
  SA73              & 08:59:54.54 & -01:32:13.4  & 0.66  &  4.3   &  KC &   Mo12 \\
  SA74              & 09:00:50.10 & -02:30:54.1  & 0.36  &  3.2   &  KC &   Mo12 \\
  SA86              & 13:56:49.33 & +55:27:07.0  & 0.46  &  3.7   &  KC &   Mo12 \\
  SA89              & 14:00:40.17 & +56:07:49.4  & 0.42  &  3.7   &  KC &   Mo12 \\
  arc81c            & 14:02:06.40 & +52:57:07.0  & 0.51  &  5.0   &  KC &   Mo12 \\
  SA93              & 14:02:47.90 & +57:08:52.0  & 1.22  &  3.2   &  KC &   Mo12 \\
  SA98              & 14:11:20.53 & +52:12:09.9  & 0.52  &  18.4  &  KC &    Mo12\\
  SL2SJ1415+5239    & 14:15:58.18 & +52:39:55.9  & 0.75  &  4.6   &  KC &   Li09 \\
  SA101             & 14:16:44.52 & +56:42:16.2  & 1.29  &  3.5   &  KC &   Mo12 \\
  SA104             & 14:21:02.56 & +52:29:42.5  & 0.18  &  11.7  &  KC &  Mo12  \\
  SL2SJ1422+5246    & 14:22:09.27 & +52:46:52.4  & 0.18  &  3.5   &  KC &   Li09 \\
  SA108             & 14:25:44.27 & +57:07:24.5  & 0.86  &  4.5   &  KC &  Mo12  \\
  SA109             & 14:26:08.04 & +57:45:23.9  & 0.39  &  3.2   &  KC &  Mo12  \\
  SA110             & 14:28:10.54 & +56:39:48.4  & 0.80  &  4.1   &  KC &  Mo12  \\
  SA111             & 14:28:34.82 & +52:13:06.4  & 0.52  &  5.0   &  KC &  Mo12  \\
  SL2SJ1431+5131    & 14:31:41.83 & +51:31:43.7  & 0.85  &  3.9   &  KC &  Li09  \\
  SA114             & 14:31:52.67 & +57:28:36.7  & 0.83  &  3.5   &  KC &   Mo12 \\
  SA116             & 14:34:34.69 & +56:59:20.2  & 0.57  &  4.1   &  KC &   Mo12 \\
  SW58              & 14:36:51.61 & +53:07:06.0  & 0.60  &  3.1   &  KC &   Mo16 \\
  SA117             & 22:01:51.79 & +04:10:08.4  & 0.43  &  7.3   &  KC &   Mo12 \\
  SA118             & 22:02:01.66 & +01:47:09.6  & 0.30  &  5.0   &  KC &   Mo12 \\
  SW39              & 22:02:15.23 & +01:21:24.0  & 0.30  &  4.6   &  KC &    Mo16\\
  arc84c            & 22:09:35.50 & +00:31:26.0  & 0.32  &  3.3   &  KC &    Ma14\\
  arc53c            & 22:10:33.10 & +00:23:51.0  & 0.58  &  4.0   &  KC &   Ma14 \\
  arc23a            & 22:15:13.40 & +01:02:41.0  & 0.69  &  7.0   &  KC &   Ma14 \\
  SA127             & 22:21:43.74 & -00:53:02.9  & 0.33  &  4.7   &  KC &   Mo12 \\
  arc55c            & 22:21:58.50 & +00:59:02.0  & 0.33  &  3.5   &  KC &   Mo14 \\                                                                 
\bottomrule
\end{longtable}                                       
{\footnotesize\flushleft
{}$^a$ IDs given to the previously known lenses in their corresponding references. \\
{}$^b$ Corresponds to the distance from the BCG of the lens system
till the average location of the arc. \\
{}$^c$ Ranking from previous studies. KL stands for known lenses, while KC stands for known candidates. \\
{}$^d$ Ca07 stands for \cite{Cabanac07}; Li09 stands for \cite{Limousin09_SL_Ggroup};
Th09 stands for \cite{Thanjavur2009_SL}; Mo12 stands for \cite{More_2012_arcfinder};
Ma14 stands for \cite{Maturi14}; and Mo16 stands for \cite{More_2016_SWI}. \\
}   
\end{center}
%%%%%%%%%%%%%%%%%%%%%%%%%%%%%%%%%%%%%%%%%%%%%%%%%%%%%%%%
%%%%%%%%%%%%%%%%%%%%%%%%%%%%%%%%%%%%%%%%%%%%%%%%%%%%%%%%

\newpage
\section{New  SL candidates} \label{app:new_SL_candidates}
In this appendix we present the 9 of the new systems found in our visual inspection analysis, 
after processing the survey with \textit{EasyCritics}. These systems are classified with a final $rank \geq 2$, 
given the scale defined in \cite{More_2016_SWI}.

%%%%%%%%%%%%%%%%%%%%%%%%%%%%%%%%
% New tentative SL candidates from EasyCritics
%%%%%%%%%%%%%%%%%%%%%%%%%%%%%%%%
\begin{figure}
\begin{center}
\begin{tabular}{c c c}
\includegraphics[width=50mm, height=50mm]{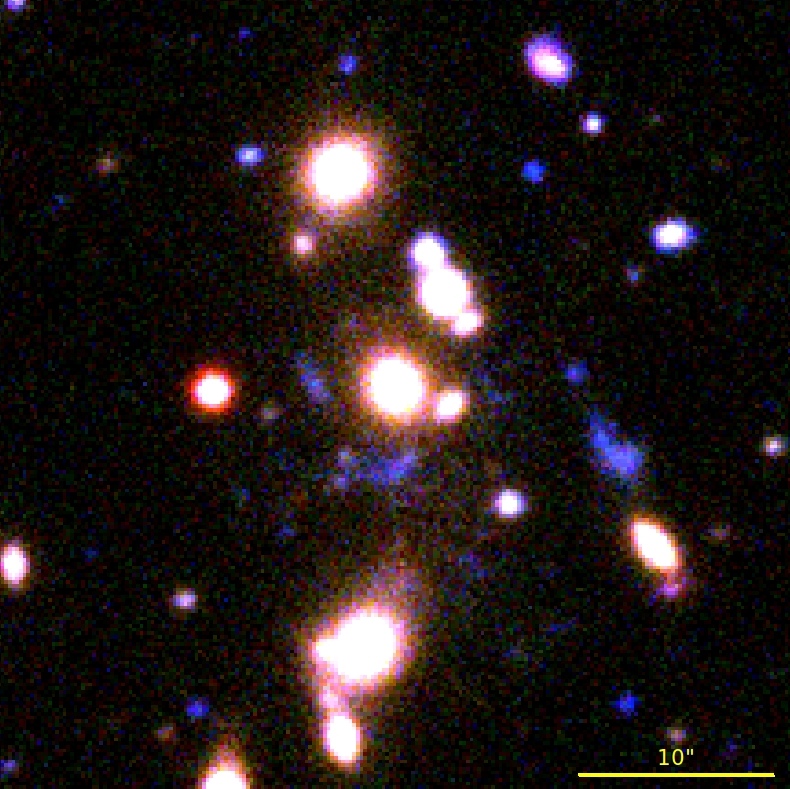} &
\includegraphics[width=50mm, height=50mm]{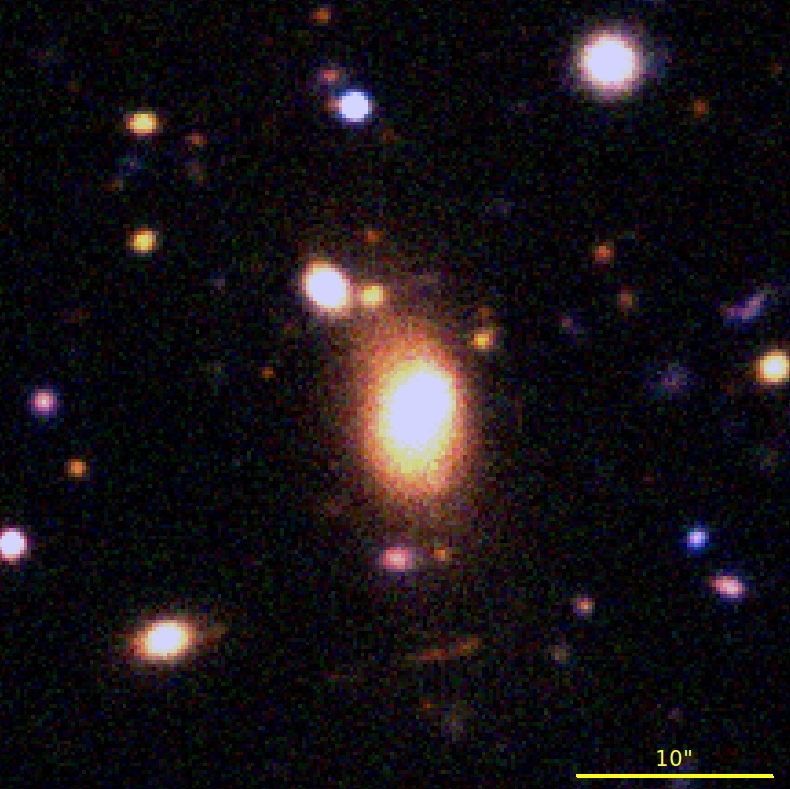} &
\includegraphics[width=50mm, height=50mm]{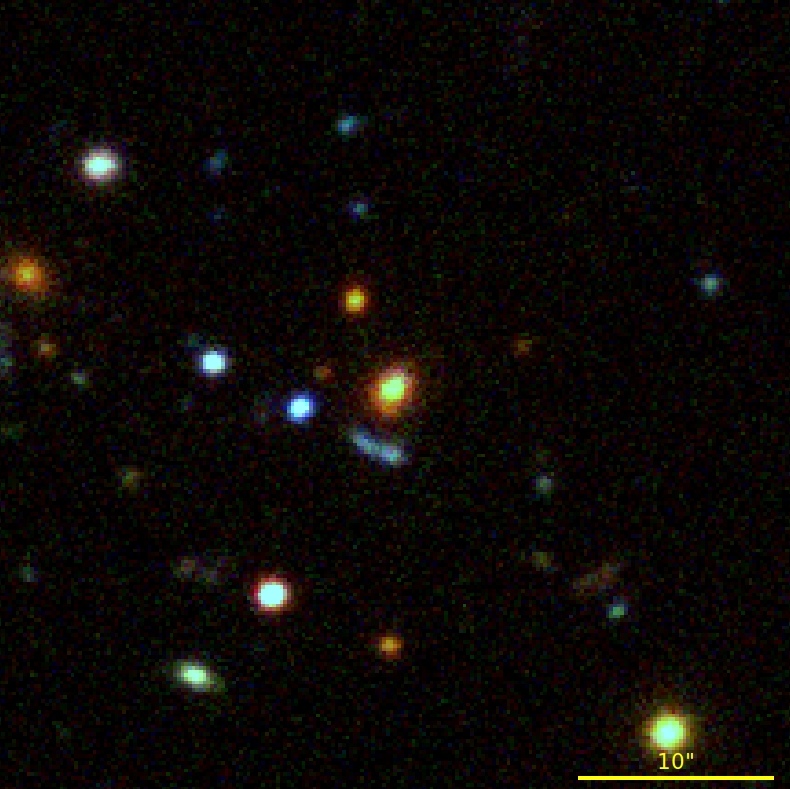} \\
\includegraphics[width=50mm, height=50mm]{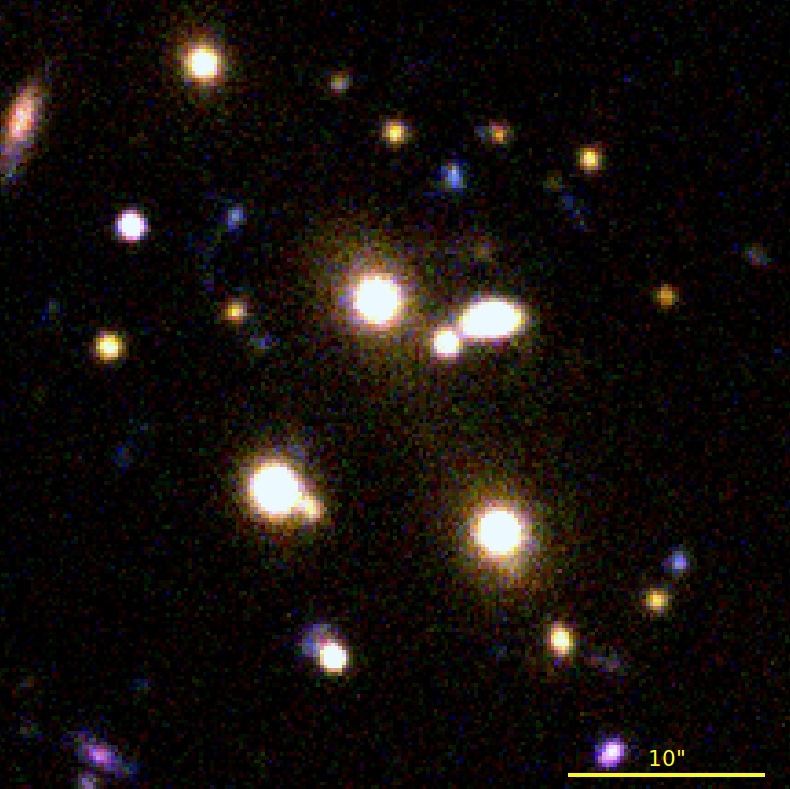} & %[3ex]
\includegraphics[width=50mm, height=50mm]{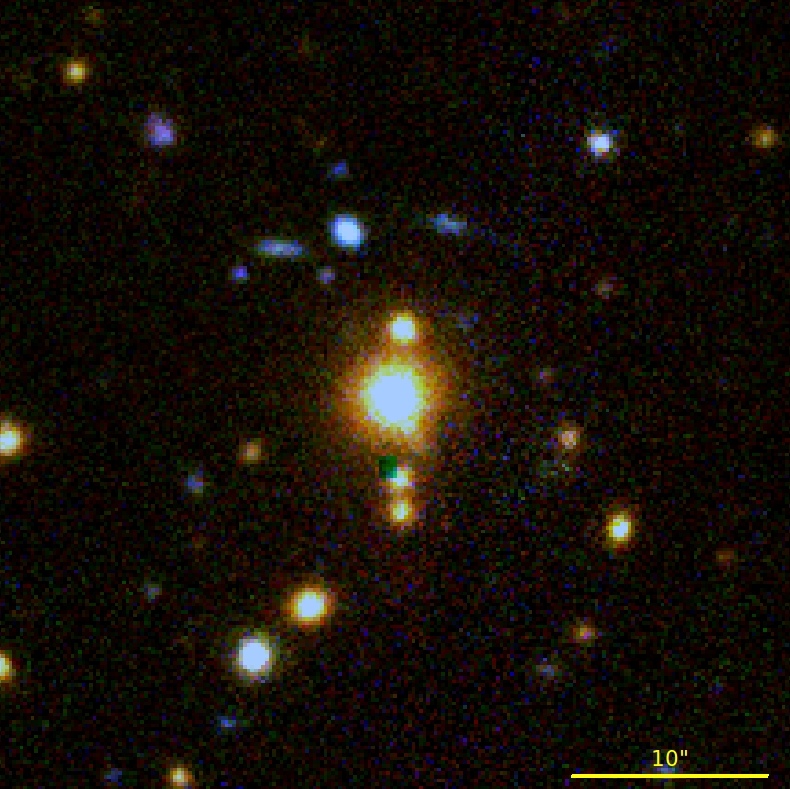} &
\includegraphics[width=50mm, height=50mm]{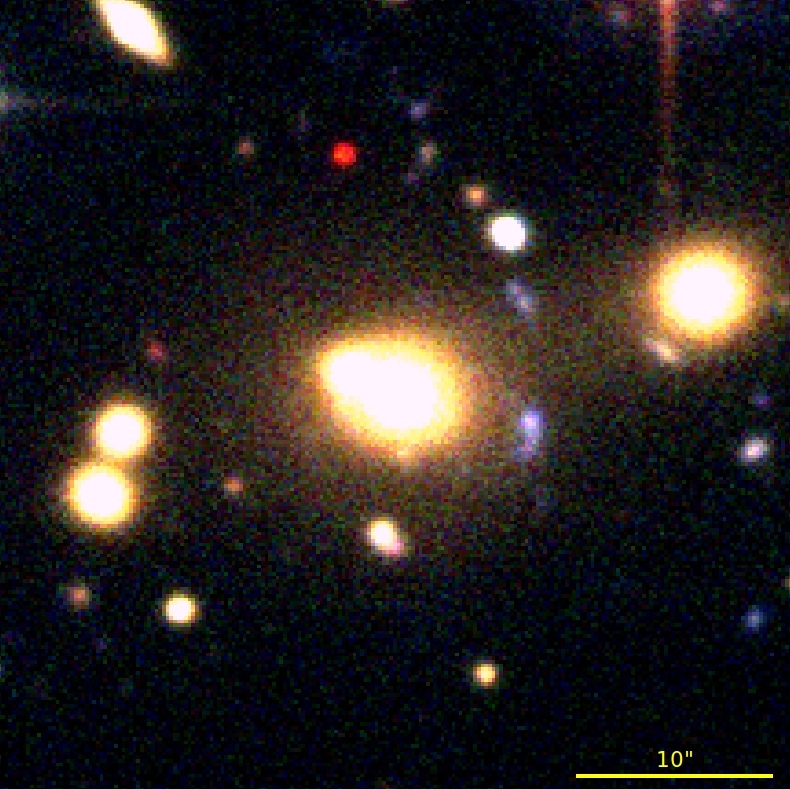} \\
\includegraphics[width=50mm, height=50mm]{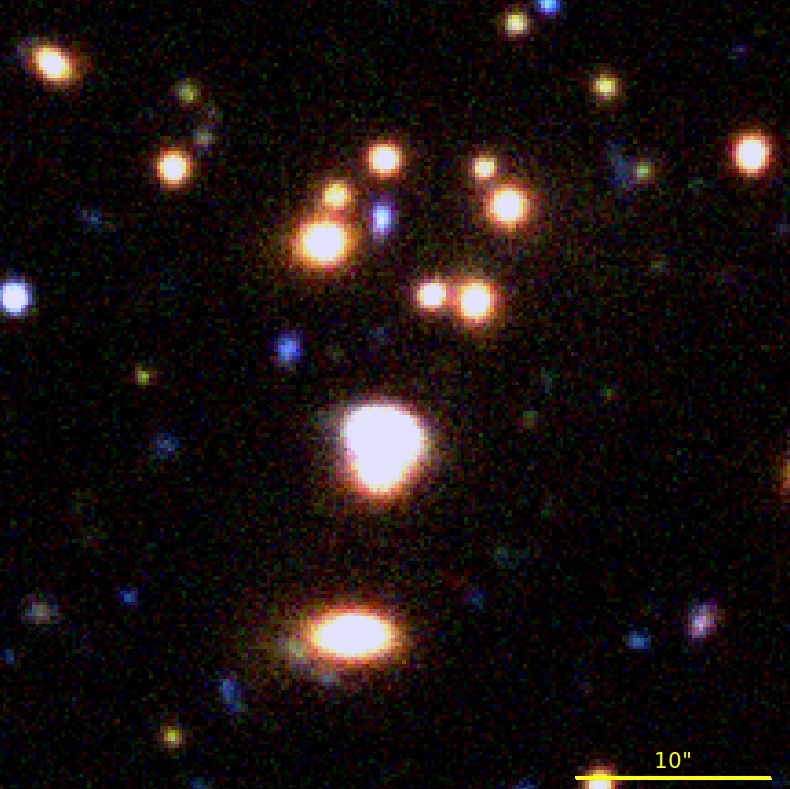} & %[3e
\includegraphics[width=50mm, height=50mm]{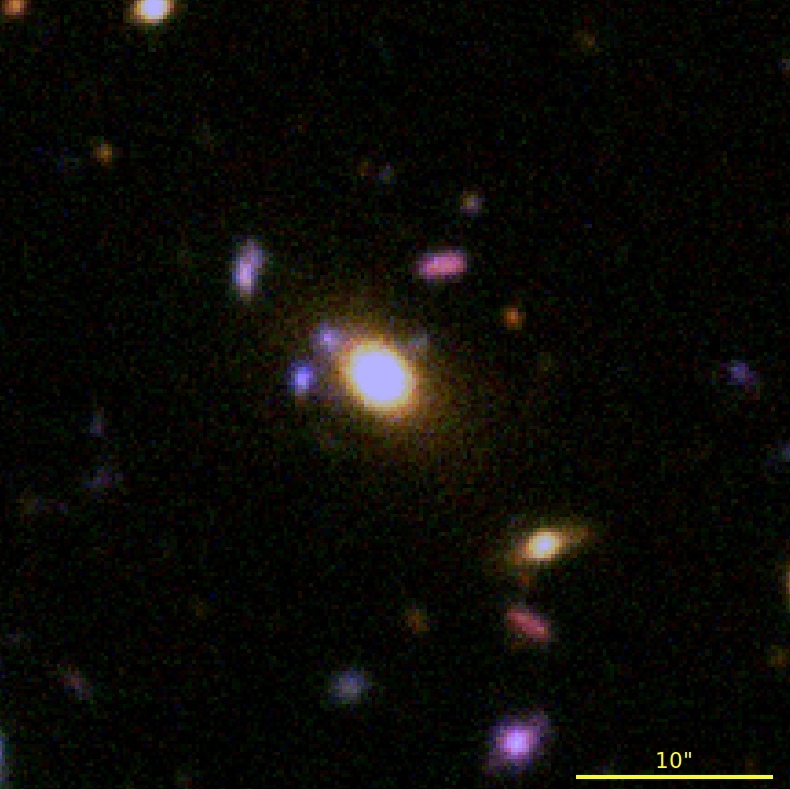} &
\includegraphics[width=50mm, height=50mm]{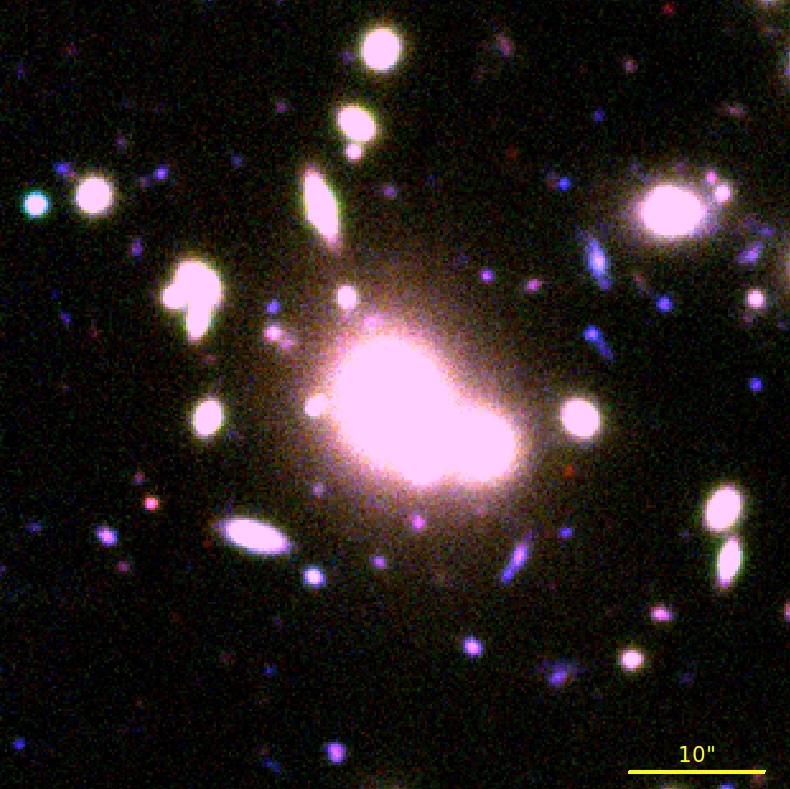} \\
\end{tabular}                                                                      
\put(-430, 210){\bf \large \color{white}  SLEC-J0211-0609}                                   
\put(-280, 210){\bf \large \color{white}  SLEC-J1405+5356}                                   
\put(-120, 210){\bf \large \color{white}   SLEC-J0211-0422}
\put(-430, 65){\bf \large \color{white}   SLEC-J0213-0951}
\put(-280, 65){\bf \large \color{white} SLEC-J0204-1017}
\put(-120, 65){\bf \large \color{white} SLEC-J2220+0058}
\put(-430, -80){\bf \large \color{white} SLEC-J0233-0530}
\put(-280, -80){\bf \large \color{white} SLEC-J0216-0558}
\put(-120, -80){\bf \large \color{white} SLEC-J0212-0820}
\caption{\label{Fig:new_SL_candidates}
New promising SL candidates pre-selected by \textit{EasyCritics} showing   SL features, which are
classified with a final $rank \geq 2$. The coordinates of these objects and the most promising candidates
are available as supplementary material; although, a portion of this
new sample is shown in Table \ref{table:new_candidates}. 
These color composite images are generated by using the $g'$-, $r'$-, and $i'$-band 
imaging data from CFHTLenS, centered on the candidate centers and 
covering different areas from $40''\times40''$ up to $60''\times60''$.
} 
\end{center}
\end{figure}
%%%%%%%%%%%%%%%%%%%%%%%%%%%%%%%% 

% Don't change these lines
\bsp	% typesetting comment
\label{lastpage}
\end{document}